\documentclass{egpubl}
\usepackage{eg2026}

\ConferencePaper
\CGFStandardLicense

\makeatletter
\def\@oddfoot{}
\def\@evenfoot{}
\def\ps@titlepage{\let\@mkboth\@gobbletwo
  \def\@oddhead{}%
  \def\@oddfoot{}%
  \let\@evenhead=\@oddhead
  \let\@evenfoot=\@oddfoot
}
\def\ps@headings{\let\@mkboth\markboth
  \def\@oddhead{\small \hfill \itshape \@shortauthor\ / \@shorttitle
                \hfill \upshape \textrm{\thepage}}%
  \def\@oddfoot{}%
  \def\@evenhead{\small \upshape \textrm{\thepage} \hfill
                 \itshape \@shortauthor\ / \@shorttitle \hfill}%
  \def\@evenfoot{}%
  \def\sectionmark##1{}%
  \def\subsectionmark##1{}%
}
\makeatother

\usepackage[T1]{fontenc}
\usepackage{dfadobe}

\BibtexOrBiblatex
\electronicVersion
\PrintedOrElectronic

\ifpdf \usepackage[pdftex]{graphicx} \pdfcompresslevel=9
\else \usepackage[dvips]{graphicx} \fi

\usepackage{egweblnk}

\usepackage{placeins}
\usepackage{booktabs}
\usepackage{tikz}
\usepackage{xspace}
\usepackage{adjustbox}
\usepackage{subcaption}
\usepackage{float}
\usepackage[percent]{overpic}
\usepackage{wrapfig}
\usepackage{multirow}

\title[Real-time Rendering with a Neural Irradiance Volume]%
      {Real-time Rendering with a Neural Irradiance Volume}

\author[Coomans \& Nazzaro et al.]
{\parbox{\textwidth}{\centering
Arno Coomans$^{*1}$\orcid{0000-0002-7396-5044} \ \
Giacomo Nazzaro$^{*1}$\orcid{0009-0002-3647-1659} \ \
Edoardo A. Dominici$^{1}$\orcid{0009-0007-9101-7279} \ \
Christian D\"oring$^{2}$\orcid{0009-0007-4763-8748} \ \ \\
Floor Verhoeven$^{1}$\orcid{0000-0003-3768-0460} \ \
Konstantinos Vardis$^{1}$\orcid{0000-0003-2282-4644} \ \
Markus Steinberger$^{3,4}$\orcid{0000-0001-5977-8536}
        }
        \\
{\parbox{\textwidth}{\centering Huawei Technologies, Switzerland$^1$, Germany$^2$, Austria$^3$\\
         Graz University of Technology, Austria$^4$
       }
}
}

\begin{document}

\teaser{
 \includegraphics[width=\linewidth]{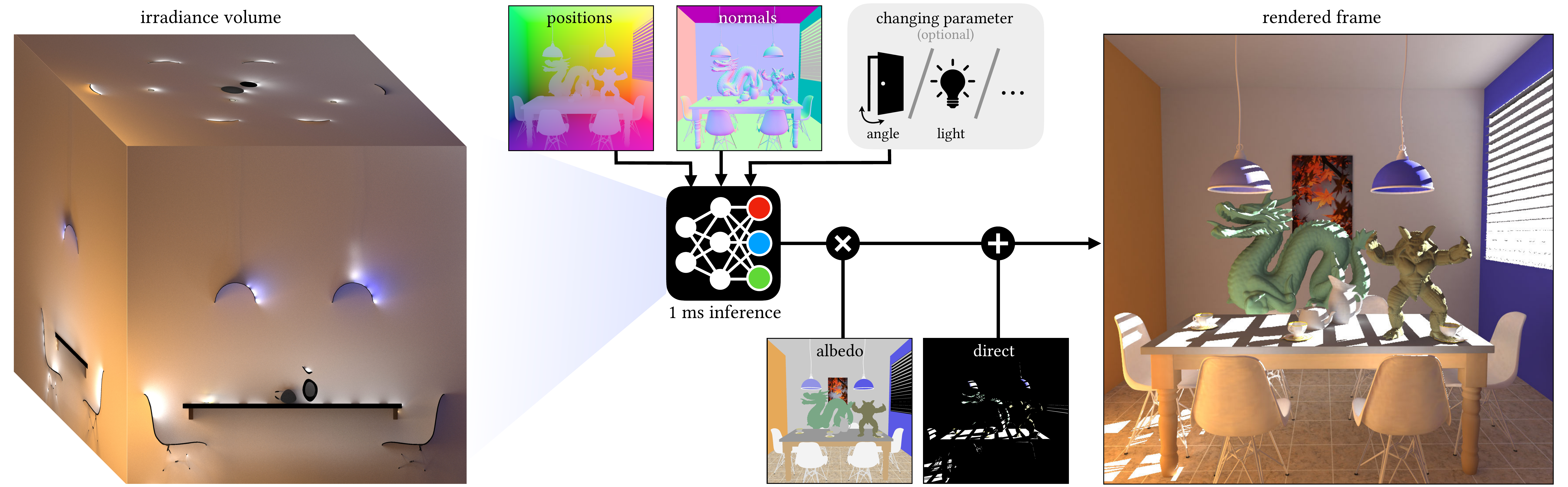}
 \centering
  \caption{Overview of rendering dynamic objects in a static scene using our \textit{Neural Irradiance Volume}. Left: The irradiance field of the scene
   is visualized through three orthogonal
   slices of its volume. Center: A lightweight neural model compresses the 5D irradiance field for the whole scene's volume.
   By querying the model with position and normal buffers as input, indirect diffuse illumination is computed at real-time rates.
   Right: The model's output is efficiently utilized to compute high-quality diffuse global illumination, not only on static surfaces but also
   on novel dynamic objects---such as the \textit{Dragon} and \textit{Armadillo} above---which are effectively immersed
   in the neural irradiance volume. Optionally, the model can be trained with changing scene parameters, obtaining non-static irradiance fields.}
\label{fig:teaser}
}

\maketitle

\begin{abstract}
Rendering diffuse global illumination in real-time is often approximated by
pre-computing and storing irradiance in a 3D grid of probes. As long as most of
the scene remains static,
probes approximate irradiance for all surfaces
immersed in the irradiance volume, including novel dynamic objects. This approach,
however, suffers from aliasing artifacts and high memory consumption.
We propose Neural Irradiance Volume (NIV), a neural-based technique that
allows accurate real-time rendering of diffuse global illumination via a compact pre-computed model, overcoming the
limitations of traditional probe-based methods, such as the expensive memory
footprint, aliasing artifacts, and scene-specific heuristics.
The key insight is that neural compression creates an adaptive and amortized
representation of irradiance, circumventing the cubic scaling of grid-based methods.
Our superior memory-scaling improves quality by at least 10x
at the same memory budget, and enables a straightforward representation of higher-dimensional
irradiance fields, allowing rendering of time-varying or dynamic effects without
requiring additional computation at runtime.
Unlike other neural rendering techniques, our method works within strict
real-time constraints, providing fast inference (around 1 ms per frame on consumer GPUs at full HD resolution), reduced memory usage (1–5 MB for
medium-sized scenes), and only requires a G-buffer as input, without expensive ray tracing or denoising.




* Both authors contributed equally to the paper.

\end{abstract}

\begin{figure}[htp]
  \centering

  \makebox[0.495\linewidth][c]{Direct Illumination}%
  \hfill
  \makebox[0.495\linewidth][c]{+ Ours (NIV)} \\[0.3em]

  \includegraphics[width=0.495\linewidth]{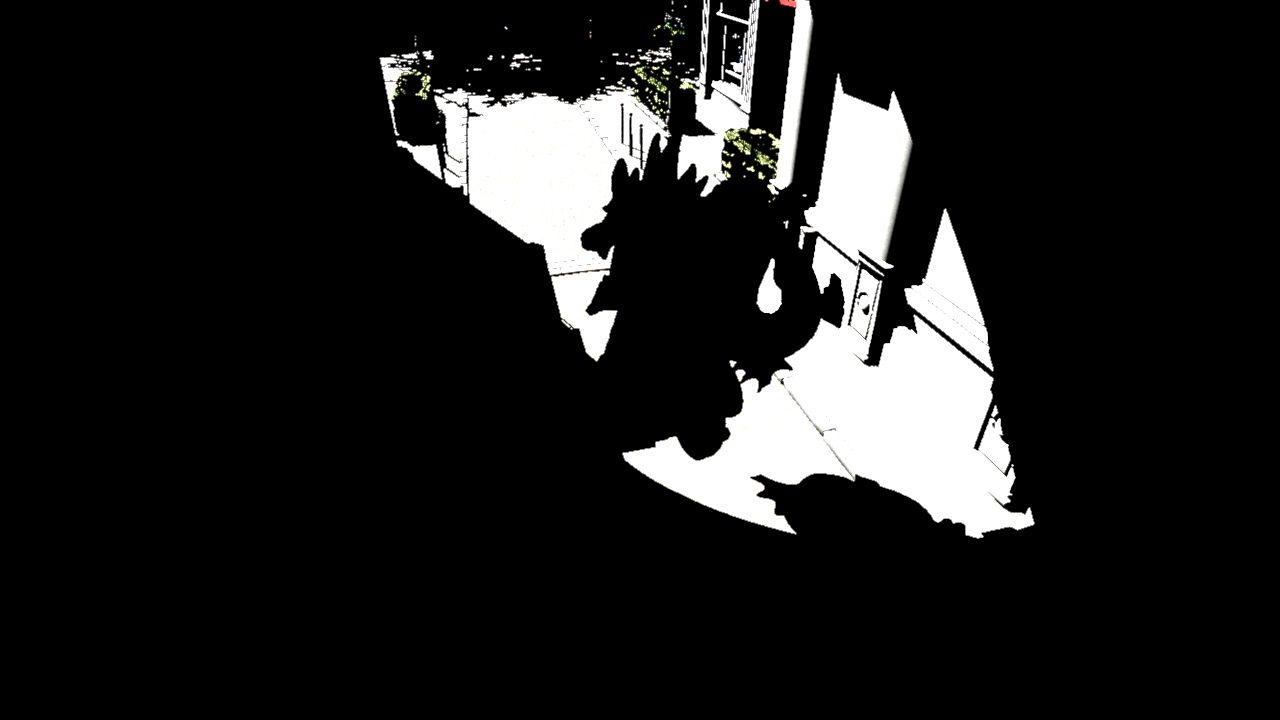} 
  \includegraphics[width=0.495\linewidth]{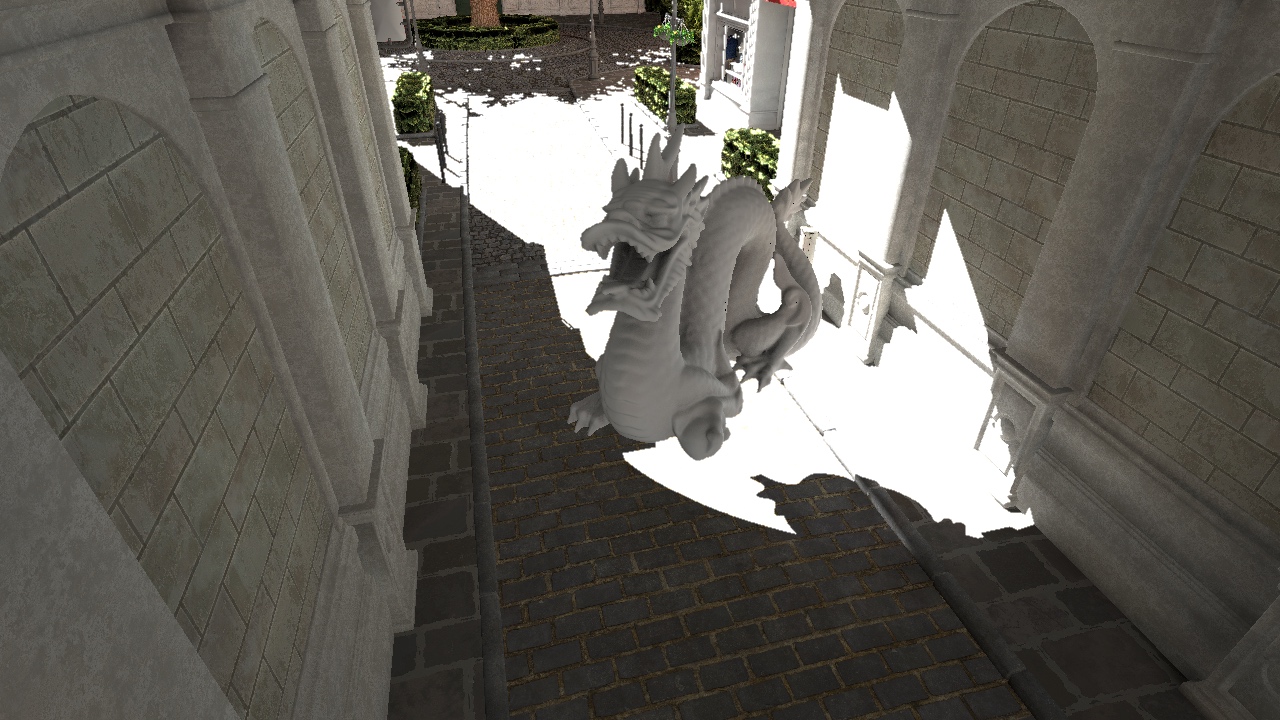} \\
  \vspace{0.1em}

  \includegraphics[width=0.495\linewidth]{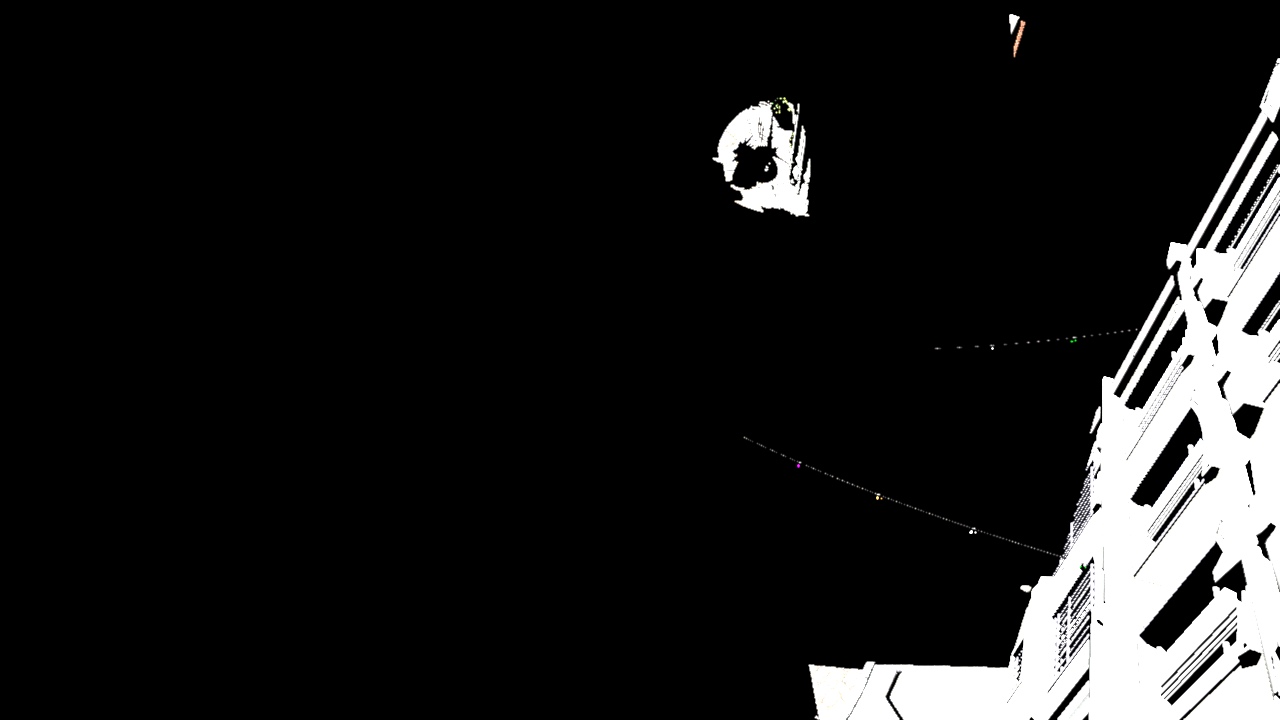} 
  \includegraphics[width=0.495\linewidth]{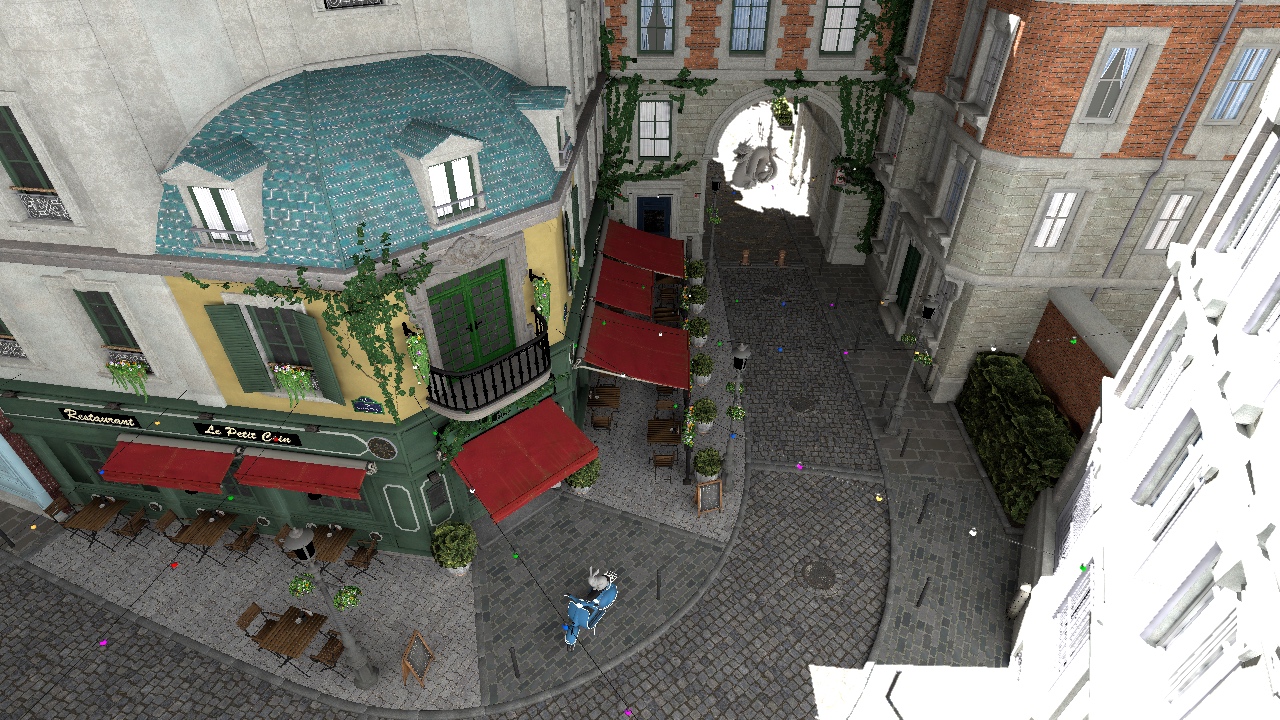}
  
  \caption{In many scenes, indirect lighting contributes the majority of visible radiance. By caching indirect lighting using an irradiance volume---such as our neural irradiance volume---the significant cost of Monte Carlo integration at runtime can be avoided.}
  \label{fig:direct_vs_ours}
\end{figure}

\vspace{-1.5em}

\section{Introduction}
Rendering moving objects within static scenes is a common requirement in real-time applications, where characters and animated entities interact with predefined environments. However, computing global illumination at high frame rates remains challenging due to the non-local and computationally expensive nature of indirect lighting.

A common solution in practice is to ``bake'' the illumination of the static scene into a grid of light probes, forming an irradiance volume that can be queried efficiently at runtime to shade unseen objects. Probe-based irradiance volumes have long constituted the de facto standard for global illumination in video games, valued for their simplicity and runtime efficiency. While not physically accurate, their rendered effects are visually convincing for all practical real-time applications. However, probe-based methods are not without limitations: placement and density must typically be determined using scene-specific heuristics, and interpolation between probes can introduce noticeable artifacts. Moreover, because the representation is inherently low-frequency, probes struggle to capture fine-scale details such as contact shadows or subtle irradiance variations. To achieve high-frequency lighting effects, the density of the probe grid needs to be significantly increased, which quickly becomes intractable due to their cubic scaling.

Beyond classical probes, recent neural methods have also demonstrated how neural models can accelerate rendering by caching radiative quantities on surfaces~\cite{hadadan2021neural, ren2013global, coomans2024real, su2024dynamic, diolatzis2022active}.
While these methods achieve impressive fidelity, they are generally unsuitable for real-time applications involving dynamic objects unseen during training, as pre-computation only accounts for known scene surfaces. Similarly, neural caches have been employed to amortize real-time path tracing costs through online learning~\cite{muller2021real}, but the reliance on expensive steps such as denoising and ray tracing makes them impractical under the strict runtime budgets of real-time applications.
However, as support for neural operations becomes increasingly widespread and performant, reframing real-time rendering problems with neural solutions allows them to benefit from the acceleration of tensor hardware performance.
In this work, we propose a Neural Irradiance Volume (NIV), which evolves the concept of irradiance volumes by leveraging recent neural advances to deliver high-quality, real-time indirect diffuse illumination at low memory, for both dynamic objects and static surfaces. NIV is inspired by the foundational work on irradiance volumes~\cite{greger1998irradiance}, adopting the same assumptions common in practical applications~\cite{barre2017certain, seyb20uberbake}. Namely, the scene is considered mainly \emph{static} at runtime, while dynamic elements—such as characters or interactive entities—are relatively small and do not significantly impact global light propagation.

Our method retains many of the runtime advantages of classical probes while enabling the superior shading quality and memory use of neural methods.
Unlike probe-based methods, our approach is heuristic-free, relying instead on backpropagation to directly optimize for reconstruction quality. 
We tackle the cubic scaling of irradiance volumes by encoding the irradiance field in a compact neural model, achieving stronger compression without sacrificing high-frequency details.
Unlike neural methods, we compute a pre-integrated and volumetric quantity, which permits noise-free shading of dynamic objects, without the use of ray tracing or denoising.

Our contributions are the following:
\begin{itemize}
    \item A precomputation scheme that bakes path tracing quality irradiance without requiring per-scene heuristics, with the ability to shade unseen moving objects.
    \item An order of magnitude lower rendering cost compared to state-of-the-art neural rendering methods.
    \item A compact representation for irradiance fields that, at a fixed memory budget, improves quality by an order of magnitude over competing real-time volumetric approaches.
    \item The ability to extend our method to non-static irradiance fields, which is a common application, e.g. a time-of-day cycle.
\end{itemize}

\section{Related Work}

\textit{Traditional Precomputation.}  Computing a noise-free solution to the rendering equation~\cite{kajiya1986rendering} of arbitrary scenes remains infeasible in real-time. 
Consequently, rendering research has long relied on preprocessing techniques, or ``baking'', to eliminate parts of this computation at runtime.
Since direct illumination is comparatively inexpensive to compute, many methods focus on precomputing the more complex, indirect component of (diffuse) light transport.

Our work draws significant inspiration from  irradiance probe volumes~\cite{greger1998irradiance},  sharing the goal of volumetrically precomputing
irradiance to enable efficient and stable shading of 
moving objects at runtime. 
A multitude of follow-up work has improved probe-based irradiance caching, focusing
on the challenges of probe placement and interpolation, to mitigate light-leaking artifacts and reconstruction errors. 
Solutions include heuristics like the use of geometric terms~\cite{silvennoinen2015multi}, normal-offsetting~\cite{hooker2016volumetric}, non-uniform probe placement~\cite{Wang2019probeplacement, zhu2025wishgi}, and  compression~\cite{Vardis2014Radiance, zhou2025gaussian}. Enhancements such as visibility information~\cite{iwanicki2017precomputed, mcguire2017real} and precomputed radiance transfer~\cite{sloan2002PRT} have further refined shading accuracy through the use of occlusion information, while extensions to glossy materials~\cite{rodriguez2020glossy} have expanded applicability. Unlike probe-based techniques, our method represents irradiance continuously, forgoing the problems of probe interpolation and placement.

\begin{figure}[t]
    \centering
    \begin{minipage}{0.325\linewidth}
        \includegraphics[width=\linewidth]{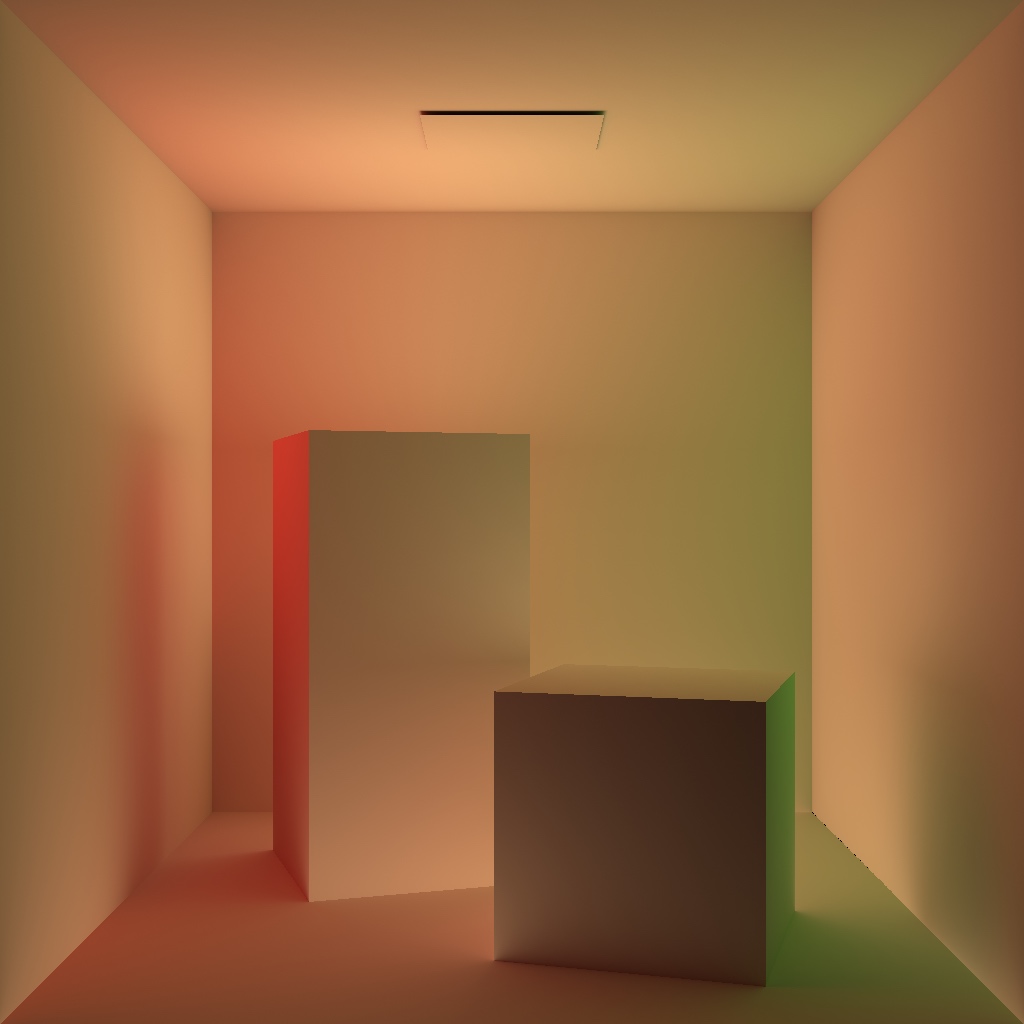}
        \centering

        {\textit{Indirect irradiance}}
        \vspace{-0.25cm}
        \begin{flushleft}
        \color{white}
        \raisebox{2.5\height}[0pt][0pt]{\makebox[0.5cm][r]{$\frac{E}{\pi}$}}
        \end{flushleft}
    \end{minipage}
    \begin{minipage}{0.325\linewidth}
        \includegraphics[width=\linewidth]{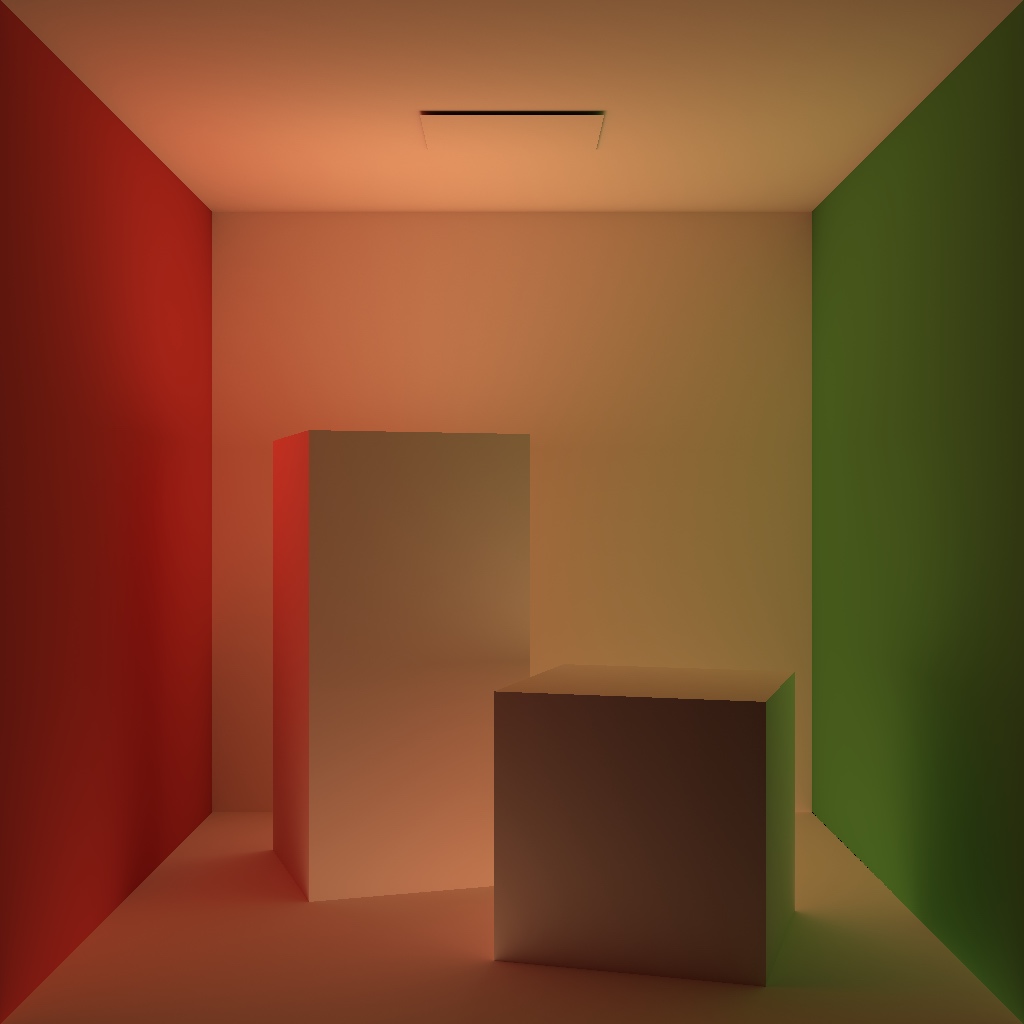}
        \centering

        {\textit{Indirect illumination}}
        \vspace{-0.25cm}
        \begin{flushleft}
        \color{white}
        \raisebox{2.5\height}[0pt][0pt]{\makebox[0.9cm][r]{$\rho \ \frac{E}{\pi}$}}
        \end{flushleft}
    \end{minipage}
    \begin{minipage}{0.325\linewidth}
        \includegraphics[width=\linewidth]{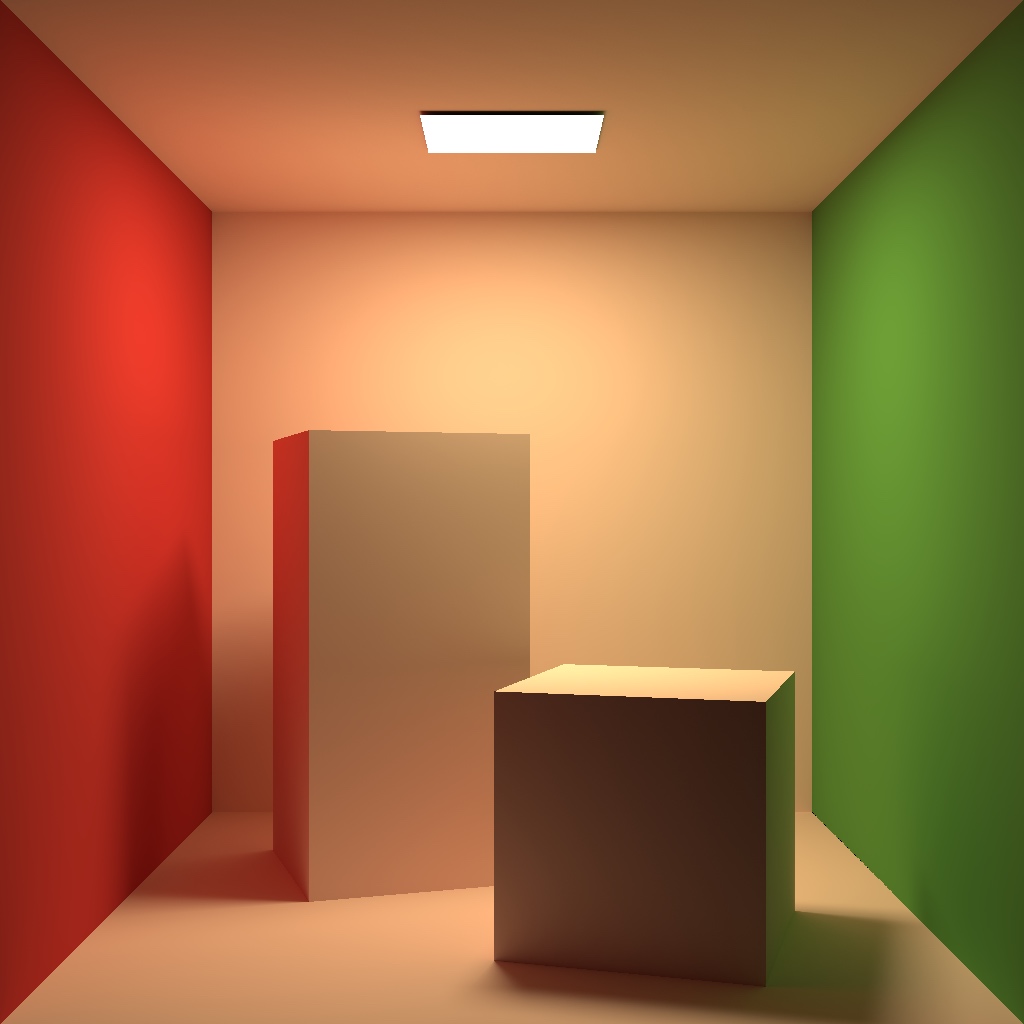}
        \centering

        {\textit{Global illumination}}
        \vspace{-0.25cm}
        \begin{flushleft}
        \color{white}
        \raisebox{2.5\height}[0pt][0pt]{\makebox[2.0cm][r]{$\rho \ \frac{E}{\pi} + D + L_e$}}
        \end{flushleft}
    \end{minipage}
    \caption{Diffuse global illumination is inexpensive to render starting from indirect irradiance $E$. 
    However, the indirect irradiance signal itself is expensive to compute, making it the most valuable component to cache. 
    Multiplying the latter by surface albedo $\rho$ results in indirect illumination, and adding direct illumination $D$ and emitted radiance $L_e$ results in diffuse global illumination.}
    \label{fig:indirect-irradiance-theory}
\end{figure}

\textit{Precomputation using Neural Networks.} 
Neural networks have been extensively studied for approximating radiance, with applications ranging from static scenes~\cite{hadadan2021neural} to those involving dynamic parameters such as scene illumination~\cite{ren2013global, rainer2022neural}, various scene configurations~\cite{su2024dynamic, diolatzis2022active} or manifolds within a set of scene parameters~\cite{coomans2024real}.  In terms of architecture, more advanced neural representations have recently been explored, such as those based on a transformer architecture~\cite{ren2024lightformer, zeng2025renderformer}. A key advantage of all the aforementioned neural methods is their ability to learn representations implicitly, eliminating the need for problem-specific heuristics required by traditional precomputation approaches. 
Neural techniques typically deliver higher quality than classical methods, though they come with longer precomputation times and higher inference costs.
More recently, attempts have been made to combine neural and traditional precomputation~\cite{iwanicki2024lightgrid}, by equipping each probe with a small neural network that computes a weighting function used for interpolation. However, in comparison to our approach, probe placement is still done manually, requiring scene-specific heuristics and/or artist input to define the precomputation volume and its spatially varying density.

\textit{Online Cache Adaptation.} Techniques that precompute parts of the rendering equation struggle with fully dynamic effects, as changes in scene parameters require a new precomputation pass to capture the updated global illumination.  To address this, methods have been developed to update cached quantities at
runtime~\cite{majercik2019dynamic, muller2021real}, transforming conventional precomputation concepts into online ones. 
However, to support frequent runtime updates, these methods rely on smaller caches to maintain efficiency,
which compromises their representational strength. \textit{Neural Radiance Caching}~\cite{muller2021real} mitigates this by delaying cache queries by tracing additional rays---which reduces bias and increases sampling variance---but requires additional denoising. 
While online cache adaptation offers robustness against arbitrary scene changes, our work focuses on a setup that does not permit the additional cost of tracing rays or denoising.

\vspace{-1.5em}

\section{Preliminaries}

By excluding view-dependent lighting effects from the rendering equation~\cite{kajiya1986rendering} and focusing solely 
on diffuse illumination, the bidirectional reflectance distribution function (BRDF) can be factored out of the reflected radiance integral.
This simplification reduces the formulation to the \textit{irradiance} function $E$: 
\begin{equation}
E(x, n) = \int_{\Omega} L_i(x, \omega_i) \langle \omega_i, n\rangle^+  d\omega_i.
\end{equation}
At a position $x$, the term $n$ can represent any direction within  the spherical domain, where $E(x, \cdot)$ defines a continuous spherical function.
Note that irradiance is defined for non-surface positions as well.
With the diffuse BRDF being a constant
proportional to the surface albedo $\rho(x)$,  diffuse reflected radiance $L_r(x)$ can be simplified to:
\begin{equation} \label{eq:scattered_radiance}
L_r(x) = \frac{\rho(x)}{\pi} \cdot E(x, n).
\end{equation} 
For simplicity, we denote the \textit{indirect} irradiance as $E$, with direct illumination $D$ added at runtime. Figure~\ref{fig:indirect-irradiance-theory} illustrates these quantities.

\section{Neural Irradiance Volume}

Our goal is to render dynamic objects that are unseen during training within known static environments using an \textit{efficient} and \textit{compact} representation that can be precomputed \textit{without the need for per-scene heuristics}. 
We achieve this by learning a neural indirect irradiance function $E_\theta$, which we term a Neural Irradiance Volume (NIV).
Given a scene, we train $E_{\theta}$ via loss-driven optimization by regressing on path traced indirect
irradiance values $E(x, n)$ for a dense sampling of position-direction pairs $(x, n)$ within the scene domain.
At runtime, NIV enables the introduction of unseen objects, inferring high quality diffuse illumination for both dynamic objects and static surfaces, running at around one millisecond per frame on consumer hardware on a full HD frame.

\subsection{Compact Irradiance Representation} \label{sec:compact}

Our approach addresses the inherent inefficiencies of traditional irradiance volumes~\cite{greger1998irradiance, majercik2019dynamic}, which are typically dense and precomputed at a fixed grid resolution. This can be wasteful as a significant portion of representational capacity may be allocated inside scene geometry or in areas where the irradiance signal varies minimally. 
To overcome these limitations, we leverage two components: a small neural network and a multi-level hash encoding~\cite{muller2022instant}.

\emph{Neural Network}. Our model uses a four-layer fully connected coordinate-based neural network~\cite{muller2021real}, with ReLU activations applied at all but the final layer, mapping an input position and direction to an irradiance value.
We train a variety of neural networks, with varying overall memory capacity (refer to Table~\ref{table:neural_runtime_data}). The smaller models add frequency encoding~\cite{mildenhall2021nerf} to the input position $x$, mitigating the struggle of neural networks to regress on high-frequency signals~\cite{tancik2020fourier}. The larger models make use of a learned input encoding discussed later in this section.

A notable difference compared to classic methods is how we structure our irradiance cache. Classic probes-based methods map each position $x$ to a spherical harmonics function, which is then evaluated at a given direction $\omega$ to return irradiance $E$, compactly described by $x \mapsto (\omega \mapsto E)$. Our model instead maps the 5D input directly to the output, i.e. $(x, \omega) \mapsto E$.
This simpler reformulation enables powerful extensions described in Section~\ref{sec:unifiedrepr} and Section~\ref{sec:higher-dimensional}, and does not require the choice of a set of basis functions, like spherical harmonics.

\begin{figure}[ht]
    \centering
    \includegraphics[width=\linewidth]{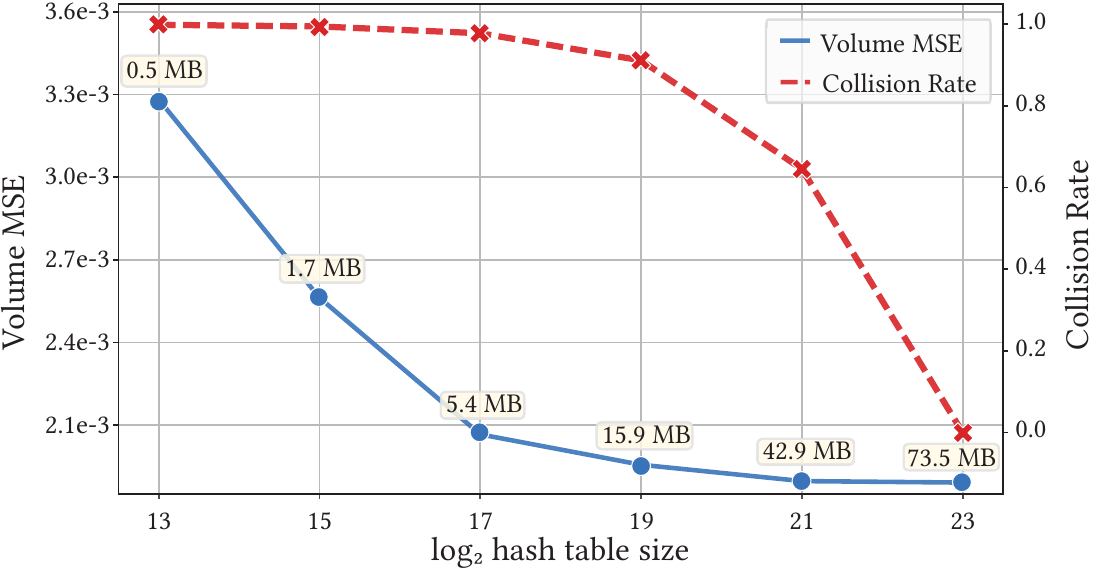}
    \caption{Varying the hash table size on \textit{Sponza} (8 levels of encoding) impacts the reconstruction error of NIV. Allowing hash collisions mildly impacts the MSE while significantly reducing the required memory.}
    \label{fig:hashmap_ablate}
\end{figure}

\emph{Learned Encodings}. To capture the irradiance field of larger scenes without requiring a more expensive neural network, we replace frequency encodings of our larger capacity models with a learned multi-level hash-grid encoding ~\cite{muller2022instant}. 
This encoding maps positions in 3D space to latent vectors at multiple resolutions, where the latents of the finer levels are stored in a hash table.
This allows adaptive use of the space, but also causes collisions of the input, as many positions may be mapped to the same latent vectors
in the hash table.
As demonstrated in prior work~\cite{muller2022instant}, the collisions are implicitly handled during optimization as gradients of more
important samples dominate the averaged gradient, enabling compressed representations. 
This prevents unimportant areas from consuming representational capacity,
unlike grid-based methods where samples inside geometry would still occupy memory. On the other hand, too many collisions results in a reduction of reconstruction quality.
As such, we control the size of the hash table to achieve a collision-rate that is optimal for our specific use case of representing
the irradiance signal, encouraging both compactness and quality of reconstruction.
Our evaluation indicates that a high collision rate only mildly affects irradiance reconstruction
while severely reducing the required memory capacity, see Figure~\ref{fig:hashmap_ablate}.
NIV is parameterized with hash table size $T=2^{17}$ for all our experiments, which strikes a balance between memory usage and representational
quality on all tested scenes.
The other parameters of the multi-level hash encoding are the following: the latent dimension size is 4,
the scaling factor between levels is $\sqrt{2}$, the size of a side on the coarsest level is 16, and
the number of levels varies between 2 and 8, depending on the target capacity (see Table~\ref{table:neural_runtime_data}),

\subsection{Unified Volume and Surface Cache} \label{sec:unifiedrepr}

Irradiance probes are a practical solution because
they offer a unified representation to render both dynamic and static surfaces~\cite{barre2017certain}. 
However, in practice, an additional representation is often needed because probes alone are prone to missing contact details and do not provide sufficiently high quality on static surfaces.
Often this additional representation is in the form of a 2D cache dedicated to static surfaces like \emph{light maps}~\cite{arvo1986backward}, which can be precomputed to provide higher quality but comes at the cost of high memory utilization, execution divergence due to branching between light maps and irradiance probes code paths, and uv-parametrization issues.

NIV provides the advantages of both approaches---high quality on surfaces and support for unseen dynamic objects---as it inherently captures a high-quality 2D representation for static surfaces embedded within the broader 5D field.
We achieve this without modifying the model architecture, but by simply adjusting the training procedure (Section~\ref{sec:training}): 
we sample a fraction of the positions explicitly on the scene's surfaces, with their directions deterministically aligned to the corresponding surface normal.
This surface domain forms a small 2D manifold of the whole 5D domain, and according to our experiments requires negligible capacity of the complete model.
We experimentally verified that allocating $20\%$ of the training samples to static surfaces strikes a good balance between volume and surface irradiance reconstruction.
This targeted adjustment significantly improves the quality of rendering, effectively capturing high-frequency details such as contact shadows, without compromising the quality of dynamic objects, as shown in the false-color visualization through the banners in \textit{Sponza} in Figure~\ref{fig:memory_error_trade-off}.
Notably, this low-dimensional specialized representation would be infeasible to model using a grid-based cache, as grids struggle to adapt resolution to arbitrary surface manifolds without significant memory overhead or resorting to separate 2D structures with their own drawbacks.

\subsection{Learning Pre-integrated Radiance} \label{sec:preintegrate}
Neural rendering techniques learn a radiative quantity on the surfaces of a scene, which can be either outgoing~\cite{muller2021real, diolatzis2022active, hadadan2021neural, coomans2024real} or incoming~\cite{dereviannykh2025neural} radiance.
In contrast to this, our method learns a volumetric radiative quantity, the irradiance $E$, for each position and normal pair in the scene volume. Since we intentionally tailor our method to shade diffuse materials, we factor out the BRDF and only learn integrated incoming radiance. Using and learning a pre-integrated and directionally smooth signal enables two important properties of our method: we do not require runtime sampling and simplify the learning task.

\textit{Sampling Variance}. Rendering with a neural representation of learned incident radiance ${L_i}_\theta$ (i.e. not pre-integrated) is computationally expensive. Namely, at runtime $\int {L_i}_\theta(x,\omega_i) \ \langle\omega_i, n \rangle^+ d\omega_i$ needs to be numerically evaluated which requires multiple samples such that the variance does not dominate the overall error. For high quality results, either multiple network evaluations per shading point would be needed or fewer but with additional denoising. 

\textit{Representational Simplicity.} Learning irradiance instead of incident radiance is a more practical choice, since it is inherently smoother and less complex to represent. We validate this claim by replacing the quantity NIV learns with $L_i$, while maintaining the same training budget. We experimentally verify that learning such a representation takes longer to converge and significantly increases the runtime cost in the supplemental material.

\begin{table}[ht] 
\centering
\caption{Inference times for a 4-layer network~\cite{muller2021real} driving Neural Irradiance Volume on full HD (1920×1080) frames using an RTX 4090 and i9-13900K. Networks without grid encoding \mbox{(-)} use eight-band positional encoding~\cite{mildenhall2021nerf}. ``Half'' renders a quarter resolution; ``full'' is full HD. All parameters in half-precision.} 

\begin{adjustbox}{width=\linewidth}
    \hspace{-1.5\tabcolsep}
    \begin{tabular}{@{}ccccc@{}}
    \hline
    width & grid levels & full (ms) & half (ms) & memory (MB) \\ \hline
    16    & -           & 0.19   & 0.029   & 0.003       \\
    32    & -           & 0.20   & 0.031   & 0.01       \\
    64    & -           & 0.25   & 0.069   & 0.03        \\
    64    & 2           & 0.31   & 0.088   & 0.16        \\
    64    & 4           & 0.67   & 0.18  & 1.20        \\
    64    & 6           & 1.06   & 0.26   & 3.30        \\
    64    & 8           & 1.35   & 0.37   & 5.40        \\ \hline
    \end{tabular}
\end{adjustbox}

\label{table:neural_runtime_data}
\end{table}

\subsection{Rendering} \label{sec:render}
After training, NIV can render frames in real-time, following a pipeline very similar to classic irradiance probes.
Starting from the G-buffer of the rendered scene, the position and normal buffers are input to the neural
model, which infers pixel-wise $E(x, n)$ in a single batch. Equation~\ref{eq:scattered_radiance} is used to obtain indirect diffuse radiance, adding the surface emission and direct illumination for the final frame, as shown visually in Figure~\ref{fig:teaser}.

\textit{Half-resolution Rendering.} Our method can be run at reduced resolution for higher performance, while still using the full size albedo buffer $\rho$ for accurate materials. Running at half-resolution significantly reduces the computation time for a frame, requiring only 0.37 ms for our largest model, refer to Table~\ref{table:neural_runtime_data} for the half-resolution runtime overview.
Rendering at half-resolution is a common approach in real-time rendering, but introduces aliasing artifacts which could be partly resolved by using industry practice upsampling methods~\cite{kopf2007joint}. Since indirect irradiance varies smoothly in screen-space, aliasing is only slightly perceivable, especially after multiplication with the full-resolution albedo buffer (see supplemental video).

\textit{Dynamic Ambient Occlusion.}
A precomputed irradiance field, cached as either probes or NIV, does not capture how dynamic objects affect indirect illumination, but this is often not relevant in practice as added geometry is expected to be small in size compared to the scene.
Some important effects might still be missed, like self-occlusion on the dynamic objects and indirect occlusion cast by them onto the scene.
A common solution to mitigating part of these missed interactions is applying an ambient occlusion pass~\cite{shanmugam2007hardware}.
Since NIV already captures local high-frequency effects on static surfaces as described in Section~\ref{sec:unifiedrepr}, we run a lighter \emph{dynamic} ambient occlusion pass,
where only dynamic geometry is considered when computing the hemisphere obstruction at the shading point.
This simulates self-occlusions when shading the dynamic object, and on static surfaces it simulates how the presence of new geometry partially occludes indirect irradiance.

\begin{figure}[ht!]
    \centering
    \includegraphics[width=\linewidth]{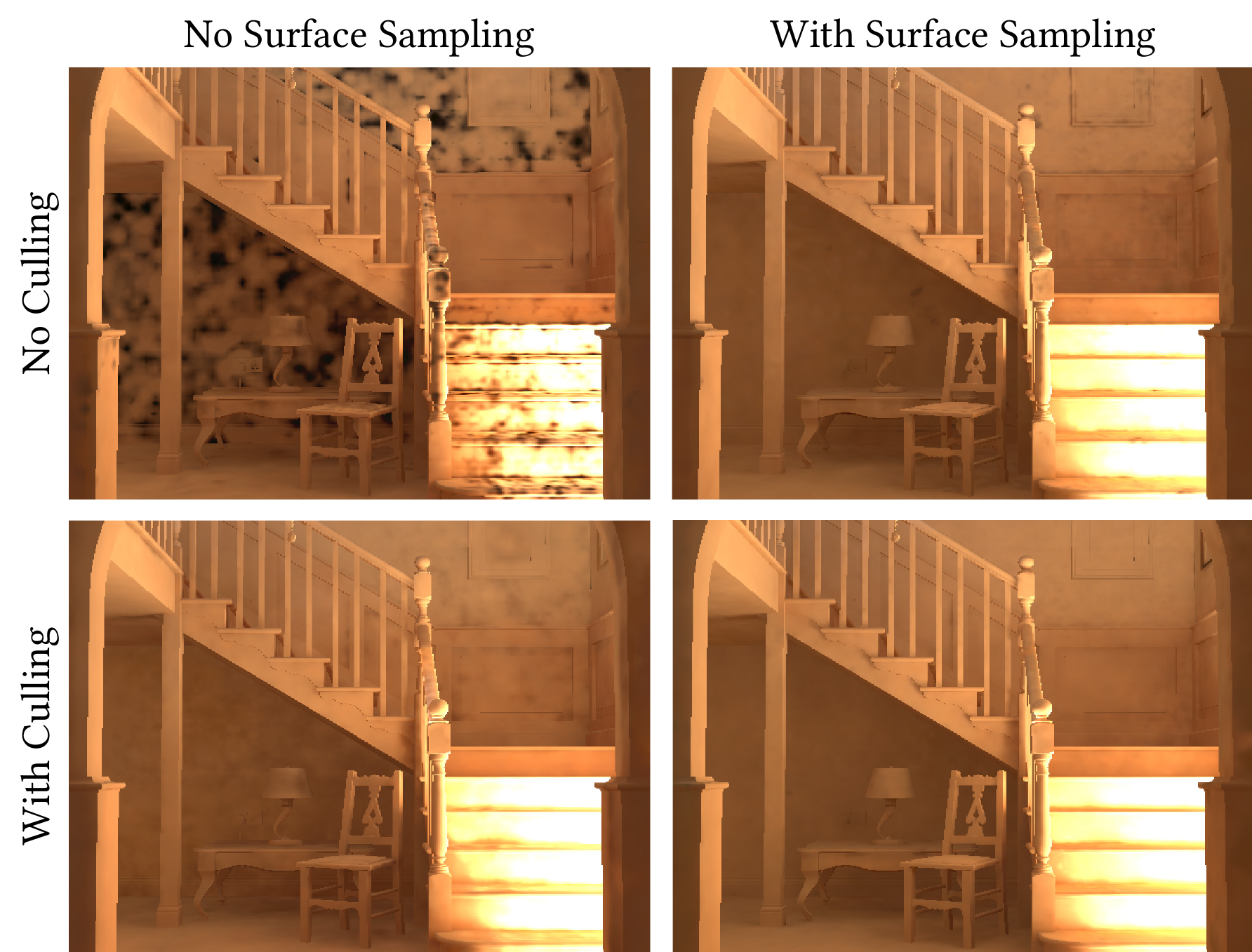}
    \caption{Culling training data inside scene geometry and allocating a portion of the training budget to sampling scene surfaces both improve reconstruction quality. 
    Combining these strategies produces the most robust results.
    }
    \label{fig:culling_sampling}
    \vspace{-1em}
\end{figure}

\subsection{Training} \label{sec:training}

NIV is trained by uniformly sampling position-direction pairs $(x, n)$ throughout the scene volume---20\% of which are uniformly sampled on the surface geometry with their direction $n$ set to the surface normal---and computing the associated ground-truth indirect irradiance $E(x, n)$ via path tracing.
Note that while any emitter type is supported during training, real-time rendering requires light sources that can be efficiently sampled to avoid sampling noise.
The model's current output $E_{\theta}$ is then used to compute the relative L2 loss, normalizing the MSE by the squared network prediction~\cite{lehtinen2018noise2noise}:
\begin{equation} \label{eq:relmse}
    \mathcal{L}_\theta( E(x,n) ,E_\theta(x,n)) = \frac{\left(E_\theta(x,n) - E(x,n)\right)^2}{sg(E_\theta(x,n)^2) + \epsilon},
\end{equation}

where $sg$ denotes a stop gradient operation
and the constant $\epsilon = 0.01$. 
During training, we discard samples that are inside of surfaces, which we identify simply by checking whether the majority of normals at the first hit are backfacing whilst estimating irradiance. 
This avoids wasting capacity on inputs that are not visible at runtime, but also prevents mild dark leaks near static surfaces. We show the impact of sampling 20\% of the training data on surfaces and culling samples with backfacing normals in Figure~\ref{fig:culling_sampling}.

We use PyTorch~\cite{paszke2017automatic} as training framework and
Mitsuba3~\cite{jakob2022mitsuba3} as renderer for ground truth data and Adam~\cite{kingma2014adam} to optimize the model parameters, using a learning rate of $10^{-2}$ and a batch size of $2^{16}$. We reduce the learning rate after the first 10k iterations by exponentially decaying to $10^{-4}$, similar to other neural rendering works~\cite{hadadan2021neural}.
All our tested scenes converge after at most 50k iterations. On a single RTX~4090 a simple scene like the \textit{Cornell Box}  converges after around five minutes or thirty minutes for a medium sized scene like \textit{Sponza}. It is worth noting that the majority of this time is spent on path tracing irradiance (e.g. 94\% of the compute budget on \textit{Cornell Box}), as opposed to optimizing the model parameters.

\begin{figure*}[h!]
    \centering
    \includegraphics[width=\linewidth]{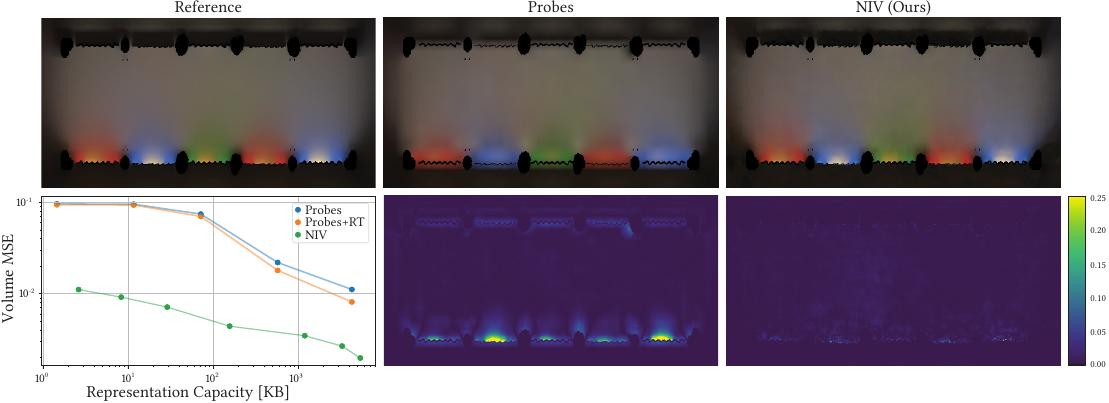}
    \caption{
    Horizontal slice through the banners of \textit{Sponza} (5.4 MB budget). NIV captures irradiance bleed and shadows better than probe-based methods.  
    NIV has $~\sim10$x higher quality across the scene's volume. Ray tracing the visibility towards probes during evaluation (``+ RT'') reduces the overall error but adds a significant performance overhead. The error map highlights the per-pixel absolute luminance error.
    }
    \label{fig:memory_error_trade-off}
\end{figure*}

\section{Results}
As our method is related to probe-based global illumination methods, neural surface caches, and neural rendering of variable scenes, we compare against the three families of approaches in detail.

\subsection{Probe-based Methods}
\label{sec:probes}

We compare NIV against a modern probe grid, similar to DDGI~\cite{majercik2019dynamic}, implemented in our  rendering system
to ensure compatibility with ground truth path tracing. See supplemental material for implementation details.
In our experiments, we follow the DDGI implementation but compute visibility using ray tracing rather than depth textures~\cite{majercik2019dynamic, mcguire2017real}. This choice favors the probe-based DDGI baseline in our quality-focused comparison, as ray-traced visibility eliminates depth-texture aliasing and avoids the additional memory overhead—where DDGI otherwise requires 1168 bytes per probe, 1024 of which are dedicated to depth buffers.

Our probe-based baseline uses second-order spherical harmonics (9 coefficients)~\cite{ramamoorthi2001efficient}, which are stored at half-precision using 54 bytes per probe. While further quantizing the probes to 28 bytes per probe is possible~\cite{roughton2024zh3}, a factor of 2 has little influence on the presented results even if the compression was lossless. 
For completeness, we also compare against the open source industry implementation of DDGI~\cite{nvidia_rtxgi_ddgi}, as shown in the supplemental material.

\textit{Memory-Error Trade-off.} 
We evaluate the MSE on randomly sampled point-direction pairs in the scene volume, as it is a good measure for the quality for shading unseen moving objects.
In all tested scenes, NIV increases quality at a given memory budget by an order of magnitude compared to probe-based methods, particularly  at lower representational capacities.
For a quantitative and qualitative example on \textit{Sponza} see Figure~\ref{fig:memory_error_trade-off}. 
While ray-traced visibility reduces the MSE of probe-based techniques, it is still unable to close the gap to NIV and incurs a significant runtime cost.  

The inferior memory scaling of probe grids is unsurprising, as the irradiance volume scales cubically with spatial discretisation and lacks the ability to adaptively allocate capacity where it is needed most.
In contrast, NIV has no such constraints and can implicitly allocate capacity based on the loss function.
This limitation of irradiance volumes manifests in two noticeable ways: the failure to capture contact shadows entirely  and the occurrence of light leaks when ray-traced visibility is not employed, see Figure~\ref{fig:memory_error_trade-off}. 
Both contact shadow accuracy and leak prevention are critical for real-world applications, which the neural representation handles automatically. 
Additional quantitative data on other scenes can be found in the supplemental material.

\textit{Performance.} The efficiency of querying a regular grid of probes is highly competitive since it consists of simple trilinear interpolation of a few spherical harmonics coefficients. 
The actual runtime cost depends on the grid size and the way coefficients are laid out in memory, which influences memory transfer and caching behavior.
Since our model uses a hashing function to map between grid coordinates and latent vectors, simple probe-based methods should always hold a performance edge over NIV, if both are optimized accordingly.
On top, NIV requires to  evaluate the input encoding and perform matrix multiplications for network evaluation~\cite{muller2021real}, which altogether incurs a cost of around 1 ms on an RTX~4090. 
As shown in Table~\ref{table:neural_runtime_data}, NIV requires a minimum of 0.19 ms at full HD resolution or 0.029 ms when rendering at half resolution. 
Relying on a multi-resolution hash grid increases runtime cost, indicating that memory transfer costs also dominate the performance of our approach.
In any case, relying on 2 or 4 grid levels (0.16 to 1.20 MB) leads to around 0.5 ms of runtime, which is 
easily fast enough for real-time applications and provides better quality compared to the probe-based approaches.

\subsection{Neural Surface Cache}
\begin{figure}[ht]
    \centering
    \includegraphics[width=\linewidth]{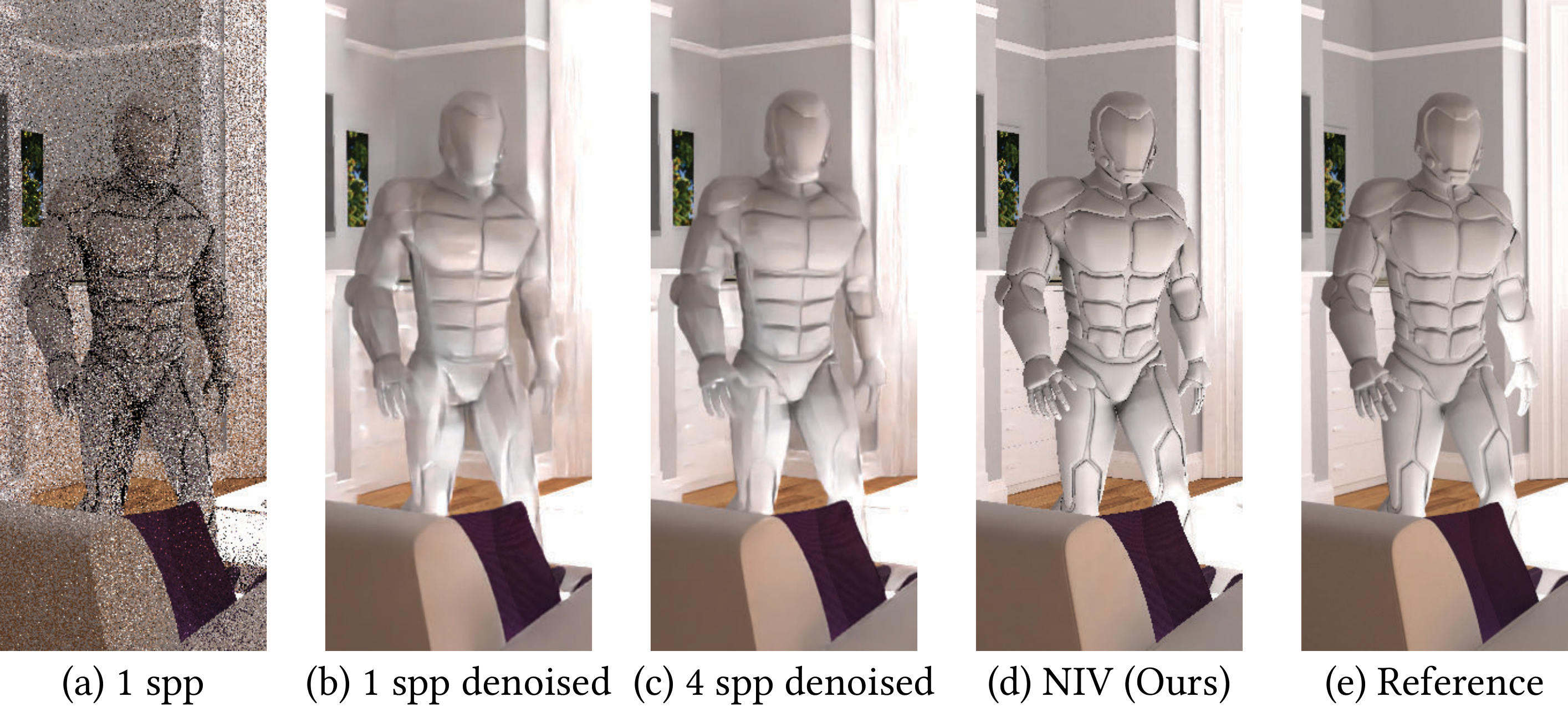}
    \caption{
    Rendering with a neural surface cache requires a single deferred surface lookup, which introduces noise (a). A neural denoiser like Optix~\cite{parker2010optix} reduces noise but produces blotchiness (b, c), resolvable only with more costly ray-traced samples. Our method avoids path tracing and denoising while significantly improving quality (d) compared to a reference image (e).
    }
    \label{fig:nrc_comparison}
    \vspace{-1.5em}
\end{figure}

Methods that rely on a neural network to store outgoing radiance along scene surfaces~\cite{ren2013global, hadadan2021neural, coomans2024real} cannot directly shade objects that are not part of the training domain. 
When a previously unseen dynamic object is introduced into the scene, ray tracing is required to defer shading computations to the trained surface locations. 
While this approach avoids introducing additional bias, it increases variance in the final shading proportional to the deferred path depth.
Caches that adapt to the scene in an online fashion~\cite{muller2021real} offer an alternative by continuously updating their neural model to account for newly introduced objects.
However, both updating the cache and deferring to mitigate the bias of their (comparatively) small cache requires ray tracing during runtime.

\begin{figure}[t]
    \centering
    \includegraphics[width=\linewidth]{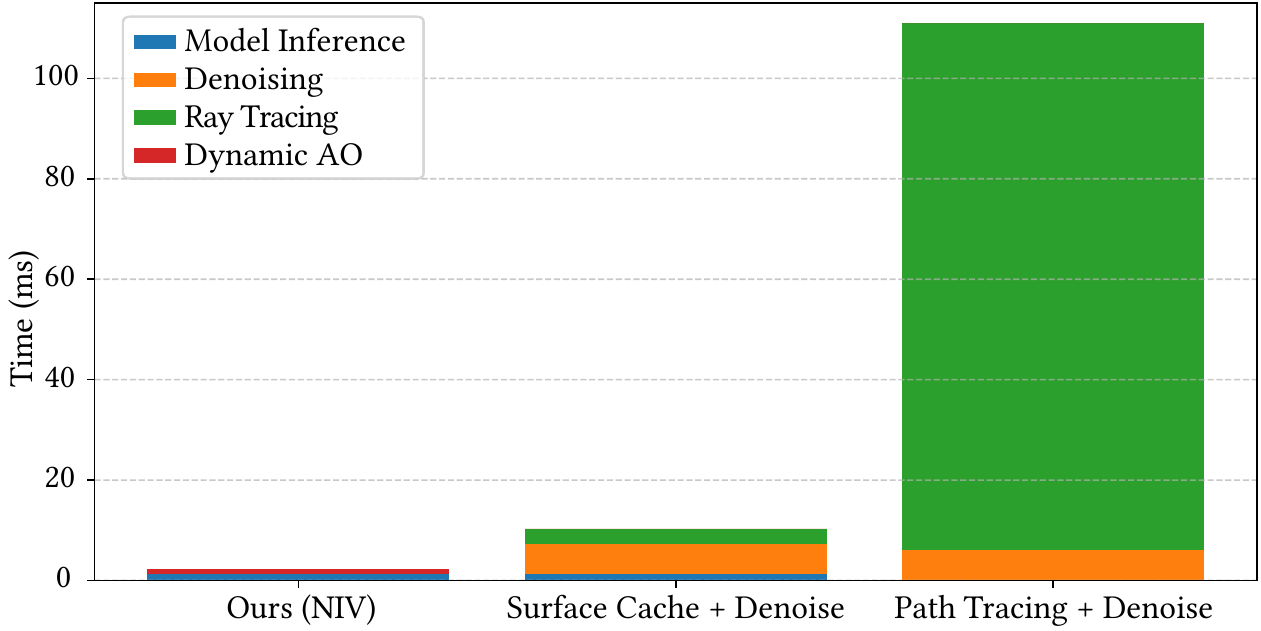}
    \caption{Runtime comparison between NIV, a denoised neural surface cache and denoised path tracing. The ray sampling budget of the methods was chosen to match the rendering quality of NIV. The surface cache only traces a single bounce (4 spp) as opposed to path tracing (16 spp), which traces full-length paths with Russian Roulette. Primary visibility is provided to all methods using a rasterization pass and is not included in the timings.}
    \label{fig:timing_niv_surface_pt}
\end{figure}

We compare against neural surface methods by training a model similar to \textit{Neural Radiance Caching}~\cite{muller2021real}, augmented with a multi-resolution hash encoding~\cite{muller2022instant} with the same representational capacity as our method. After training, this model captures the outgoing radiance at any point of the static scene, but not on dynamics objects. 
By deferring cache queries by one bounce (or more when the deferred bounce lands on a dynamic object again) the neural surface cache can be used to estimate indirect illumination on dynamic objects.
To remove the resulting variance, a denoiser must be used.
Both ray tracing and denoising add a significant runtime overhead of around 5-10 ms per frame, even when using hardware supported ray tracing and highly efficient neural denoising, see Figure~\ref{fig:timing_niv_surface_pt}.  

\begin{figure}[t]
    \centering
    \includegraphics[width=\linewidth]{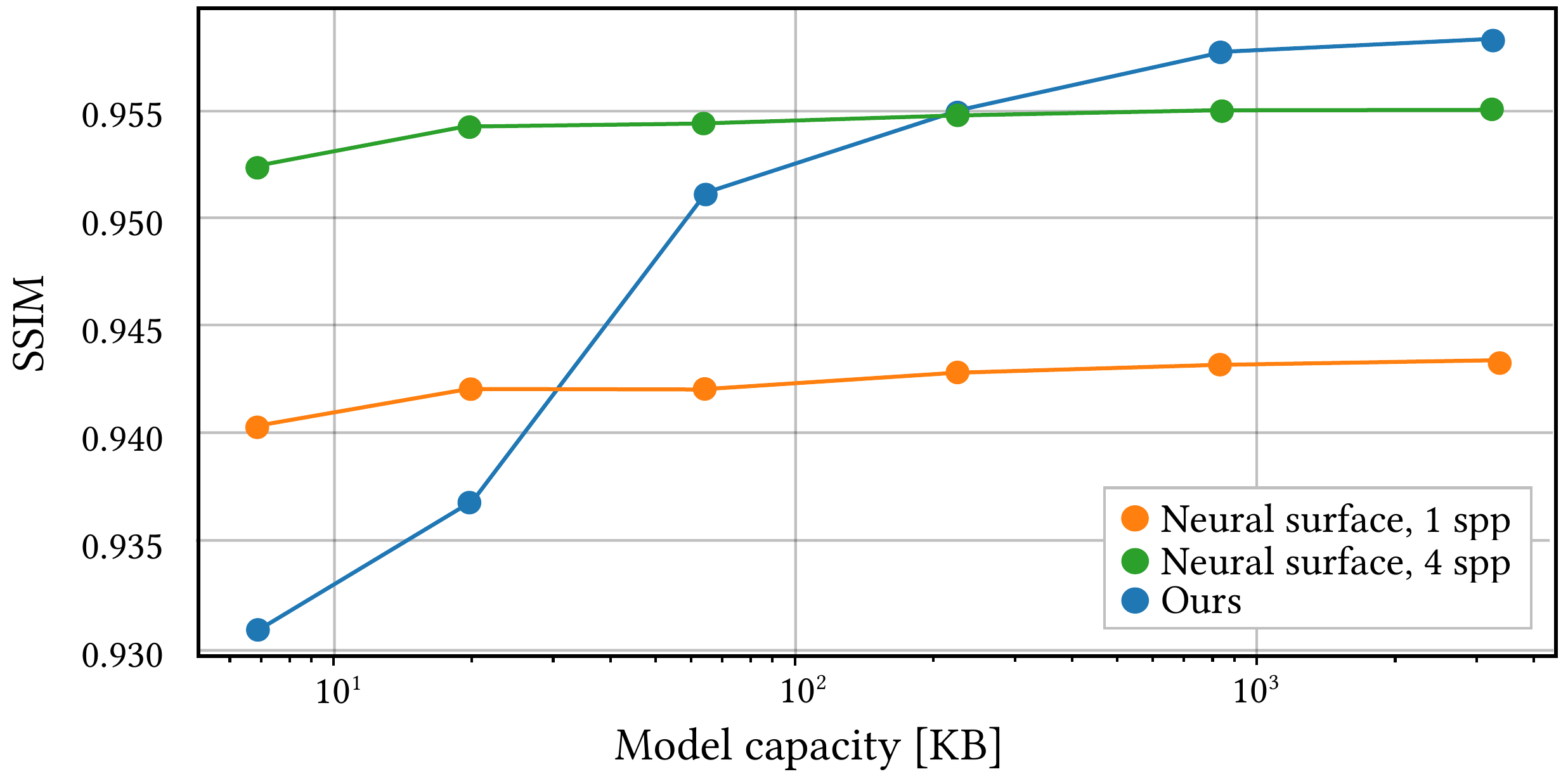}
    \vspace{-2em}
    \caption{Rendering error of a neural surface cache and ours compared against reference on \textit{Sponza}. Since the majority of the error in the neural surface caching is the variance from deferring, increasing the model capacity---and thus reducing its bias---does not improve the overall quality. Since NIV does not require sampling, its quality scales with the allocated memory.
    }
    \label{fig:nrc_plot}
\end{figure}

\begin{table}[t]
\centering
\caption{Comparison between a neural surface cache and NIV. Scenes from our dataset were rendered with the addition of dynamic objects and compared to path traced references. Both methods used a capacity of 5.40 MB. We use the HDR variant of FLIP~\cite{Andersson2021a}.}

\begin{tabular}{ccccc}
\toprule
Scene       & \multicolumn{2}{c}{FLIP ($\downarrow$)} & \multicolumn{2}{c}{MSE ($\downarrow)$}   \\ \cmidrule(l){2-5} 
            &  \ Surface  \   &  \ \ \ Ours   \ \ \        &  \ \ \ Surface  \ \ \ &  \  Ours \             \\ \cmidrule(l){2-5}
Sponza & 0.25 & \textbf{0.10} & 2.32e-3 & \textbf{9.09e-6} \\ 
Cornell box & 0.14 & \textbf{0.05}  & 5.57e-4 & \textbf{2.53e-5} \\ 
Dining room & 0.27 & \textbf{0.24}  & 3.07e-2 & \textbf{3.31e-3}  \\ 
Bathroom & 0.15 & \textbf{0.06}  & 3.22e+0 & \textbf{2.76e-3}  \\ 
White room & 0.17 & \textbf{0.09}  & 2.76e-3 &\textbf{ 8.13e-4} \\ 
Living room & 0.10 & \textbf{0.04}  & 1.14e-2 &\textbf{ 2.36e-4}  \\ 
\bottomrule
\end{tabular}

\label{tab:flip_mse}
\end{table}

\textit{Results.} We render scenes from our dataset with NIV and with a neural surface cache, and evaluate the rendering error with respect to path traced references.
All test scenes contain dynamic objects which cover between 5\% and 10\% of the output image, Figure~\ref{fig:nrc_comparison} shows a crop of one such 
dynamic object.
As shown in Table \ref{tab:flip_mse}, NIV achieves higher quality results for all inputs, while using the same memory budget and running 5-10 ms faster since ray tracing and denoising are not required.
Figure~\ref{fig:nrc_comparison} exemplifies the bias-variance tradeoff on the \emph{White room} scene: the majority of artifacts from neural surface caching come from high Monte Carlo variance, hence the denoiser (Optix) does not have enough information to reconstruct an artifact-free frame.
Figure~\ref{fig:nrc_plot} shows that increasing the surface model capacity has little effect on quality. Since variance is the major source of error, increasing the sample count is a more effective way of improving quality, but linearly raises the runtime cost.

\subsection{Neural Rendering of Variable Scenes}
\label{sec:variable_neural}

\begin{figure}[ht]
  \centering

  \begin{subfigure}[b]{0.49\linewidth}
    \centering
    \begin{overpic}[width=0.49\linewidth]{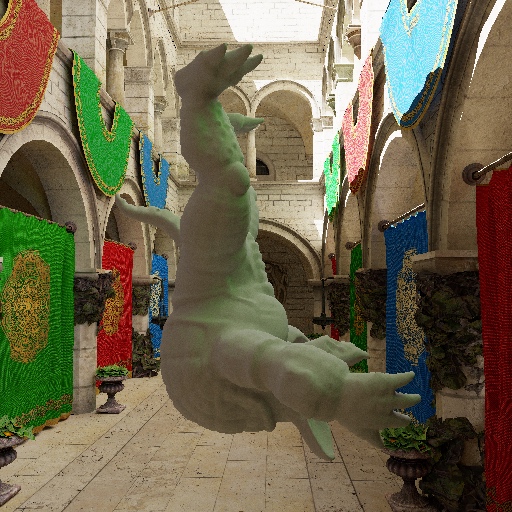}
      \put(20,5){\small \color{white} Reference}
    \end{overpic}%
    \hfill%
    \begin{overpic}[width=0.49\linewidth]{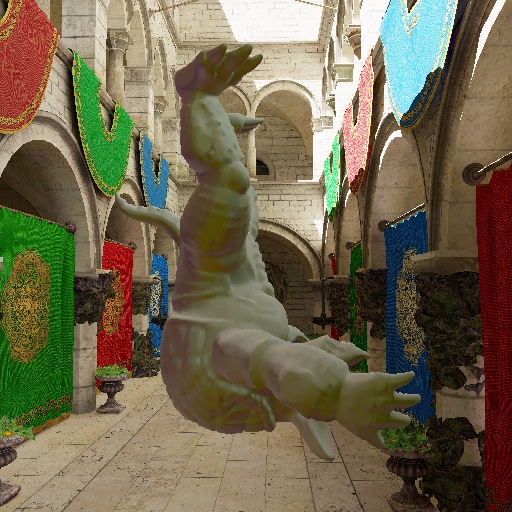}
      \put(22,5){\small \color{white} \cite{diolatzis2022active}}
    \end{overpic}\\[0.5mm]
    \begin{overpic}[width=0.49\linewidth]{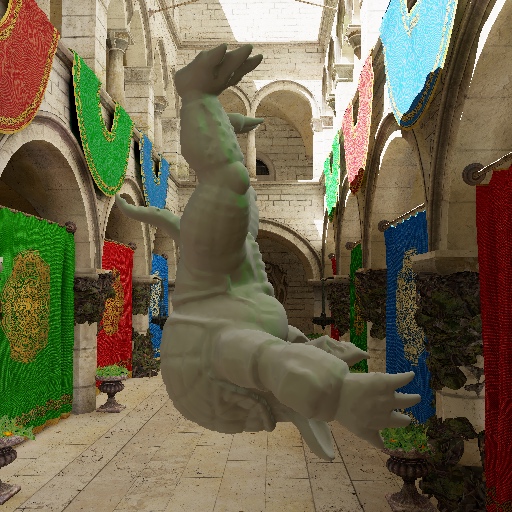}
      \put(35,5){\small \color{white} Ours}
    \end{overpic}%
    \hfill%
    \begin{overpic}[width=0.49\linewidth]{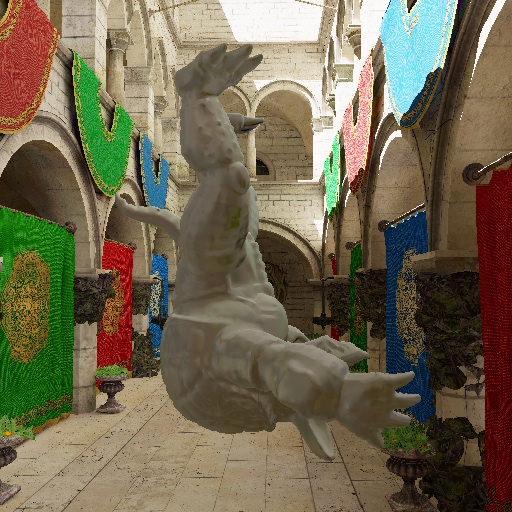}
      \put(20,5){\small \color{white} \cite{su2024dynamic}}
    \end{overpic}
    \caption{5 Variable Parameters}
    \label{fig:grid_a}
  \end{subfigure}
  \hfill
  \begin{subfigure}[b]{0.49\linewidth}
    \centering
    \begin{overpic}[width=0.49\linewidth]{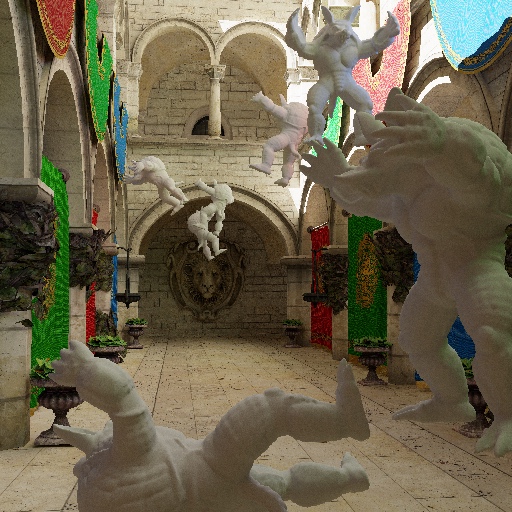}
      \put(20,5){\small \color{white} Reference}
    \end{overpic}%
    \hfill%
    \begin{overpic}[width=0.49\linewidth]{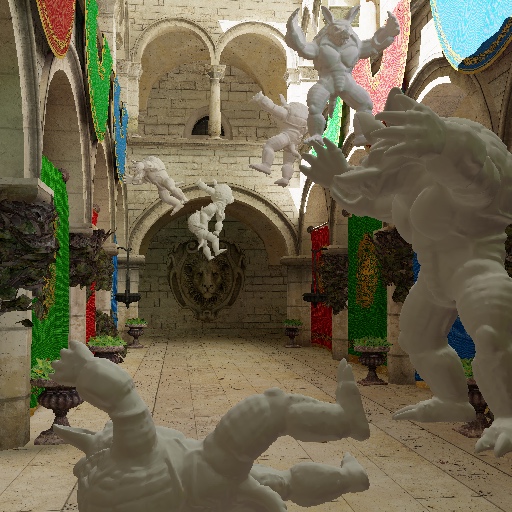}
      \put(22,5){\small \color{white} \cite{diolatzis2022active}}
    \end{overpic}\\[0.5mm]
    \begin{overpic}[width=0.49\linewidth]{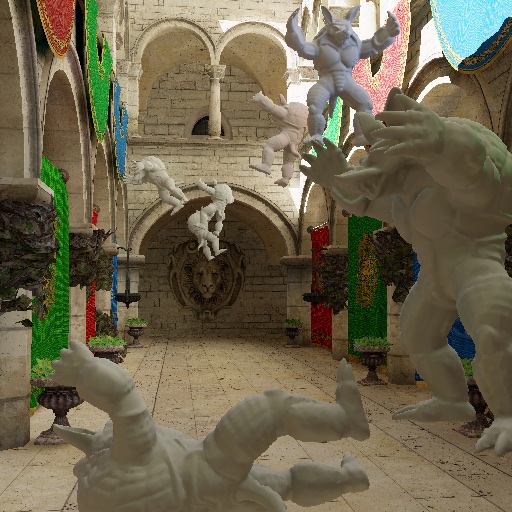}
      \put(35,5){\small \color{white} Ours}
    \end{overpic}%
    \hfill%
    \begin{overpic}[width=0.49\linewidth]{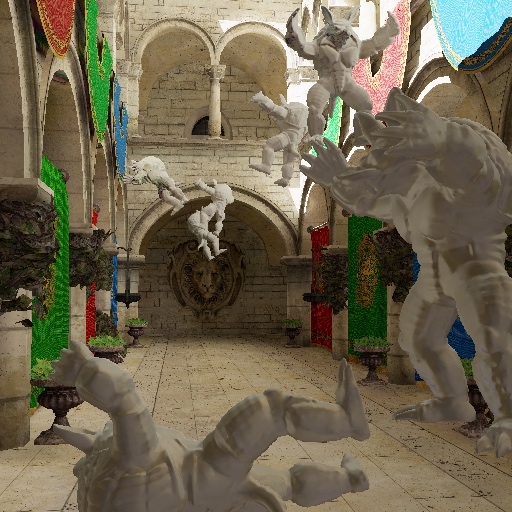}
      \put(20,5){\small \color{white} \cite{su2024dynamic}}
    \end{overpic}
    \caption{35 Variable Parameters}
    \label{fig:grid_b}
  \end{subfigure}

  \caption{At a model capacity of 1~MB NIV captures color bleed from the static scene onto the variable scene elements---the floating armadillos---while the variable scene methods~\cite{diolatzis2022active, su2024dynamic} do not. As opposed to the compared methods, NIV has not seen dynamic objects during training, and can be queried with any amount of them without re-training. See Figure~\ref{fig:variable_vs_niv} for quantitative data.}
  \label{fig:renders_variable}
\end{figure}

While our goal is to shade moving objects without retraining, other methods explicitly take scene variation into account during the training phase. These approaches generally assume a predefined range for each variable scene element, which is encoded into a structured representation---usually a vector. The representation can then be used directly as an input for the neural model~\cite{diolatzis2022active}, or processed through a learned encoding~\cite{su2024dynamic}. Such strategies attempt to learn a surface radiance field for all possible permutations of scene variables, in contrast to techniques that instead constrain learning to a single manifold within these parameters~\cite{coomans2024real}.

Despite correctly modeling the light transport induced by complex variable-scene interactions, these approaches present several drawbacks when compared to (neural) irradiance volumes. 
First, they require retraining whenever the set of scene parameters changes, such as when objects are added, removed, or modified. 
Second, they fail to achieve real-time performance, with inference times exceeding 100 ms in our experiments, see the supplemental material. 
Third, the inference time scales with the number of dynamic objects present in the scene, which limits the use of primitives with many scene variables like animated meshes.
These limitations highlight challenges in deploying such methods for practical applications. 
Nevertheless, a natural question is how the rendering quality of these methods compares to NIV when they are similarly configured to only learn indirect diffuse global illumination. 
NIV does not explicitly model higher-order interactions between dynamic objects and the surrounding scene. In contrast, methods based on variable scene encodings need to devote part of their representational capacity to learning these interactions on top of learning the outgoing radiance on the static scene geometry. In our results we find that for a fixed model size, NIV often achieves comparable or lower rendering error, particularly as the number of scene variables increases.

\begin{figure}[t]
    \centering
\includegraphics[width=\linewidth]{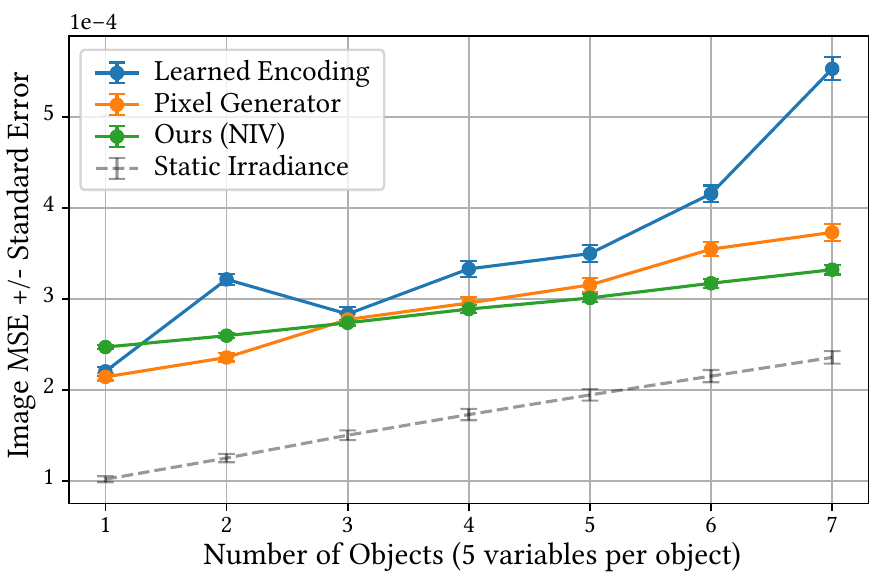}
    \caption{
We compare NIV with variable-scene methods that encode changing scene parameters via a PixelGenerator~\cite{diolatzis2022active} or a learned encoding~\cite{su2024dynamic}. Unlike these approaches, which require training for each scene configuration, NIV is trained only once on the static irradiance of the empty scene. Yet it achieves lower rendering error at higher object counts, underlining the effectiveness of the irradiance assumption. All methods use the same 1 MB capacity, with each object adding five degrees of freedom (35 scene variables at the rightmost data point). }
    \label{fig:variable_vs_niv}
\end{figure}

As illustrated in  Figure~\ref{fig:renders_variable} and Figure~\ref{fig:variable_vs_niv}, the rendering error achieved by NIV lies within the same range as methods that were trained with explicit knowledge of scene objects. This suggests that the irradiance volume formulation, combined with dynamic ambient occlusion, is highly effective at generalizing to unseen configurations. 
However, there exists a theoretical lower bound to irradiance volumes in this setup, represented in Figure~\ref{fig:variable_vs_niv} by the dotted line. This bound corresponds to the error level of a path-traced static irradiance field computed without the dynamic objects. 
As the model capacity of variable-scene methods grows, their rendering error can drop below this threshold, allowing them to produce shading effects that more closely resemble path traced global illumination, including the higher-order interactions between the variable parameters and the scene. 
In the supplemental material, we show an additional figure similar to Figure~\ref{fig:variable_vs_niv}, where the variable scene methods are ran using the recommended model capacities as described in the respective works~\cite{diolatzis2022active, su2024dynamic}, which improves their rendering quality at the cost of extra model capacity and an even slower runtime.

\section{Application: Higher-dimensional Irradiance Fields} \label{sec:higher-dimensional}

In many real-world applications, only a known set of scene parameters is expected to change at runtime~\cite{seyb20uberbake}, e.g. rotating emitters that mimic a time-of-day cycle.
Given the prevalence of such behavior we extend pre-computing a neural irradiance volume with respect to such changing scene parameters, by introducing new inputs to the model, e.g. the current position of the moving emitters. We train the irradiance field while randomly sampling the variable scene parameter, similar to the neural variable scene methods in Section~\ref{sec:variable_neural}. 
After training, this removes the need to update the representation in real-time and avoids potential temporal artifacts and updating costs. 

\begin{figure}[ht] 
    \centering
    {\includegraphics[width=0.32\linewidth]{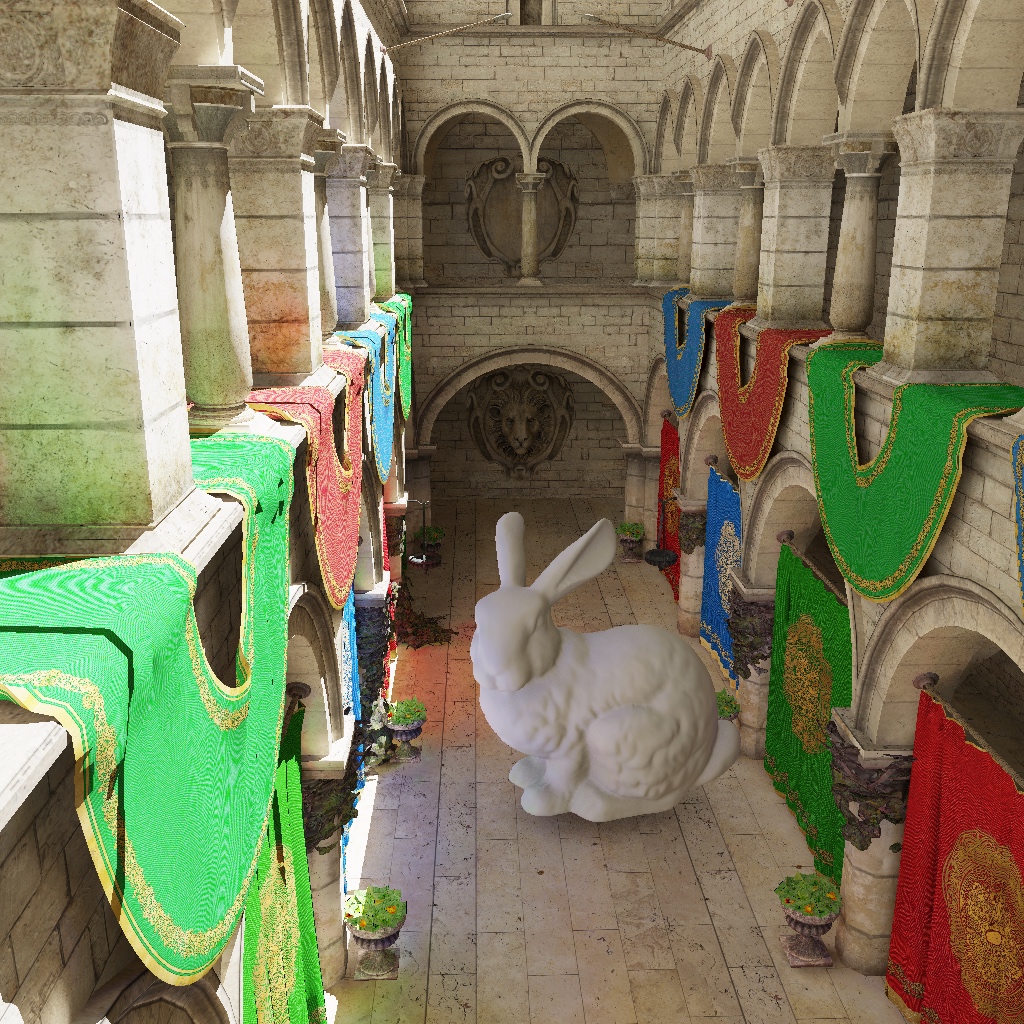}}\hfill
    {\includegraphics[width=0.32\linewidth]{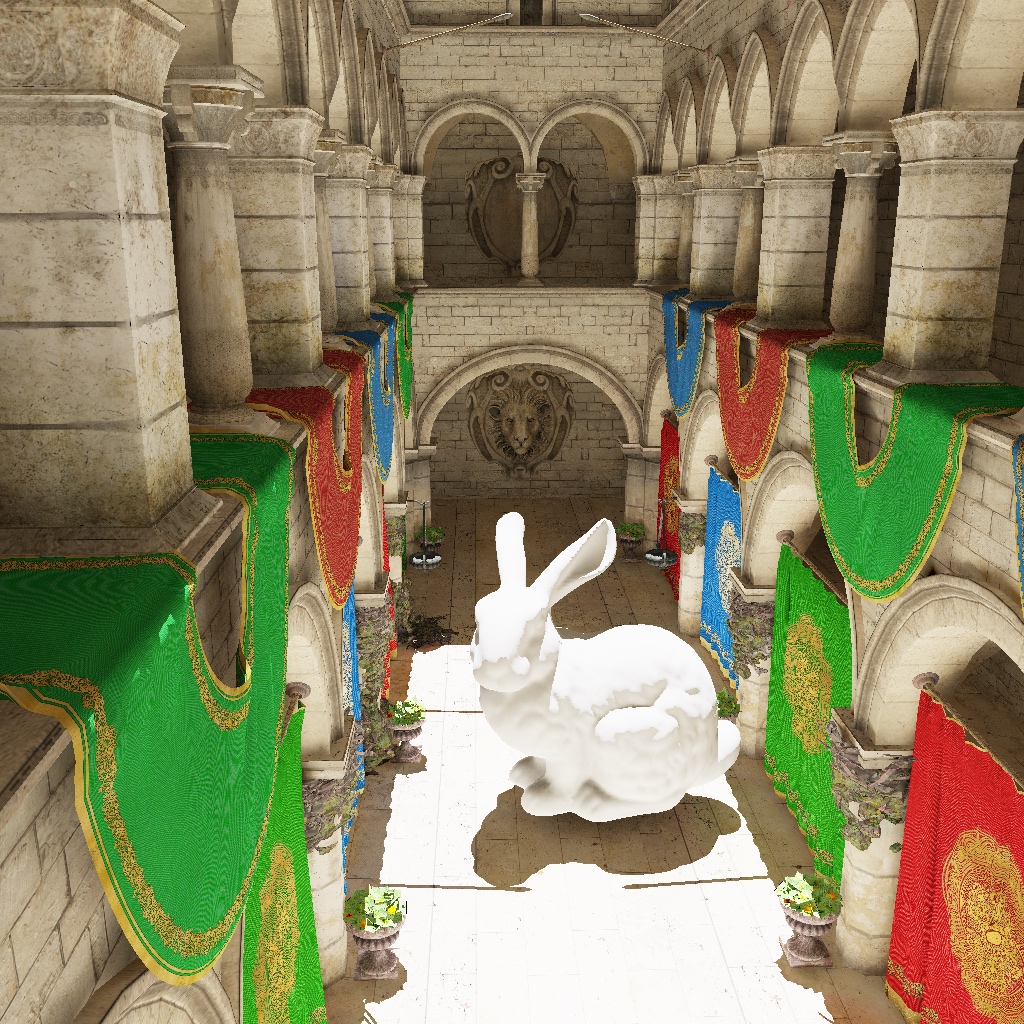}} \hfill
    {\includegraphics[width=0.32\linewidth]{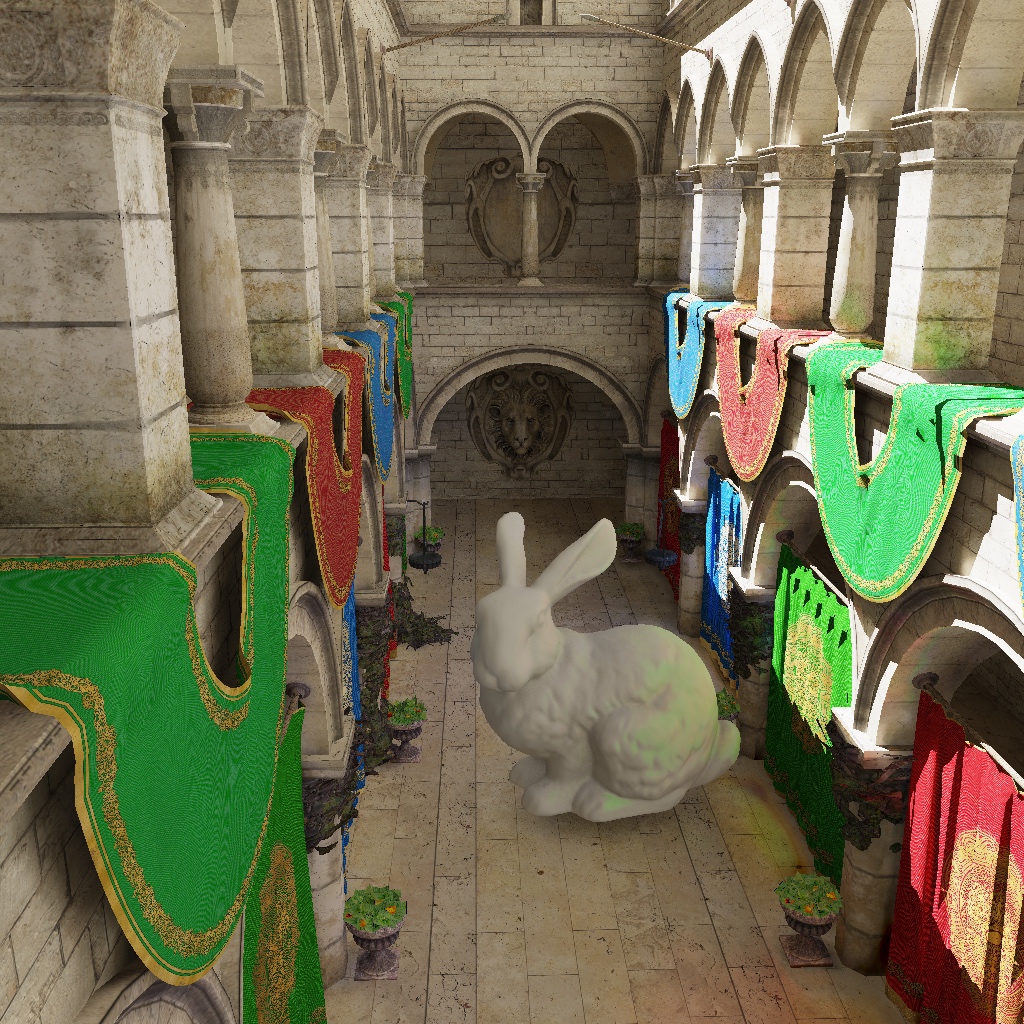}}\hfill
    \caption{
    Adding the angle of a directional light to NIV lets us model time-of-day changes in Sponza
    ---a common use case in game development---
    without requiring further training at runtime. The Bunny is unseen during training, but receives high-quality color bleed from the scene.
    }
    \label{fig:time_of_day}
\end{figure}

Training with many variable parameters, as highlighted before, significantly impacts the performance of a neural representation. To maintain performance we targeted learning 1-2 variables that model dynamic phenomena useful in practice.
We find it to be sufficient to apply frequency encoding ~\cite{mildenhall2021nerf} to such  changes, which only adds little inference overhead. Scaling to more than two scene variables at the same representational capacity causes visible reconstruction error, which could be resolved through the use of learned encodings, with the same disadvantages shown in Section~\ref{sec:variable_neural}. We implemented two use cases: a rotating directional light source simulating a time-of-day cycle and a moving occluding object in the scene. See results in Figure~\ref{fig:time_of_day} and the supplemental video.
To ensure artifact-free results, we double the size of the hash table that encodes the scene position for this application (i.e. $\sim10$MB).

\section{Limitations \& Future Work}
\textit{Direct Illumination.} 
A fundamental assumption of NIV is that direct illumination can be estimated efficiently at runtime without introducing sampling-related noise.
However, this assumption only holds for simple lighting and low numbers of lights, and may require additional render passes, e.g., shadow maps~\cite{williams1978casting}.
While many real-time techniques exist to handle the many lights problem~\cite{bitterli2020spatiotemporal}, they often exhibit significant noise levels themselves and thus efficient noise-free shading is impossible.

\textit{Irradiance Volume Assumptions.} Traditional probe based illumination assumes the impact of moving objects on the scene's light transport is small---or artist controlled~\cite{seyb20uberbake}---which holds true for many real-life applications and does not hinder its widespread use in games~\cite{barre2017certain}. We inherit this assumption, and do not capture the moving object's impact on the scene's light transport. As with probe based illumination, extensions would be needed for glossy materials~\cite{rodriguez2020glossy} or higher-order self-occlusions~\cite{sloan2002PRT}.
\textit{Compression.}
Further compressing multi-level hash encoding and MLP architectures is an active research topic~\cite{takikawa2023compact}. Exploiting such compression might further improve the memory-error trade-off and runtime cost of NIV. It is worth noting that such compression can introduce artifacts that might require problem-specific regularization terms or training parameter tuning~\cite{datta2023efficient}. Our early experiments show that using a strong Laplacian regularizer and a lower learning rate significantly helps reconstruction.

\textit{Glossy materials.} 
NIV can be used with non-diffuse materials by deferring queries to diffuse intersections, at the cost of sampling-related noise. To maintain noise-free rendering, alternative solutions could be explored. A promising direction incorporates surface roughness as a control parameter in the neural model~\cite{verbin2022ref}, possibly allowing the method to natively handle glossy materials.

\textit{Online Learning.}
Updating a learned cache at runtime has been shown to be possible for surface-based representations~\cite{muller2021real}.
Extending NIV with online learning works, but is inherently slower to learn since it has a larger input domain---the scene volume, as opposed to a subset of the scene surfaces---and requires a low bias representation as it avoids deferring queries to other scene locations.
We experimentally verified that starting from a learned NIV significantly improves convergence as opposed to learning from scratch. However, to update it in real-time, further extensions are required such as only sampling the camera's view frustum and/or employing loss-driven sampling~\cite{diolatzis2022active}.

\textit{Larger Scenes.} When scaling NIV to larger scenes, we found that model capacity eventually becomes a limiting factor, leading to increased reconstruction error. One way to address this is to partition the scene into a grid of smaller regions~\cite{reiser2021kilonerf} and learn a different NIV for each of them, which allows the method to retain accuracy by distributing representational capacity.  An interesting direction for future work is striking a balance between traditional spatial subdivision and the neural network capacity as explored in related work~\cite{weier2024n}. Another promising avenue is the integration of level-of-detail techniques, allocating capacity adaptively or selecting a lower frequency model based on the position of the camera.

\section{Conclusion}
We introduced a Neural Irradiance Volume (NIV), a novel approach that modernizes real-time global illumination for dynamic objects. 
By compressing the indirect irradiance field in a compact neural model, NIV overcomes the limitations of traditional probe-based methods, offering superior visual quality and an order of magnitude improvement in memory efficiency. 
Our technique provides a unified, noise-free, and high-quality solution that is practical for real-time applications. 
By eliminating the need for expensive runtime operations such as ray tracing or denoising, which are common in neural rendering techniques, NIV shows that high-quality indirect lighting is achievable without sacrificing performance.
We believe that NIV has the potential to benefit existing applications that utilize irradiance volumes, such as game engines, by reducing memory requirements and complexity, and improving visual quality. Furthermore, its differentiable nature makes it a promising building block for inverse rendering pipelines and other neural scene representations.

\newpage
\begin{figure*}[ph!]
    \centering
    \begin{tikzpicture}
    \node[align=right]{

   \begin{minipage}[t]{0.605\textwidth}
        \vspace{0pt}
        \centering
        \includegraphics[width=0.320\linewidth]{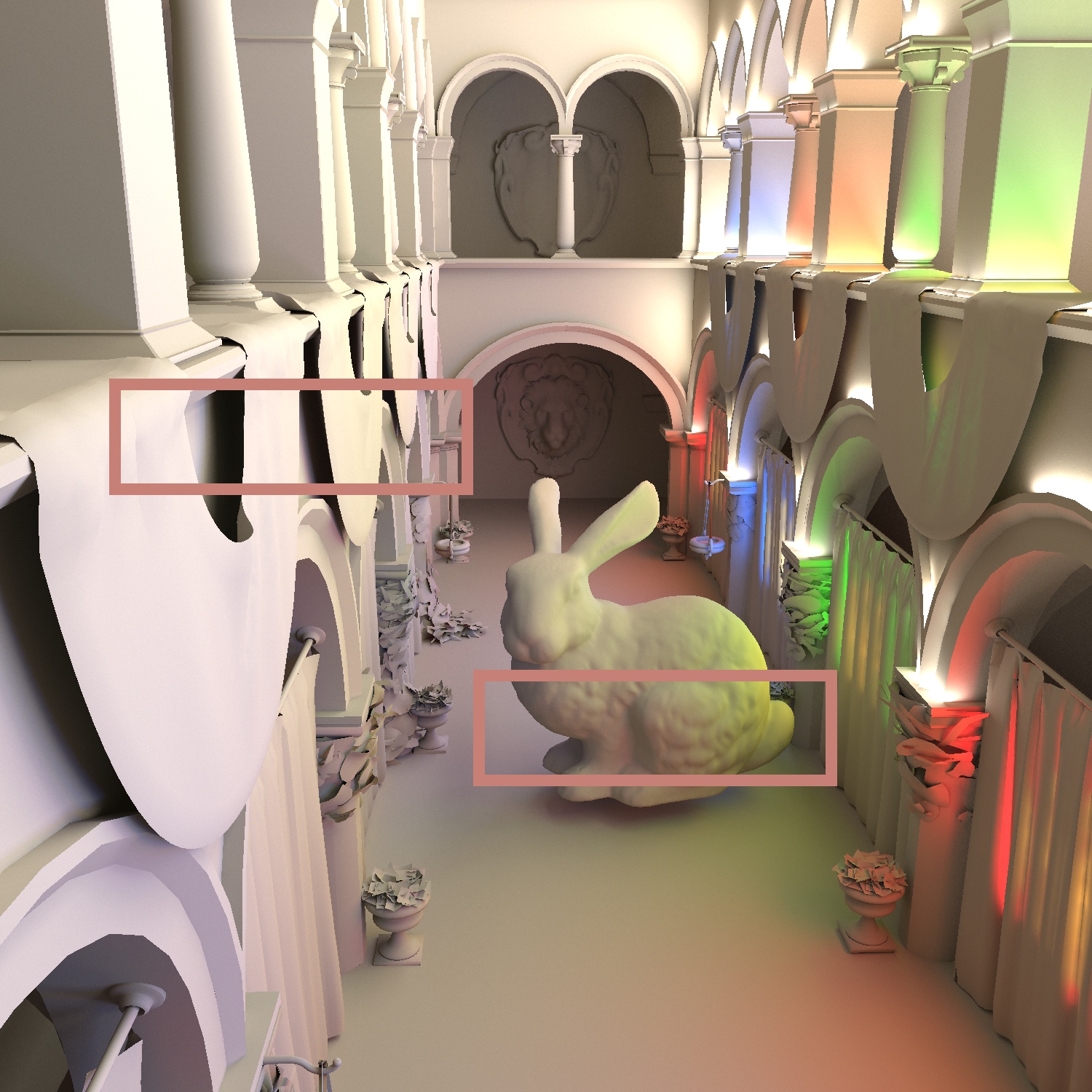}
        \hfill
        \includegraphics[width=0.320\linewidth]{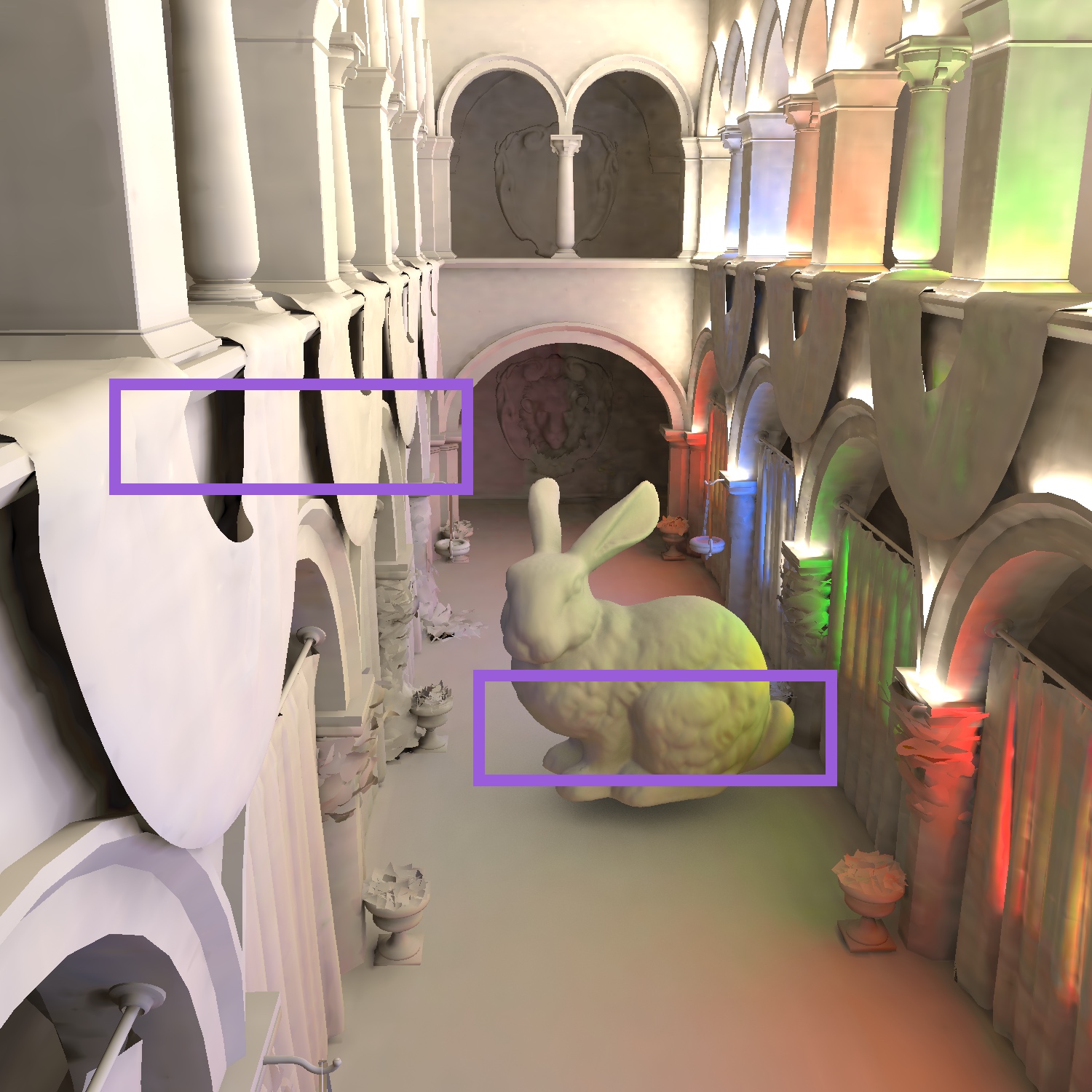}
        \hfill
        \includegraphics[width=0.320\linewidth]{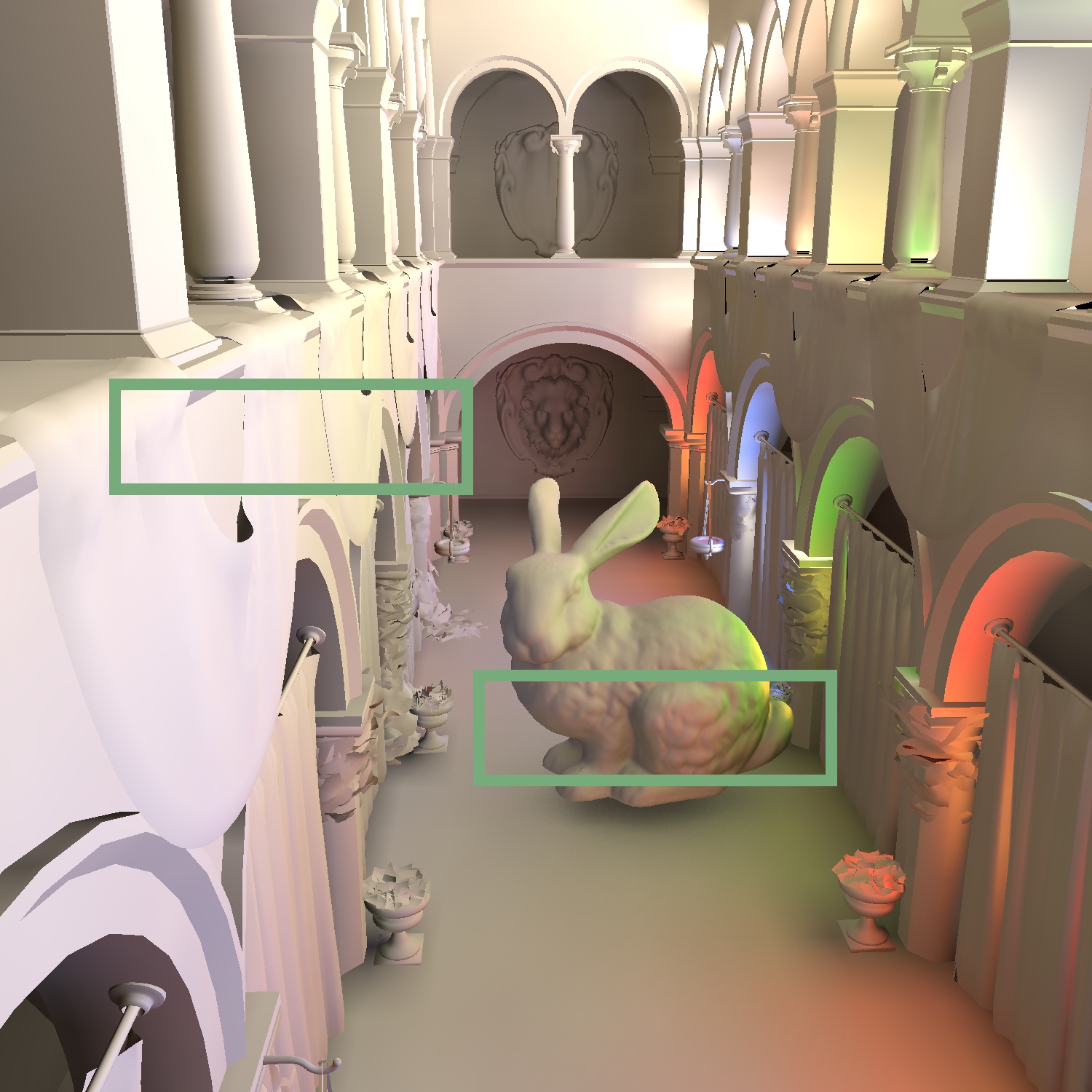}
    \end{minipage}
    \hfill
    \begin{minipage}[t]{0.195\textwidth}
        \vspace{0pt}
        \centering
        \includegraphics[width=\linewidth]{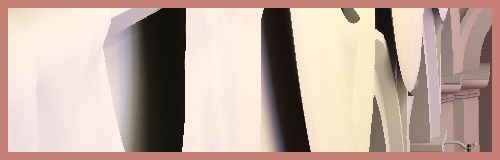} \\
        \vspace{0.6px}
        \includegraphics[width=\linewidth]{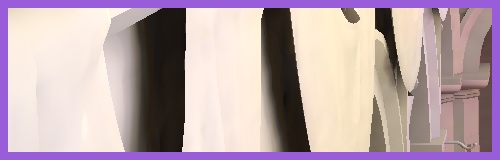} \\
        \vspace{0.6px}
        \includegraphics[width=\linewidth]{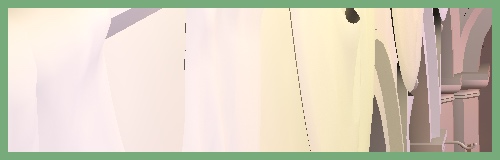} \\
    \end{minipage}
    \begin{minipage}[t]{0.195\textwidth}
        \vspace{0pt}
        \centering
        \includegraphics[width=\linewidth]{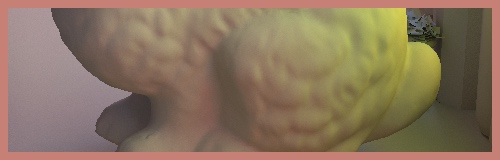} \\
        \vspace{0.6px}
        \includegraphics[width=\linewidth]{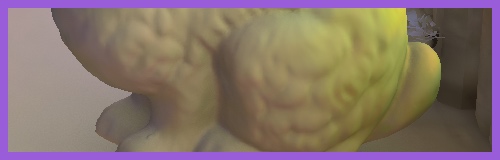} \\
        \vspace{0.6px}
        \includegraphics[width=\linewidth]{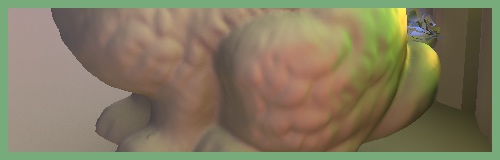} \\
    \end{minipage}

    \\ \mbox{} \\

   \begin{minipage}[t]{0.605\textwidth}
        \vspace{0pt}
        \centering
        \includegraphics[width=0.320\linewidth]{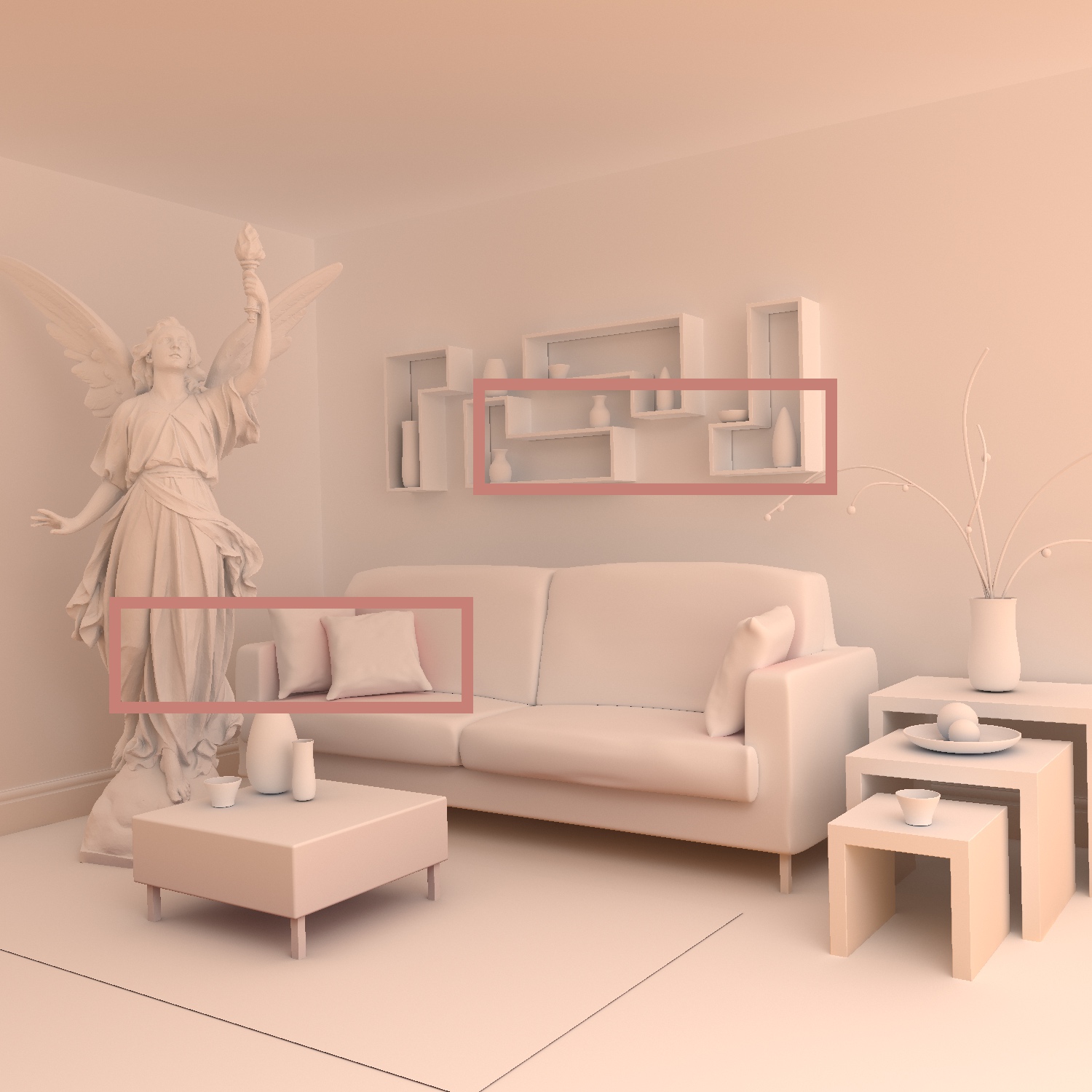}
        \hfill
        \includegraphics[width=0.320\linewidth]{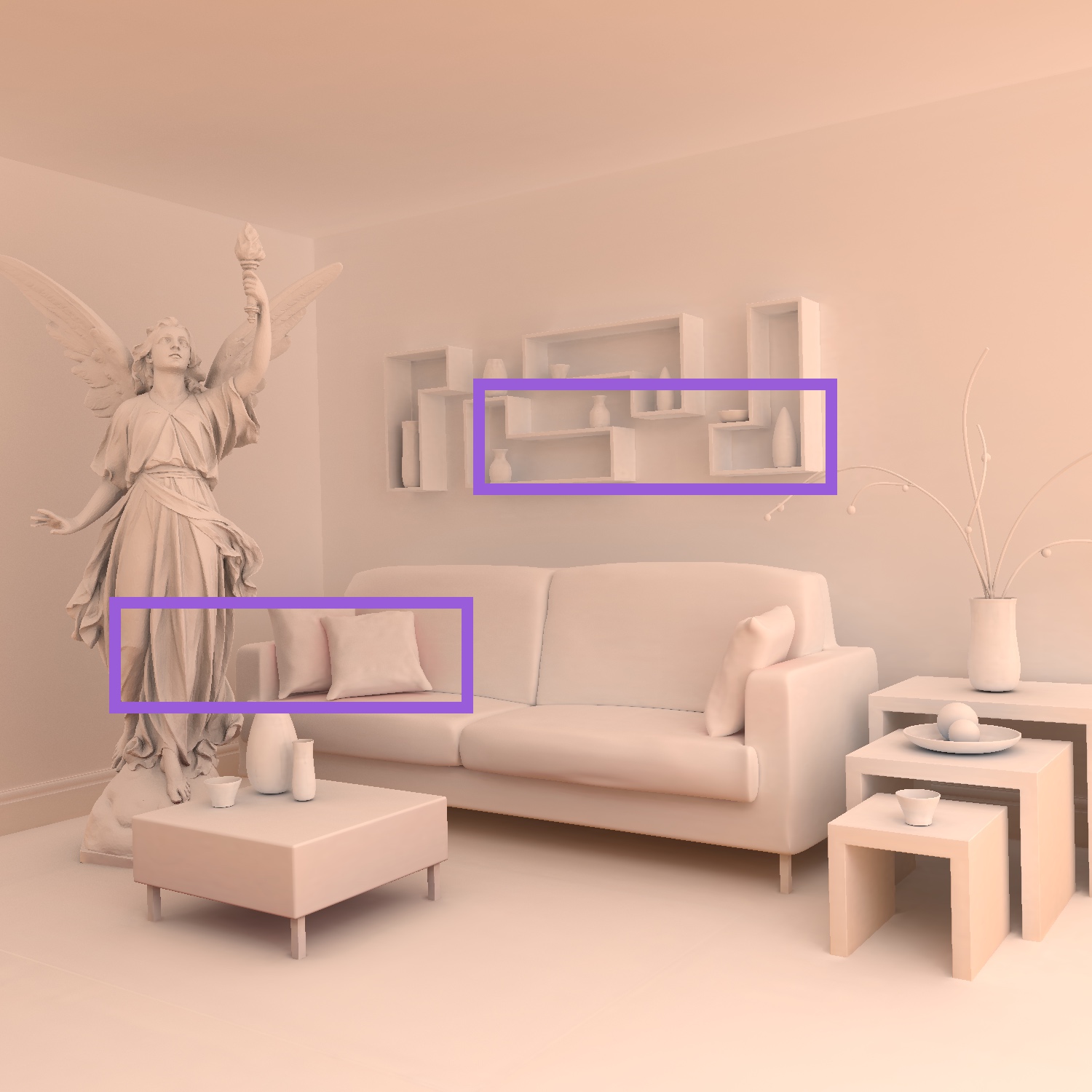}
        \hfill
        \includegraphics[width=0.320\linewidth]{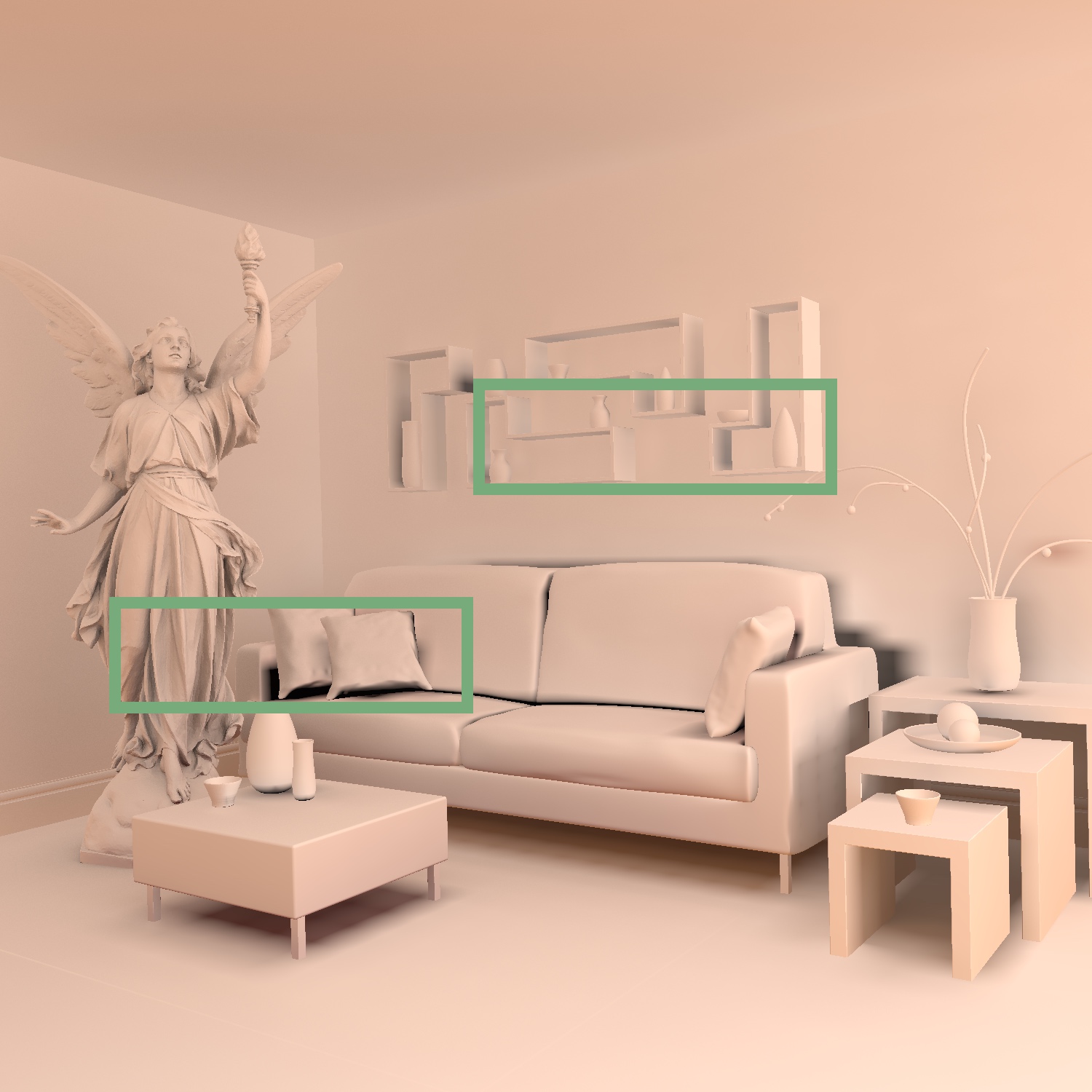}
    \end{minipage}
    \hfill
    \begin{minipage}[t]{0.195\textwidth}
        \vspace{0pt}
        \centering
        \includegraphics[width=\linewidth]{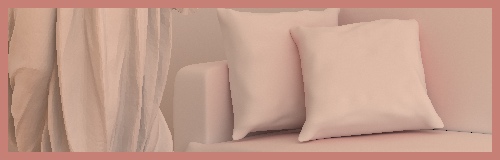} \\
        \vspace{0.6px}
        \includegraphics[width=\linewidth]{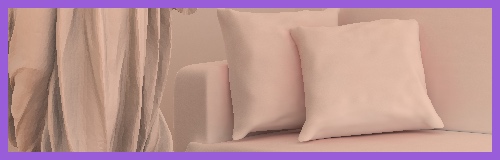} \\
        \vspace{0.6px}
        \includegraphics[width=\linewidth]{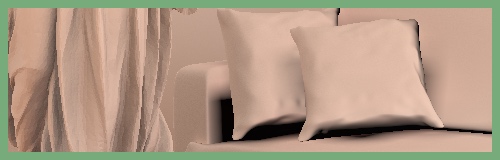} \\
    \end{minipage}
    \begin{minipage}[t]{0.195\textwidth}
        \vspace{0pt}
        \centering
        \includegraphics[width=\linewidth]{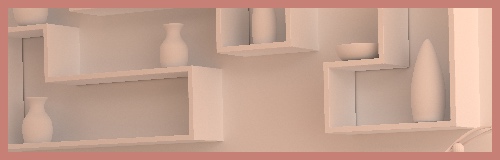} \\
        \vspace{0.6px}
        \includegraphics[width=\linewidth]{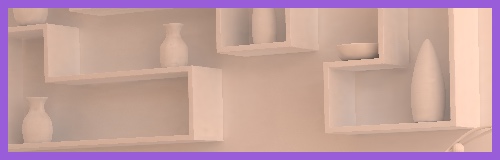} \\
        \vspace{0.6px}
        \includegraphics[width=\linewidth]{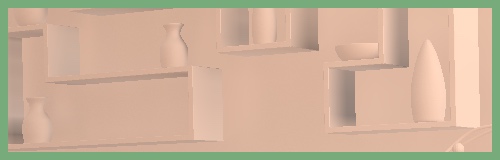} \\
    \end{minipage}

    \\ \mbox{} \\

   \begin{minipage}[t]{0.605\textwidth}
        \vspace{0pt}
        \centering
        \includegraphics[width=0.320\linewidth]{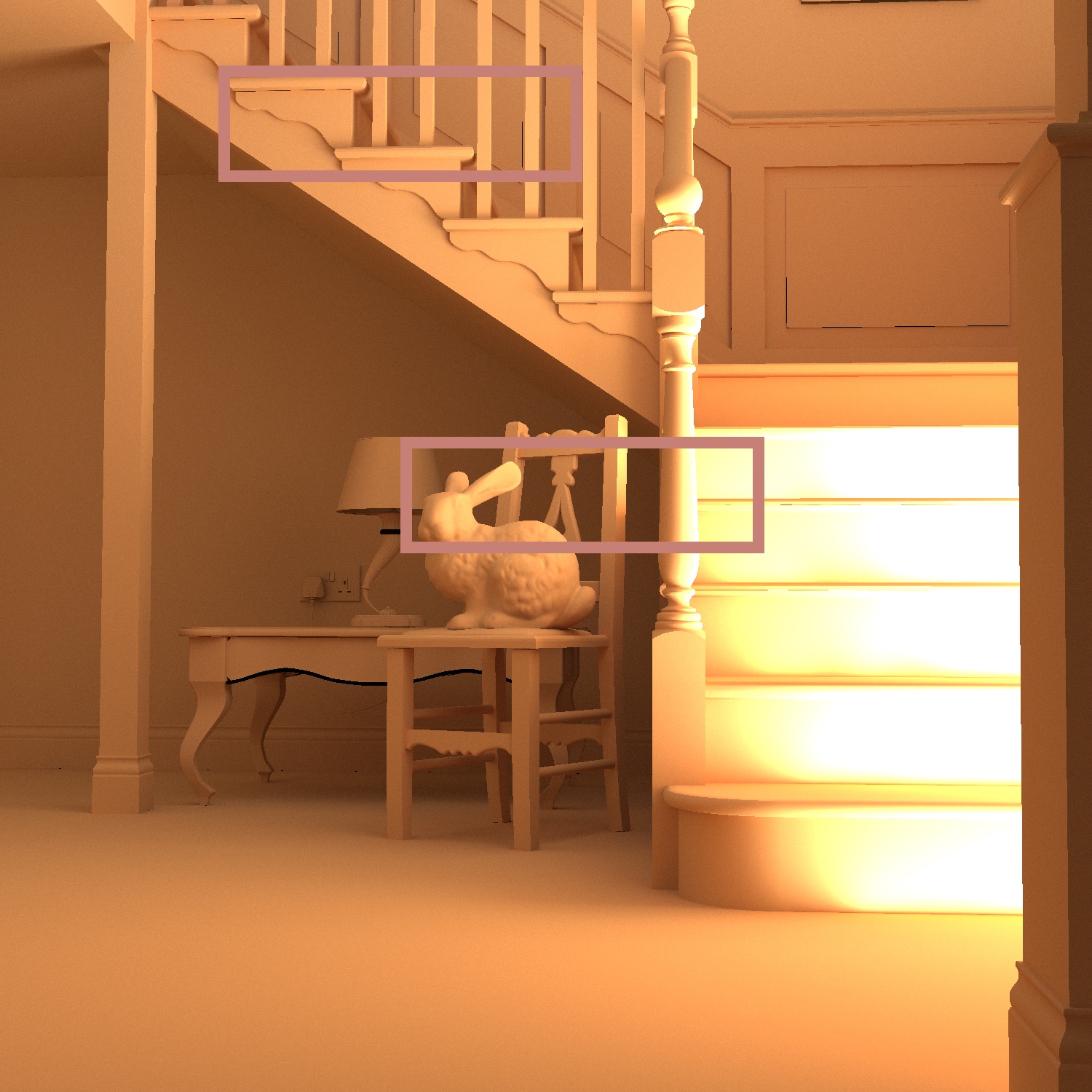}
        \hfill
        \includegraphics[width=0.320\linewidth]{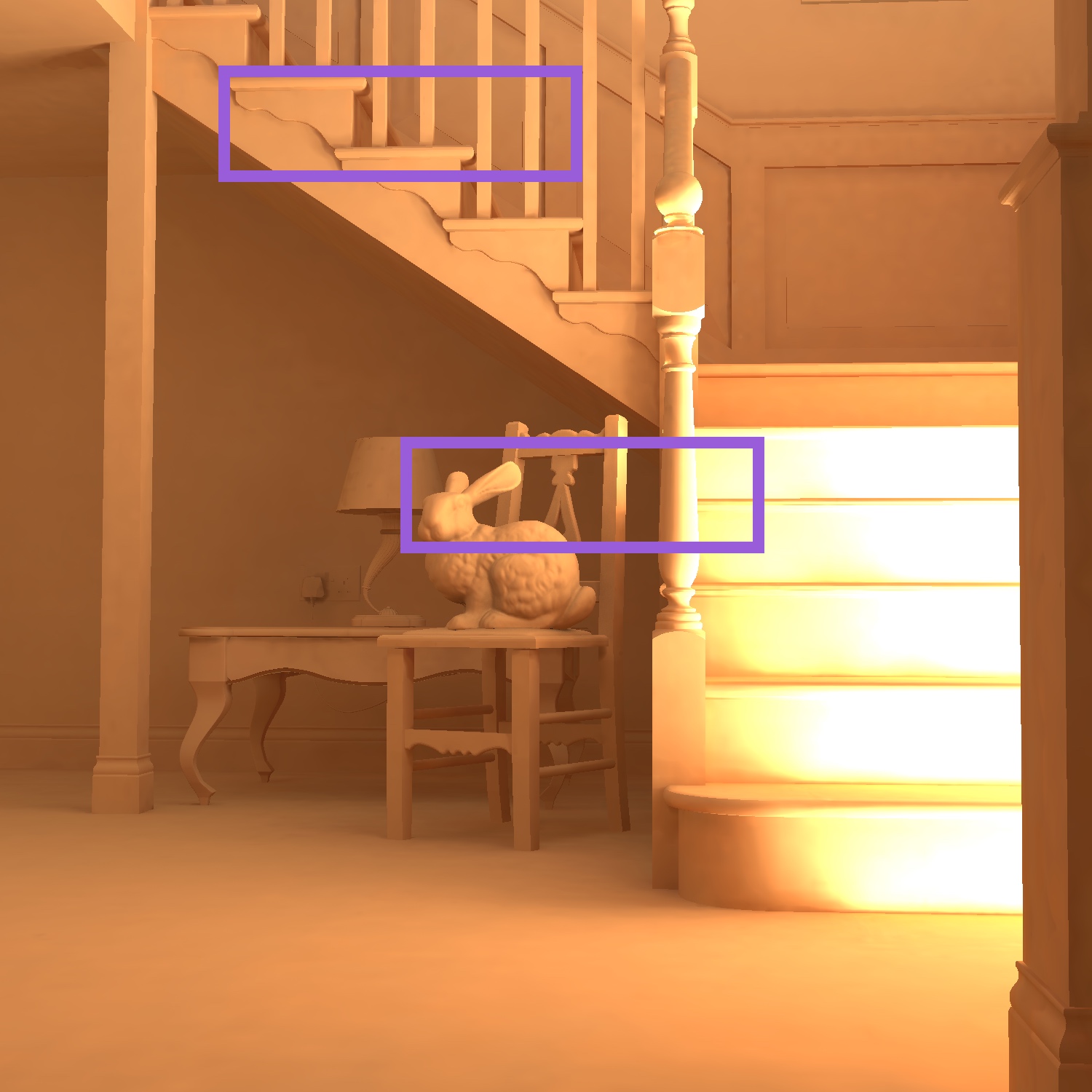}
        \hfill
        \includegraphics[width=0.320\linewidth]{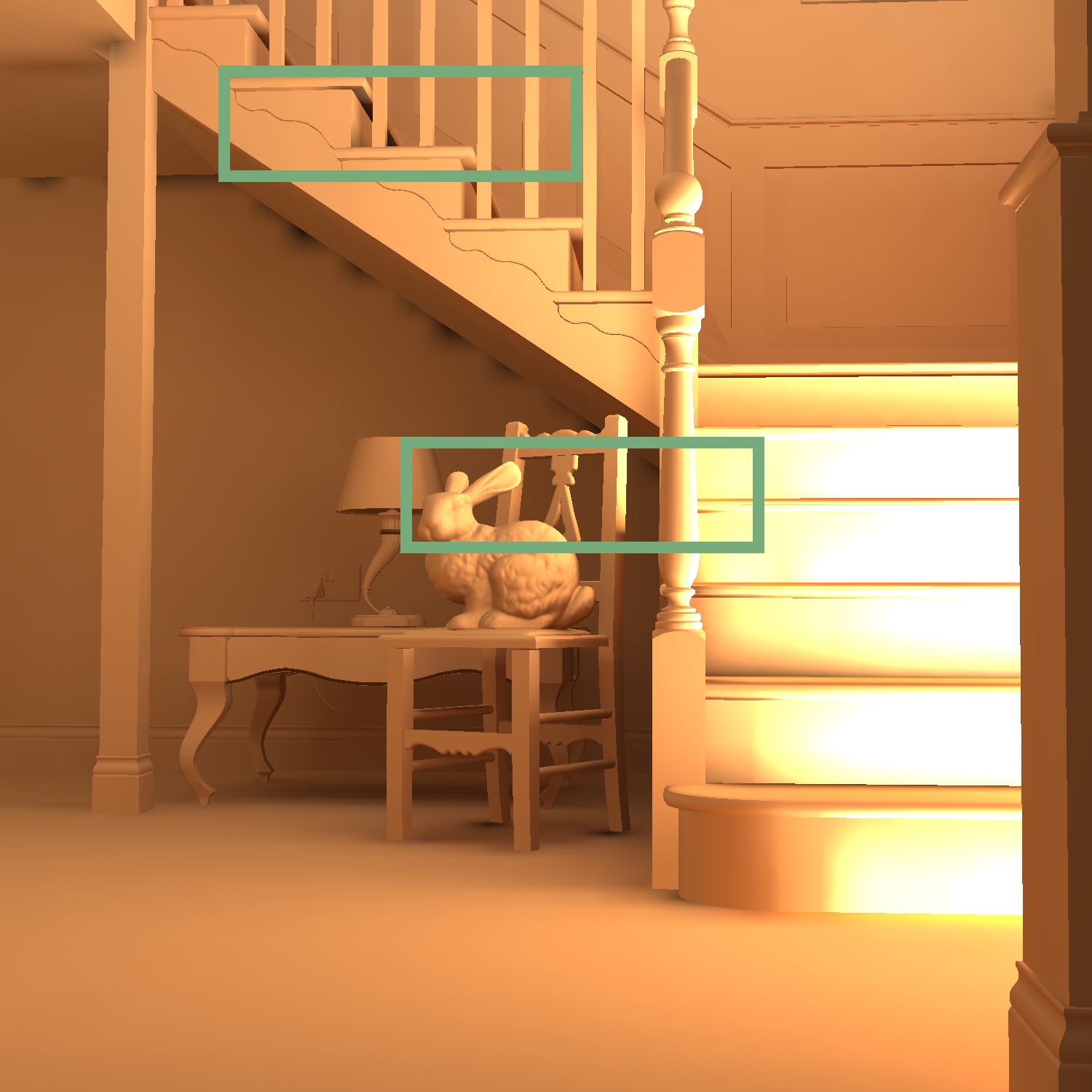}
    \end{minipage}
    \hfill
    \begin{minipage}[t]{0.195\textwidth}
        \vspace{0pt}
        \centering
        \includegraphics[width=\linewidth]{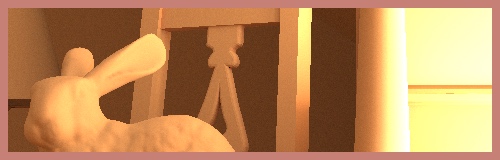} \\
        \vspace{0.6px}
        \includegraphics[width=\linewidth]{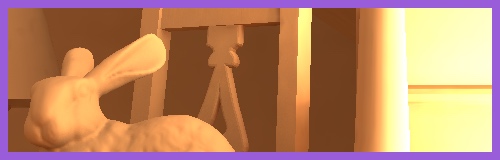} \\
        \vspace{0.6px}
        \includegraphics[width=\linewidth]{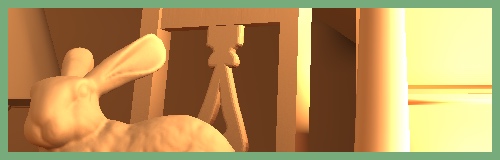} \\
    \end{minipage}
    \begin{minipage}[t]{0.195\textwidth}
        \vspace{0pt}
        \centering
        \includegraphics[width=\linewidth]{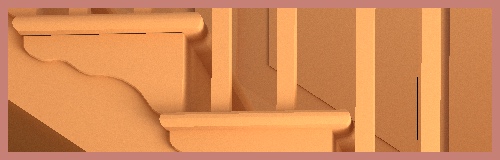} \\
        \vspace{0.6px}
        \includegraphics[width=\linewidth]{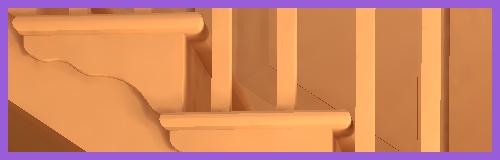} \\
        \vspace{0.6px}
        \includegraphics[width=\linewidth]{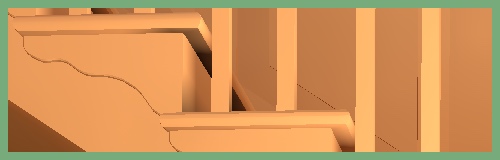} \\
    \end{minipage}

    \\ \mbox{} \\

   \begin{minipage}[t]{0.605\textwidth}
        \vspace{0pt}
        \centering
        \includegraphics[width=0.320\linewidth]{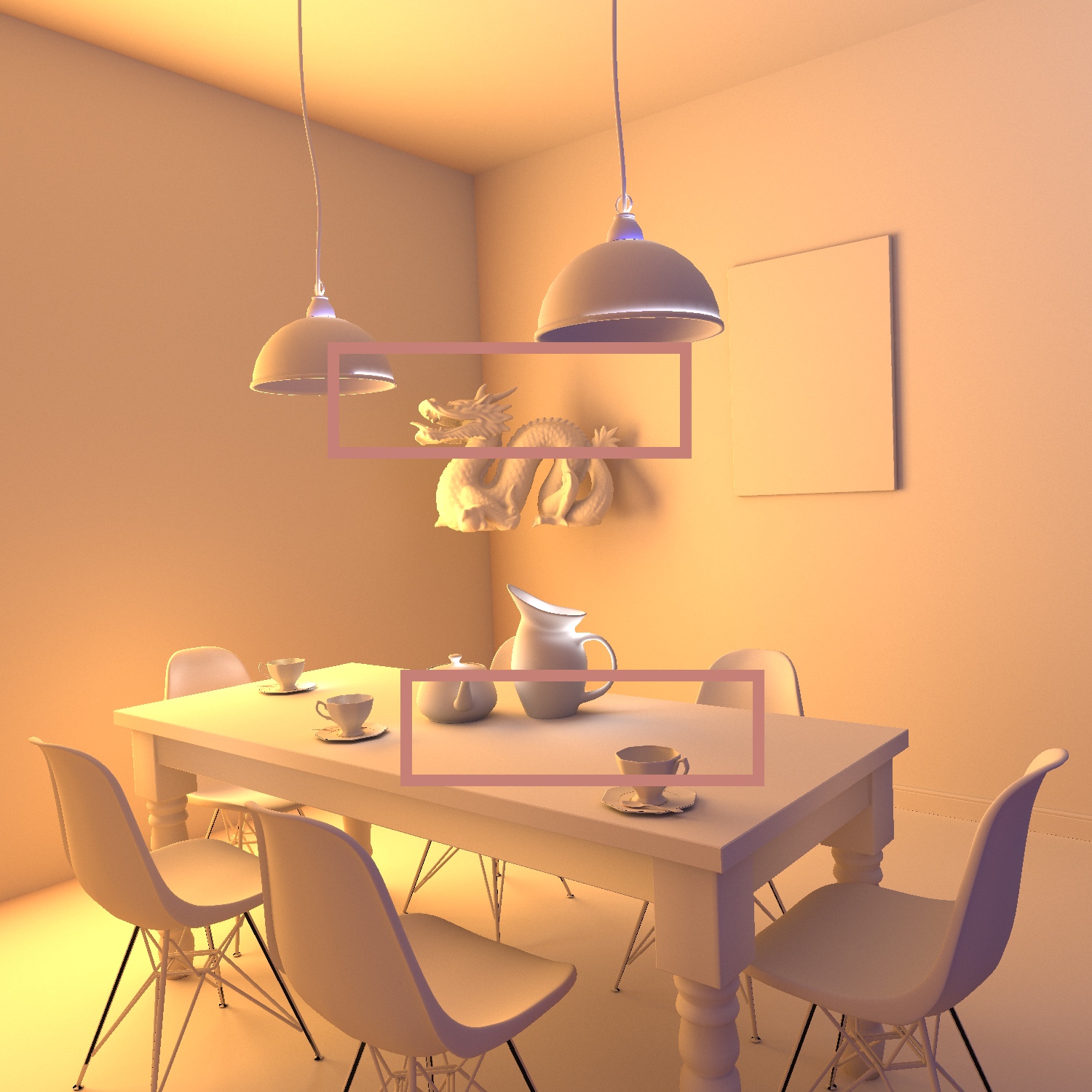}
        \hfill
        \includegraphics[width=0.320\linewidth]{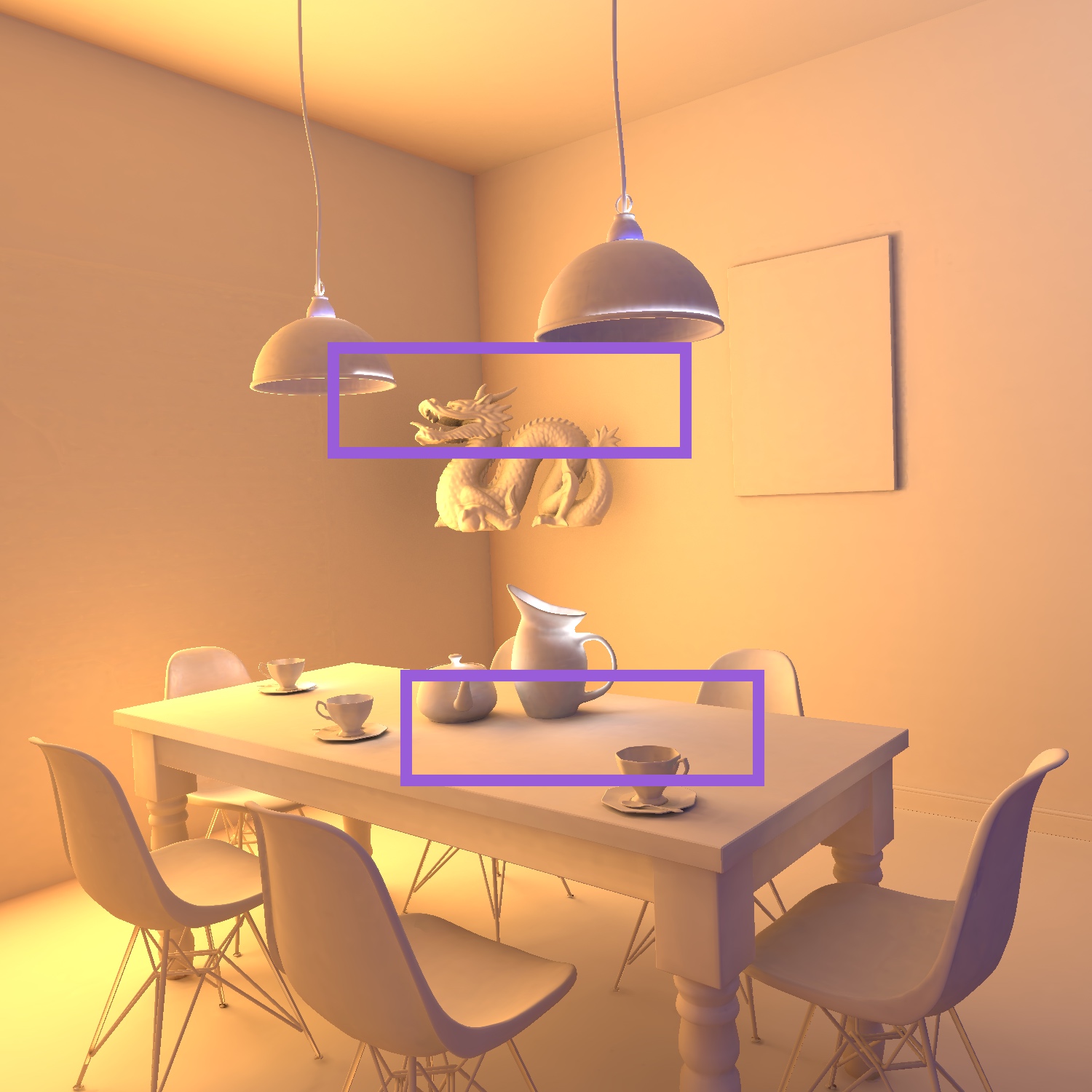}
        \hfill
        \includegraphics[width=0.320\linewidth]{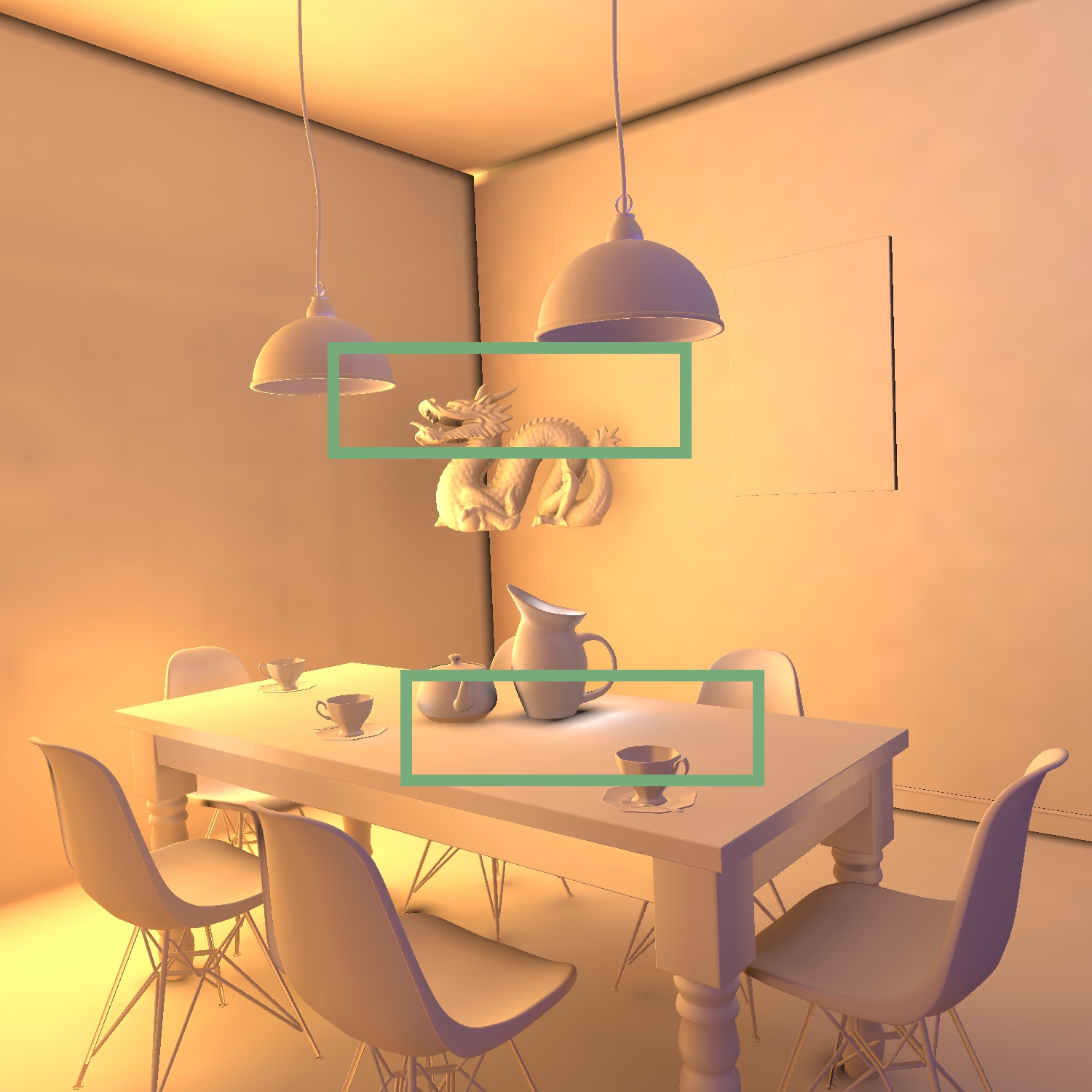}
    \end{minipage}
    \hfill
    \begin{minipage}[t]{0.195\textwidth}
        \vspace{0pt}
        \centering
        \includegraphics[width=\linewidth]{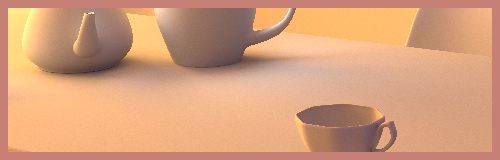} \\
        \vspace{0.6px}
        \includegraphics[width=\linewidth]{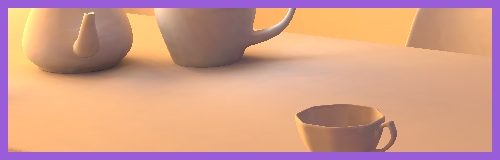} \\
        \vspace{0.6px}
        \includegraphics[width=\linewidth]{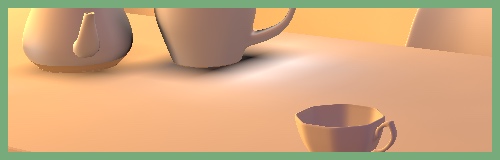} \\
    \end{minipage}
    \begin{minipage}[t]{0.195\textwidth}
        \vspace{0pt}
        \centering
        \includegraphics[width=\linewidth]{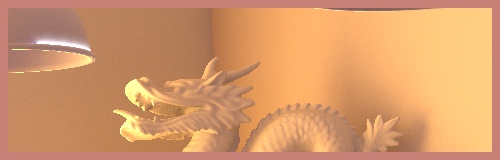} \\
        \vspace{0.6px}
        \includegraphics[width=\linewidth]{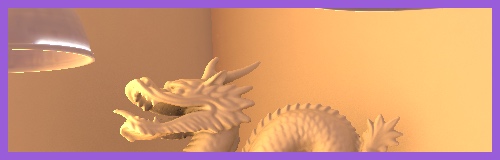} \\
        \vspace{0.6px}
        \includegraphics[width=\linewidth]{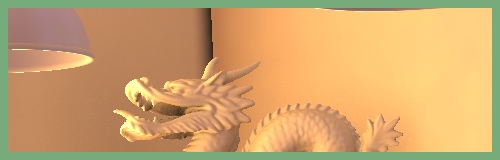} \\
    \end{minipage}

  \\ \mbox{} \\ 

   \begin{minipage}[t]{0.605\textwidth}
        \vspace{0pt}
        \centering
        \includegraphics[width=0.320\linewidth]{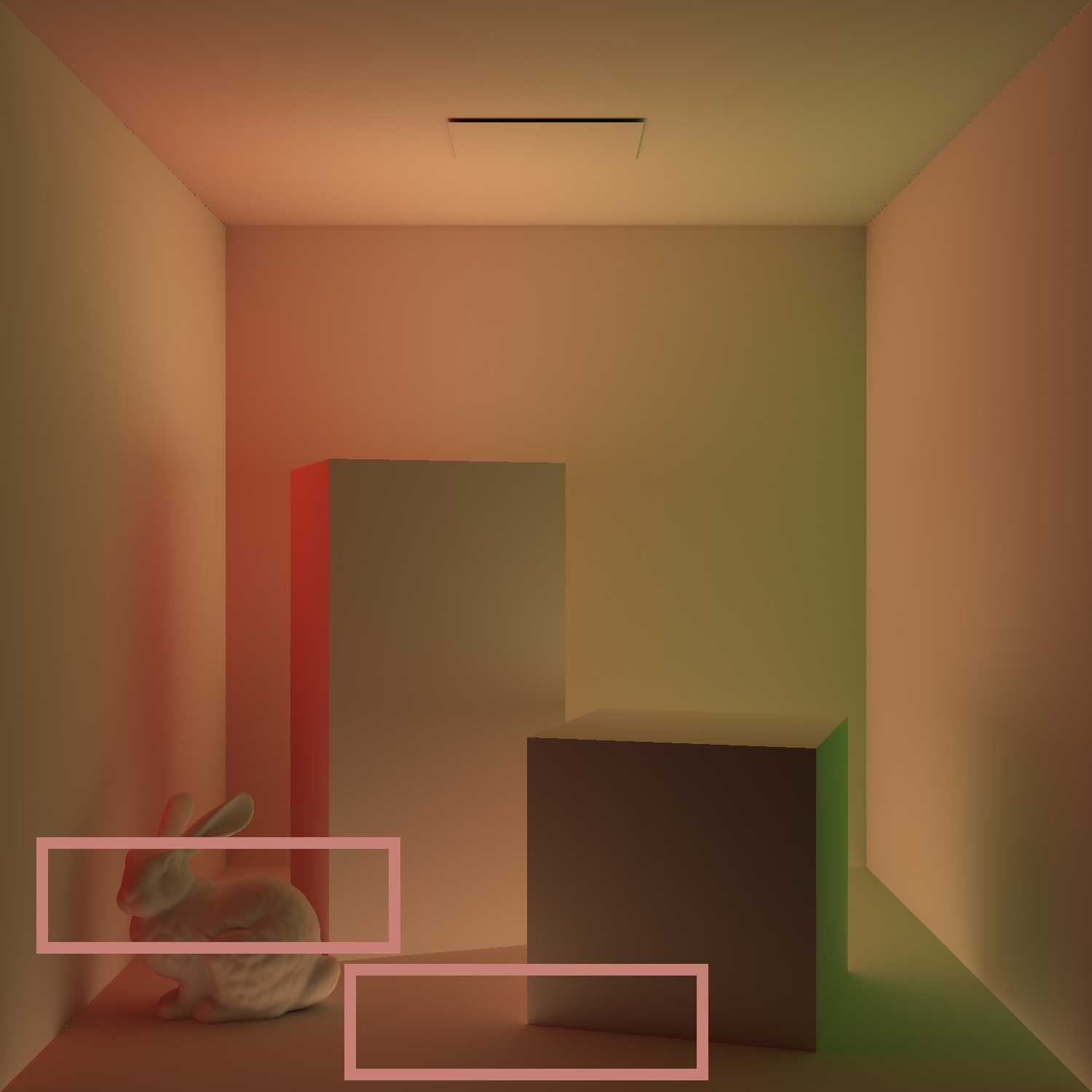}
        \hfill
        \includegraphics[width=0.320\linewidth]{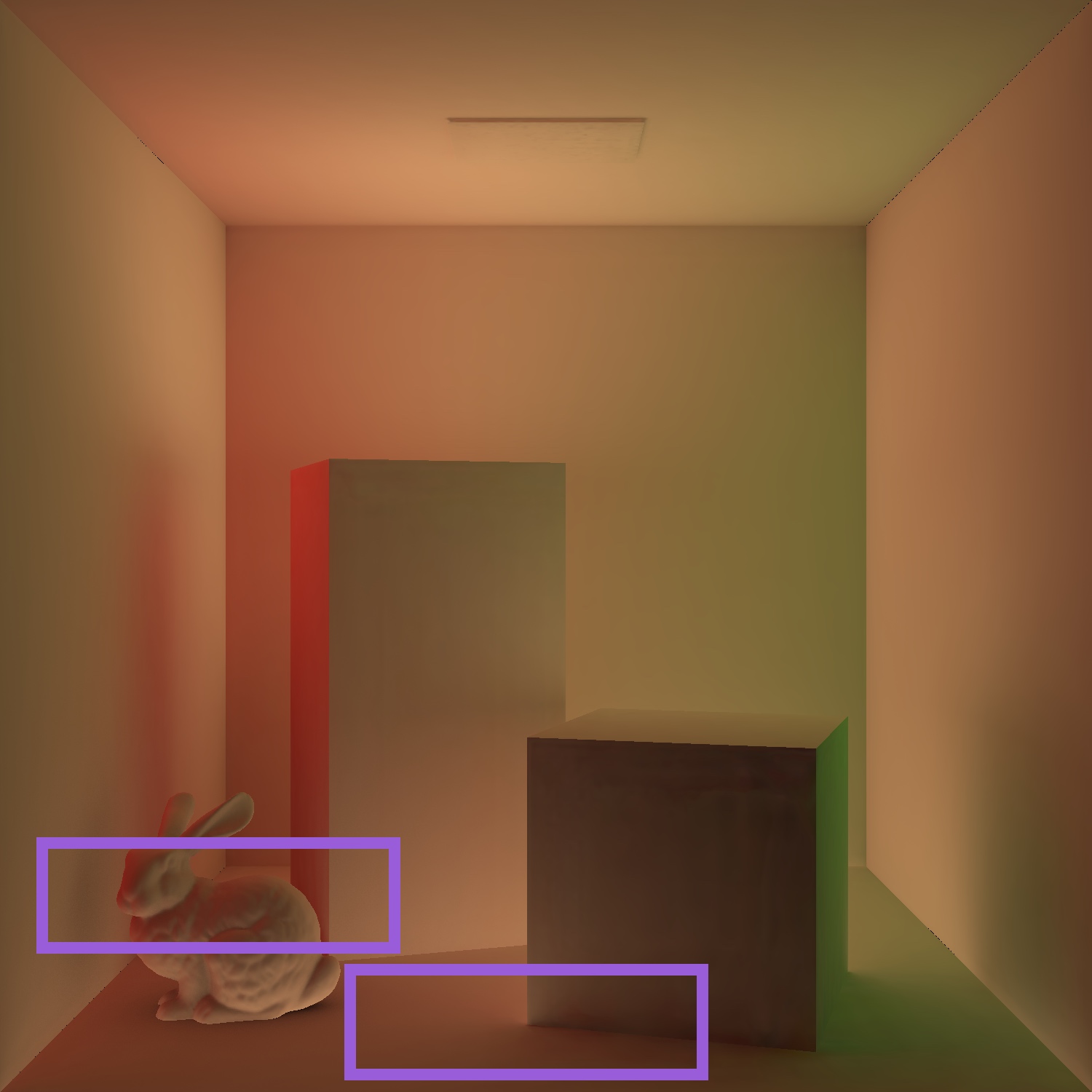}
        \hfill
        \includegraphics[width=0.320\linewidth]{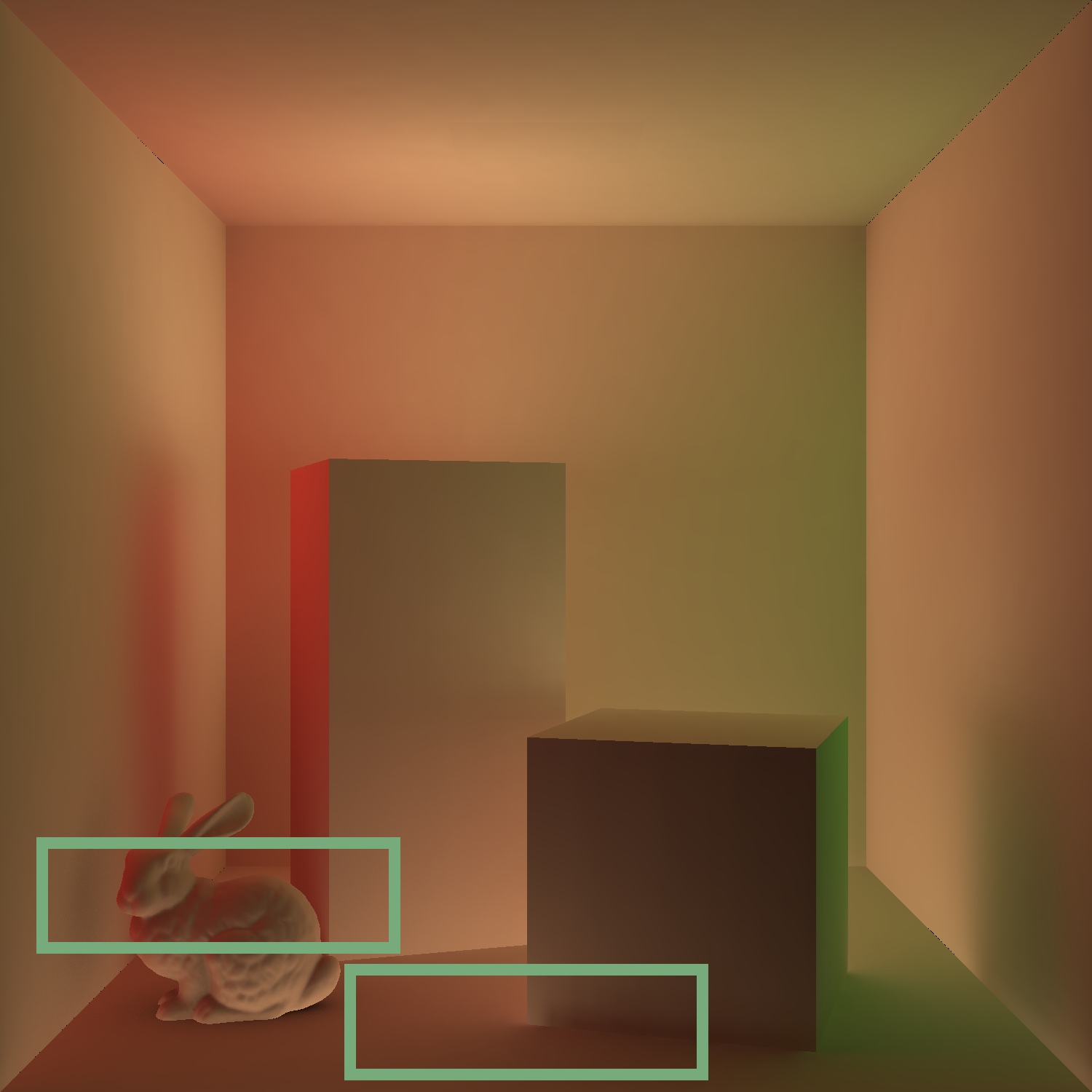}
    \end{minipage}
    \hfill
    \begin{minipage}[t]{0.195\textwidth}
        \vspace{0pt}
        \centering
        \includegraphics[width=\linewidth]{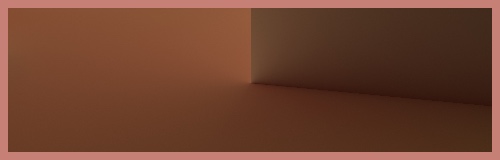} \\
        \vspace{0.6px}
        \includegraphics[width=\linewidth]{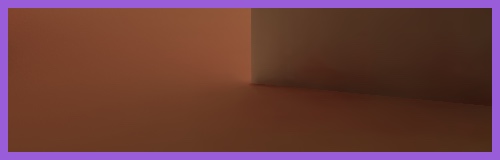} \\
        \vspace{0.6px}
        \includegraphics[width=\linewidth]{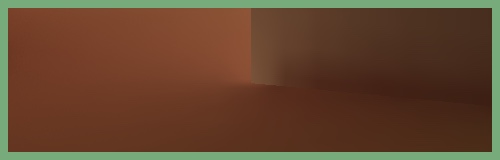} \\
    \end{minipage}
    \begin{minipage}[t]{0.195\textwidth}
        \vspace{0pt}
        \centering
        \includegraphics[width=\linewidth]{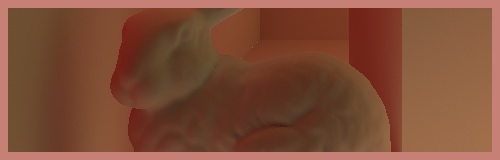} \\
        \vspace{0.6px}
        \includegraphics[width=\linewidth]{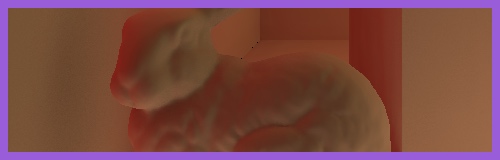} \\
        \vspace{0.6px}
        \includegraphics[width=\linewidth]{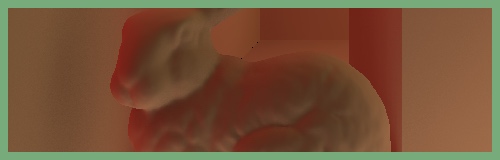} \\
    \end{minipage}

    };

    \draw (-7, 9.7) node {Reference};
    \draw (-3.5, 9.7) node {NIV (Ours)};
    \draw (0, 9.7) node {Probes};
    \draw (5.25, 9.7) node {Insets: \textbf{\color[HTML]{95382F}Reference}, \textbf{\color[HTML]{531CB3}NIV (Ours)}, \textbf{\color[HTML]{306B34}Probes}  };
    \end{tikzpicture}

    \caption{Overview of the indirect irradiance---i.e. without albedo for clarity---using our Neural Irradiance Volume. The movable objects \textit{Lucy}, \textit{Bunny} and \textit{Dragon}  are placed in the scene at render time and are not seen during training. Since NIV and probe volumes capture only the indirect irradiance of the known scene, dynamic ambient occlusion is also added as described in Section~\ref{sec:render}. All insets can be compared with a high sample count path traced reference. Compared to NIV, probe grids noticeably struggle with several kinds of light leaks and incorrect color bleed.}  
    \label{fig:niv_scenes}
\end{figure*}

\FloatBarrier

\bibliographystyle{eg-alpha-doi}
\bibliography{bibliography}

\clearpage

\appendix

\section{Probe Base Implementation}
We implement a probe-based technique with similar design choices as DDGI~\cite{majercik2019dynamic}. To allow for quantitative comparisons using the same reference images computed using mitsuba3~\cite{jakob2022mitsuba3}, we implemented DDGI in the same code base as NIV.
As mentioned in the main text, our probe implementation supports visibility based probe rejection using ray tracing, which replaces the cosine fall-off heuristic mentioned below. The choices with respect to shading-related heuristics we made while implementing DDGI are:
\begin{itemize}
    \item the directional indirect irradiance is represented using second-order spherical harmonics (9 coefficients)~\cite{ramamoorthi2001efficient} which are quantized to half-precision after baking for memory efficiency (54 bytes / probe).
    \item a cosine fall-off~\cite{hooker2016volumetric} is applied  such that probes behind the sampled normal do not contribute to trilinear interpolation
    \item the interpolation weights are clamped to a small number ($1e-6$) to avoid no probes contributing to a shading point, similar to DDGI's implementation~\cite{majercik2019dynamic}.
    \item interpolation weights are re-normalized after the above heuristics
\end{itemize}

\section{Open Source DDGI Comparison}

\begin{figure}[H]

    \centering
    \subfloat[Ours]{%
        \includegraphics[width=0.46\linewidth]{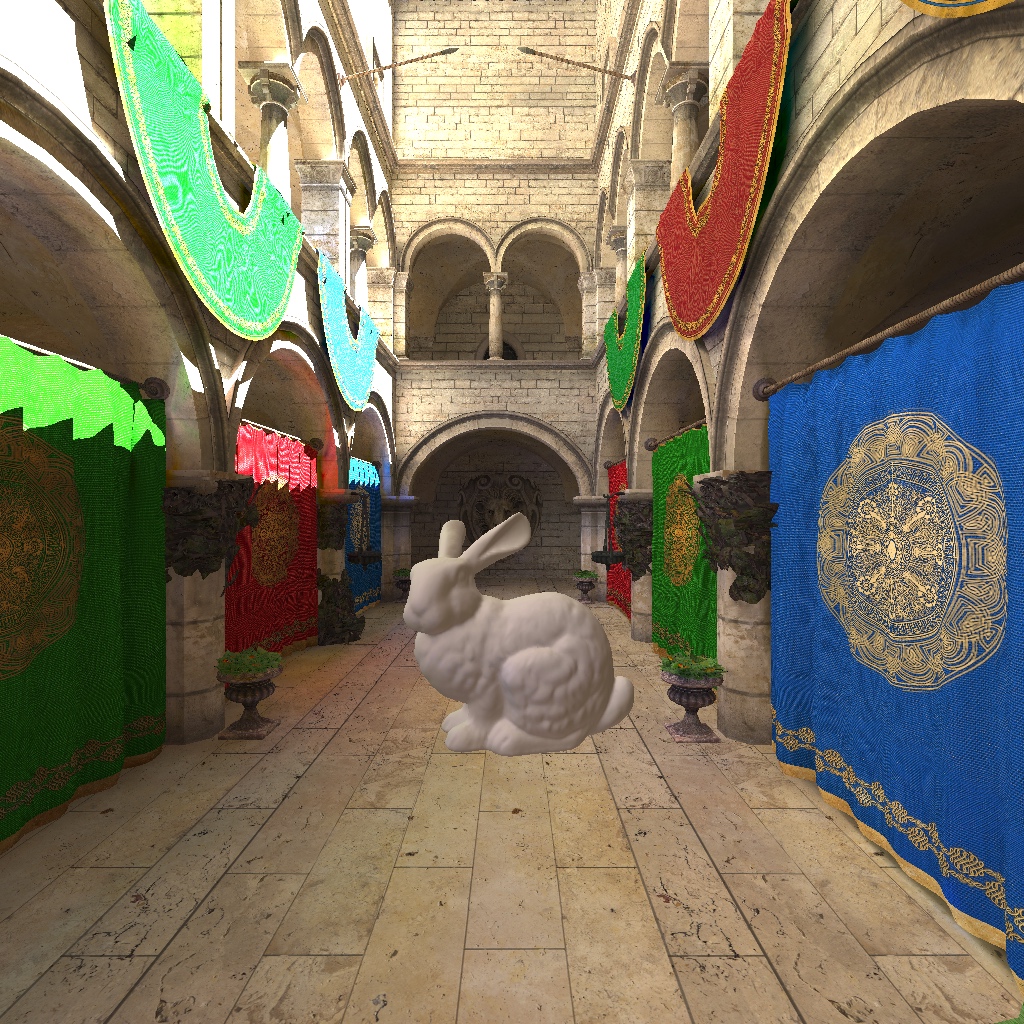}%
        \label{fig:ddgi_supp}
    }
    \hfill
    \subfloat[Our Reference]{%
        \includegraphics[width=0.46\linewidth]{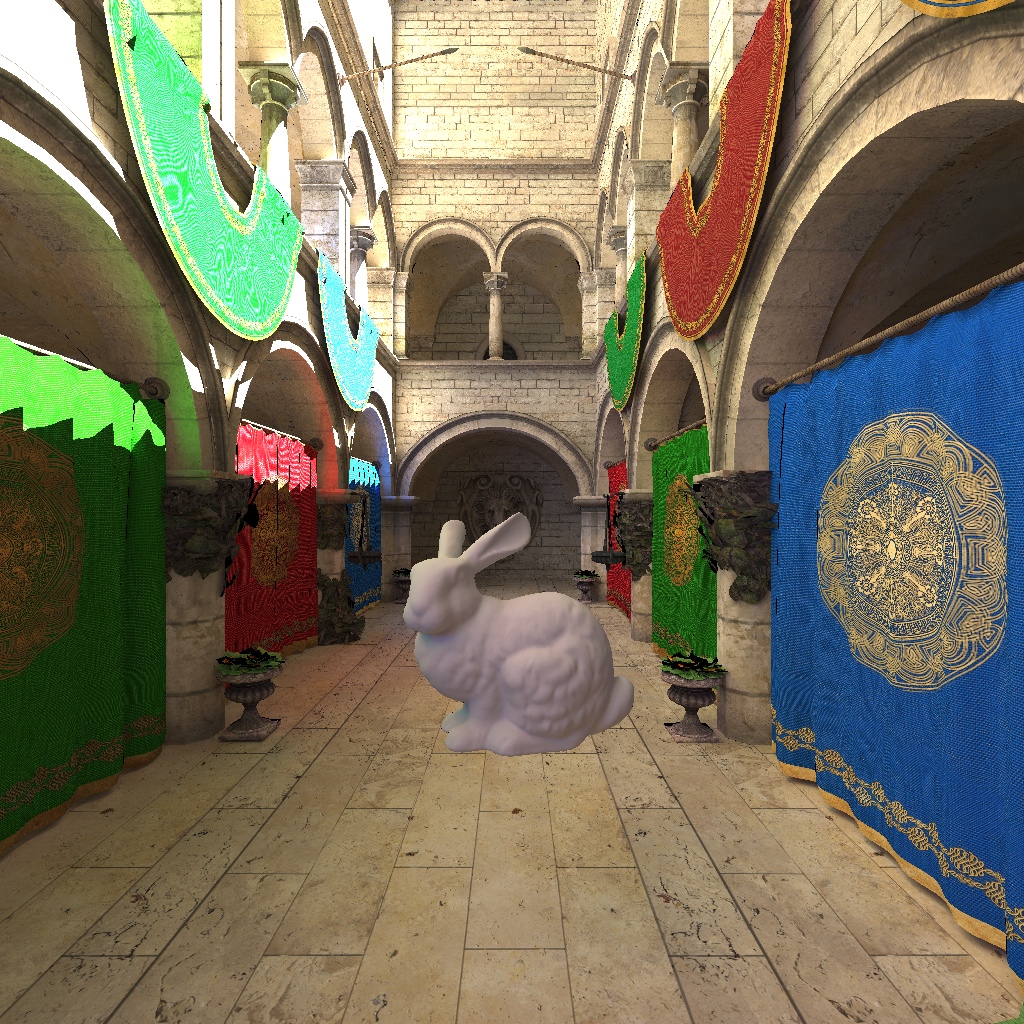}%
        \label{fig:ours_supp}
    }
    \\
    \subfloat[DDGI]{%
        \includegraphics[width=0.46\linewidth]{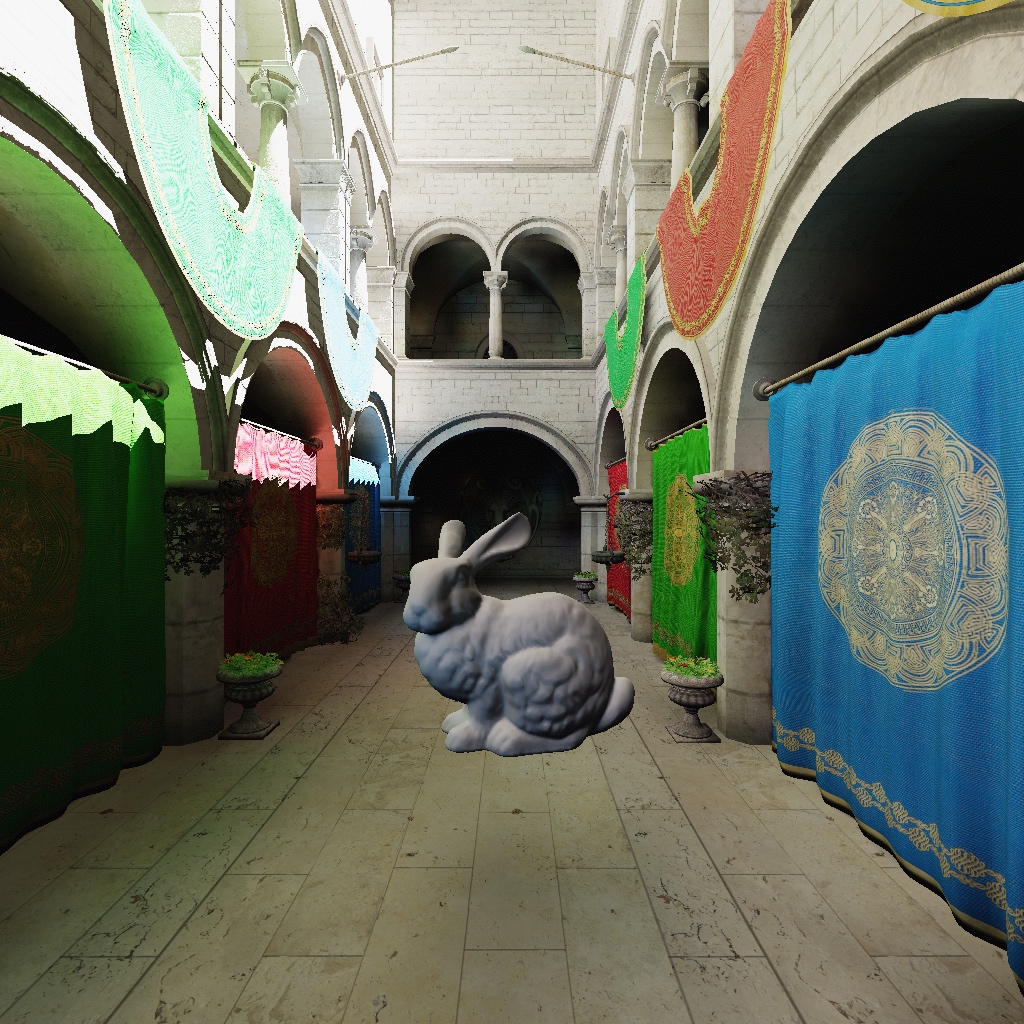}%
    }
    \hfill
    \subfloat[DDGI Reference]{%
        \includegraphics[width=0.46\linewidth]{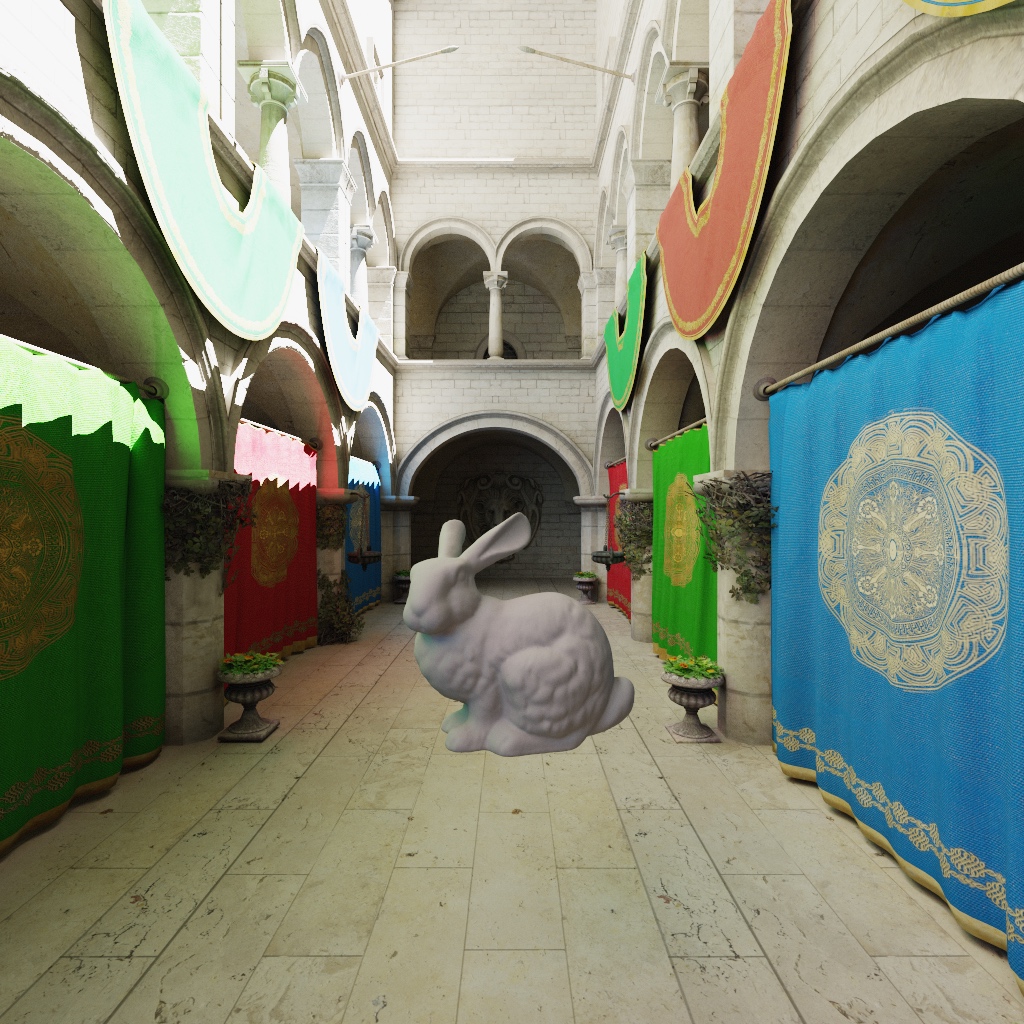}%

    }
        \vspace{-1em}
    \caption{
    Comparison with the open-source DDGI implementation~\cite{nvidia_rtxgi_ddgi}. Both approaches employ their respective path tracers for the reference images, resulting in differences due to varying material models. DDGI uses its default settings for \textit{Sponza} with a $22^3$ probe grid, consuming 35.5~MB, whereas our method requires only about 0.25~MB. Neither approach uses ambient occlusion, and both rely on a single directional light. Notice the overall shift in shading tone and the absence of contact shadows on the banners along the right side of \textit{Sponza} when using DDGI. Furthermore, DDGI erroneously positions a probe inside the bunny, which, combined with their runtime updates, produces a dark smudge on the floor.}
    \label{fig:sponza_reference_supp}
\end{figure}

\section{Memory-Error Trade-off Curves}
\begin{figure}[H]
    \includegraphics[width=0.9\linewidth]{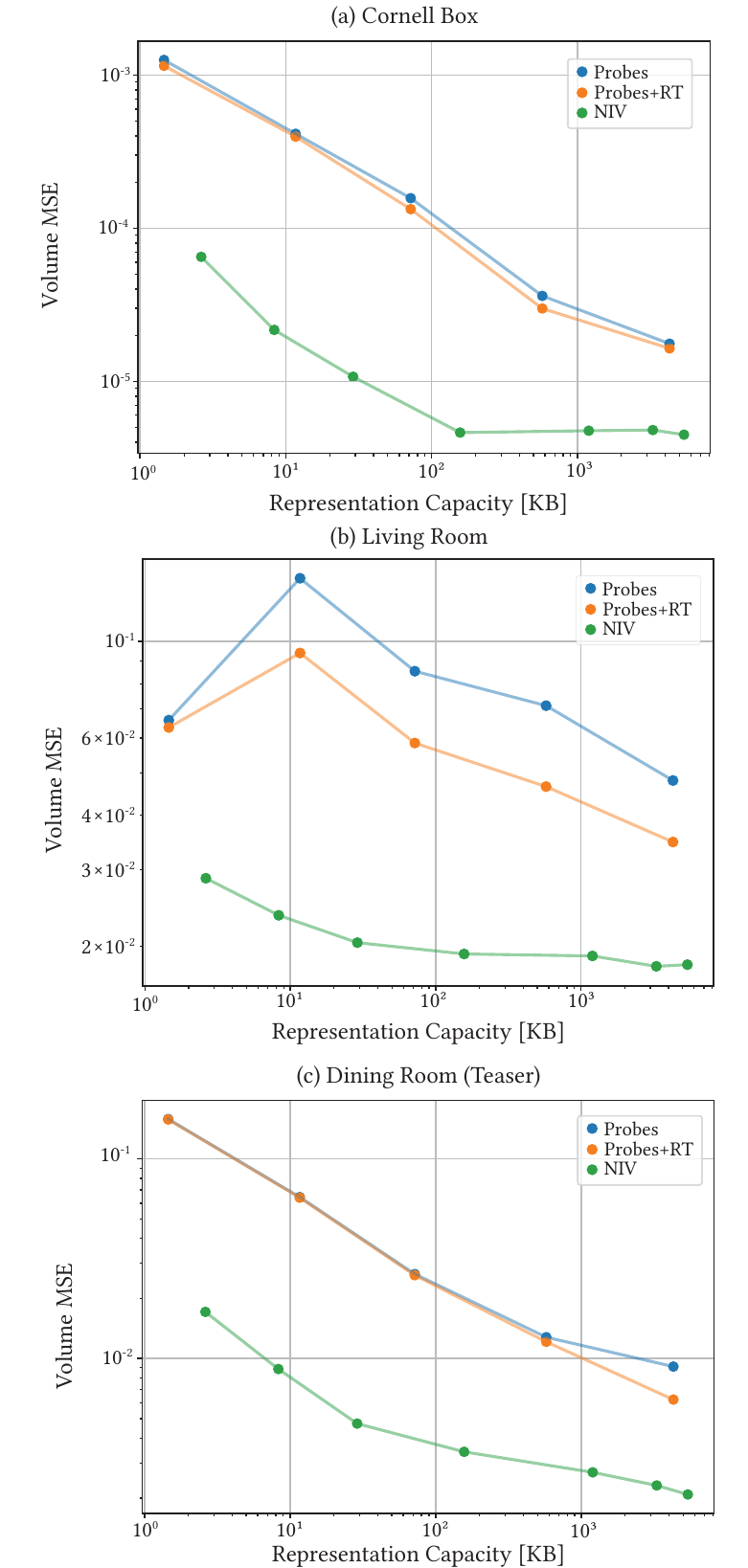}
    \caption{Additional results comparing probe-based techniques to our method. Since probes are placed on a regular grid, certain grid configurations can increase its volumetric error. All three scenes are from Bitterli's Rendering resources~\cite{resources16}, the \textit{Dining Room (Teaser)} scene is adapted by changing the color of a wall as shown in the teaser image of the main text.}
\end{figure}

\section{Learning Incoming Radiance}
\begin{figure}[H]
    \centering
    \subfloat[Pre-integrated irradiance $E$]{\includegraphics[width=0.48\linewidth]{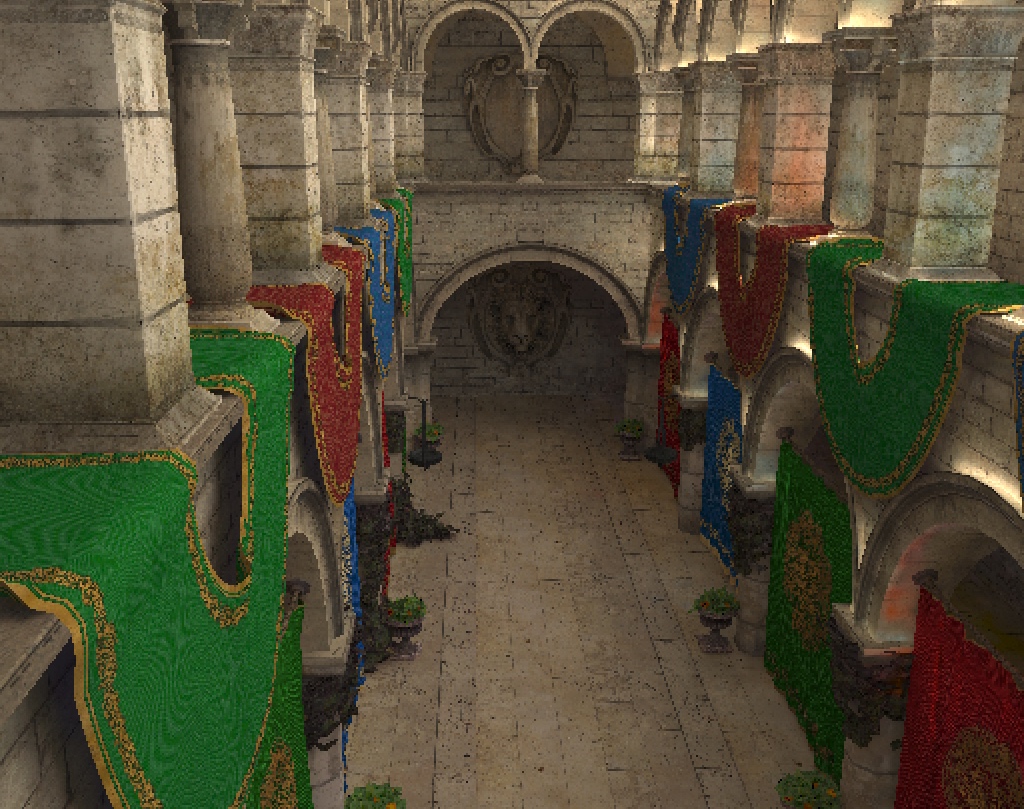}}
    \hfill
    \subfloat[Sampled incident radiance $L_i$]{\includegraphics[width=0.48\linewidth]{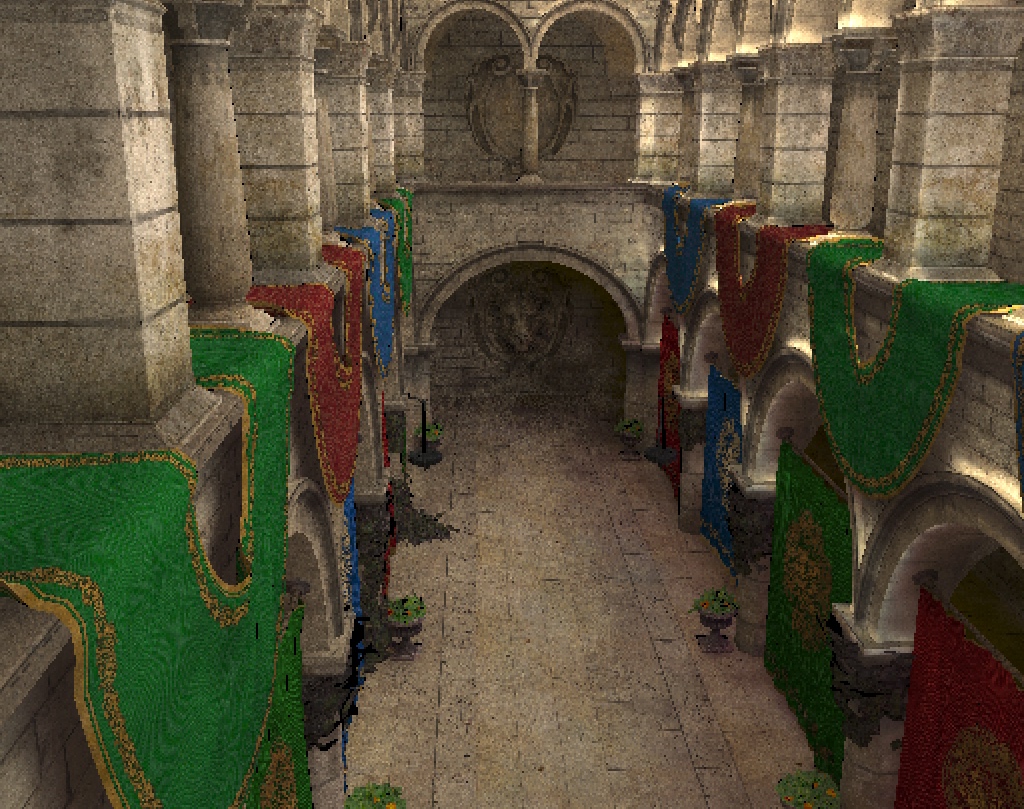}}
    \caption{Learning incident radiance at the same training budget reduces the training progression compared to training on comparatively smooth irradiance. Additionally, since incident radiance is not pre-integrated, it requires several network invocations (64 samples per pixel in the right image) which can be seen by the remaining sampling variance.}
    \label{fig:incoming_vs_irradiance}
\end{figure}

\begin{figure}[H]
    \captionsetup[subfigure]{labelformat=empty}
    \centering
    \subfloat[Incident Radiance]
    {\includegraphics[width=0.32\linewidth]{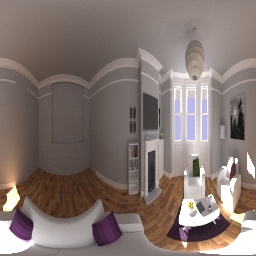}}\hfill
    \subfloat[Irradiance $E(x, \cdot)$]{\includegraphics[width=0.32\linewidth]{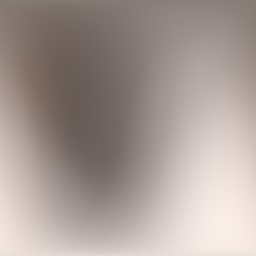}}\hfill
    \subfloat[Our $E_\theta(x, \cdot)$]{\includegraphics[width=0.32\linewidth]{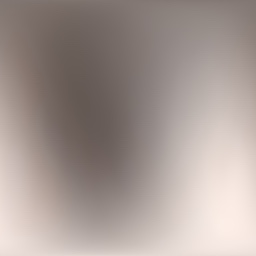}} \hfill
    \caption{Comparison of the ground truth indirect irradiance to point-sampling our Neural Irradiance Volume $E_\theta$ on the spherical domain. NIV represents the directionally continuous irradiance with high fidelity. The incident radiance is shown with the same spherical mapping for illustrative purposes.}
    \label{fig:sh_vs_neural}
\end{figure}

Changing the quantity NIV encapsulates from irradiance $E_\theta$ to the incoming radiance ${L_i}_\theta$ highlights the importance of learning a pre-integrated quantity. Figure~\ref{fig:incoming_vs_irradiance} provides a qualitative example, showing that when using incoming radiance, the amount of training data and network capacity required is higher than when using the pre-integrated irradiance. This is intuitive, as irradiance is a directionally smoother quantity, as illustrated in Figure~\ref{fig:sh_vs_neural}.
Nevertheless, in certain applications, learning the incident radiance directly could be advantageous---for instance, to perform importance sampling proportional to the learned incoming radiance.

\section{Lower-end Devices}
\begin{table}[H]
\caption{The performance of NIV is primarily bound by the GPU's  matrix multiplication throughput. In the table below, we evaluate our method on an older-generation RTX 2080 Ti. Although this hardware increases runtime costs by approximately a factor of four, our method remains within the real-time regime.}
\begin{adjustbox}{width=\linewidth}
    \hspace{-1.5\tabcolsep}
    \begin{tabular}{@{}ccccc@{}}
    \hline
    width & grid levels & full (ms) & half (ms) & memory (MB) \\ \hline
    16    & -           & 0.47   & 0.13   & 0.003       \\
    32    & -           & 0.47   & 0.14   & 0.01       \\
    64    & -           & 0.83   & 0.20   & 0.03        \\
    64    & 2           & 1.63   & 0.42   & 0.16        \\
    64    & 4           & 2.75   & 0.71  & 1.20        \\
    64    & 6           & 3.83   & 0.98   & 3.30        \\
    64    & 8           & 5.16   & 1.32   & 5.40        \\ \hline
    \end{tabular}
\end{adjustbox}
\end{table}

\begin{figure*}[ht!]
    \centering
    \begin{tikzpicture}
        \draw (0, 0) node[inner sep=0] {\includegraphics[width=\linewidth]{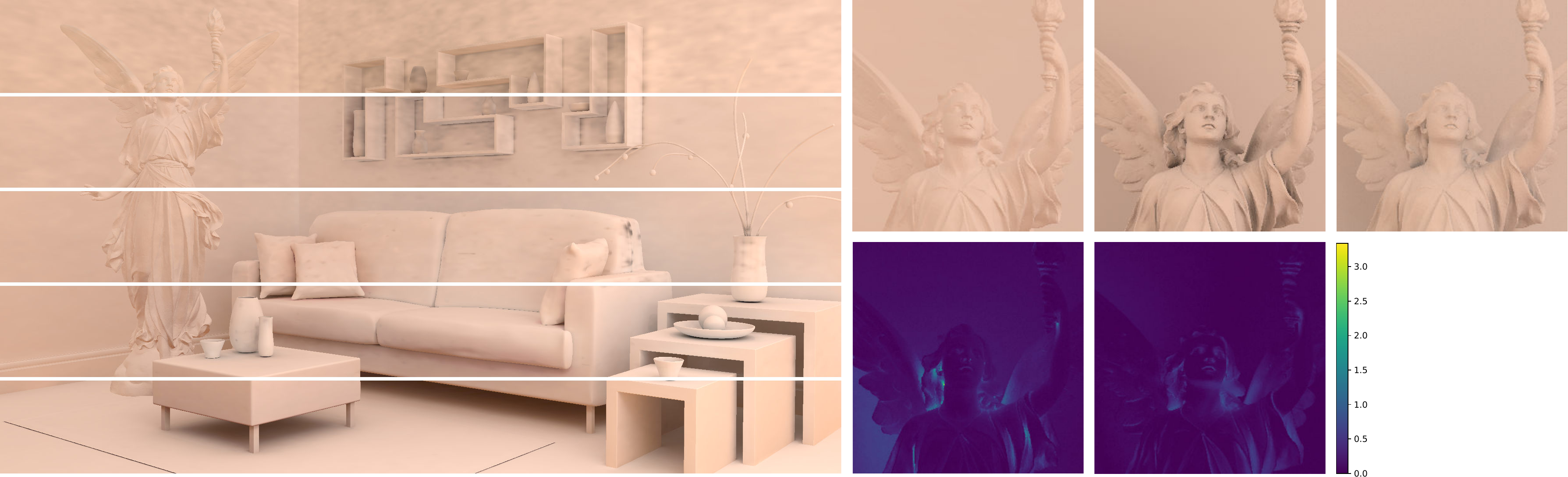}};
        \draw (-4, 3) node {Progressive Training};
        \draw (-8.4, 2.1) node {256};
        \draw (-8.4, 1.0) node {1024};
        \draw (-8.4, -0.1) node {4096};
        \draw (-8.4, -1.1) node {16384};
        \draw (-8.4, -2.1) node {65536};
        \draw (2.1, 3) node {NIV};
        \draw (4.75, 3) node {NIV + AO};
        \draw (7.5, 3) node {Reference};
    \end{tikzpicture}
    \caption{Progressive training of the indirect irradiance $E_\theta$ with a moving \textit{Lucy}-statue which is unseen during training. NIV represents the indirect irradiance at high quality, which can be seen from the sharp contact shadows on the known geometry. The impact of ambient occlusion is especially noticeable on \textit{Lucy}, since the mesh has strong self-occlusion.
    }
\end{figure*}

\section{Neural Variable Scenes}
\begin{figure}[H]
    \centering
    \includegraphics[width=\linewidth]{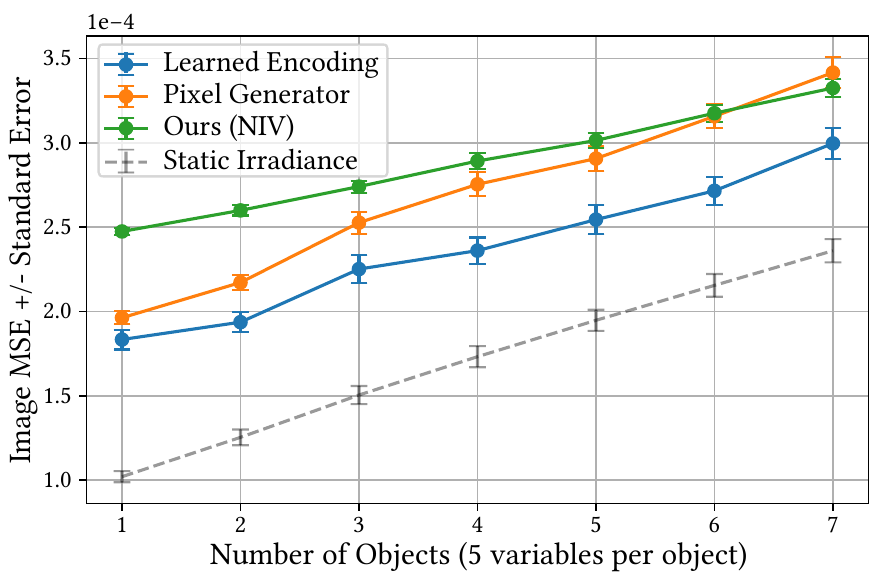}
    \caption{Increasing the capacity of neural variable scene methods (\cite{diolatzis2022active} $\sim 5$MB  and~\cite{su2024dynamic} $\sim 100$ MB) to the network architecture described in the respective works---as opposed to constraining them to the same capacity---significantly reduces their rendering error.
    NIV (using 1 MB) still performs similarly to variable scene methods while using significantly less capacity.}
    \label{fig:placeholder_supp}
\end{figure}

\begin{table}[H]
\centering
\caption{Runtime in milliseconds of variable scene methods at the same network capacity of 1 MB where all models use half-precision floats. The table denotes the compute time of a single forward pass on a 1920x1080 frame. Identity encoding~\cite{diolatzis2022active} hardly influences the runtime as the size of the base model is larger than the added trainable parameters. Learned encodings~\cite{su2024dynamic} add three 2-dimensional hash encodings per added variable which impacts the runtime significantly as the amount of variables goes up. NIV is not conditioned on the variable parameters, and has the same performance in each setting. The input buffer generation (position, normals) is assumed to be the same for each method, and is not included in the timings.}
\begin{tabular}{cccc}
\toprule
Variables & \cite{su2024dynamic} & \cite{diolatzis2022active} & Ours (NIV) \\
\midrule
5 & 38.38 & 116.35 & 0.67 \\
10 & 79.12 & 106.93 & 0.67 \\
15 & 126.14 & 118.62 & 0.67 \\
20 & 140.85 & 119.82 & 0.67 \\
25 & 194.80 & 123.37 & 0.67 \\
30 & 256.86 & 119.61 & 0.67 \\
35 & 306.11 & 123.65 & 0.67 \\
\bottomrule
\end{tabular}
\label{tab:neural_runtime_comparison}
\end{table}

\section{{Non-diffuse Objects}}

\begin{figure}[h!]
    \centering
    \includegraphics[width=0.6\linewidth]{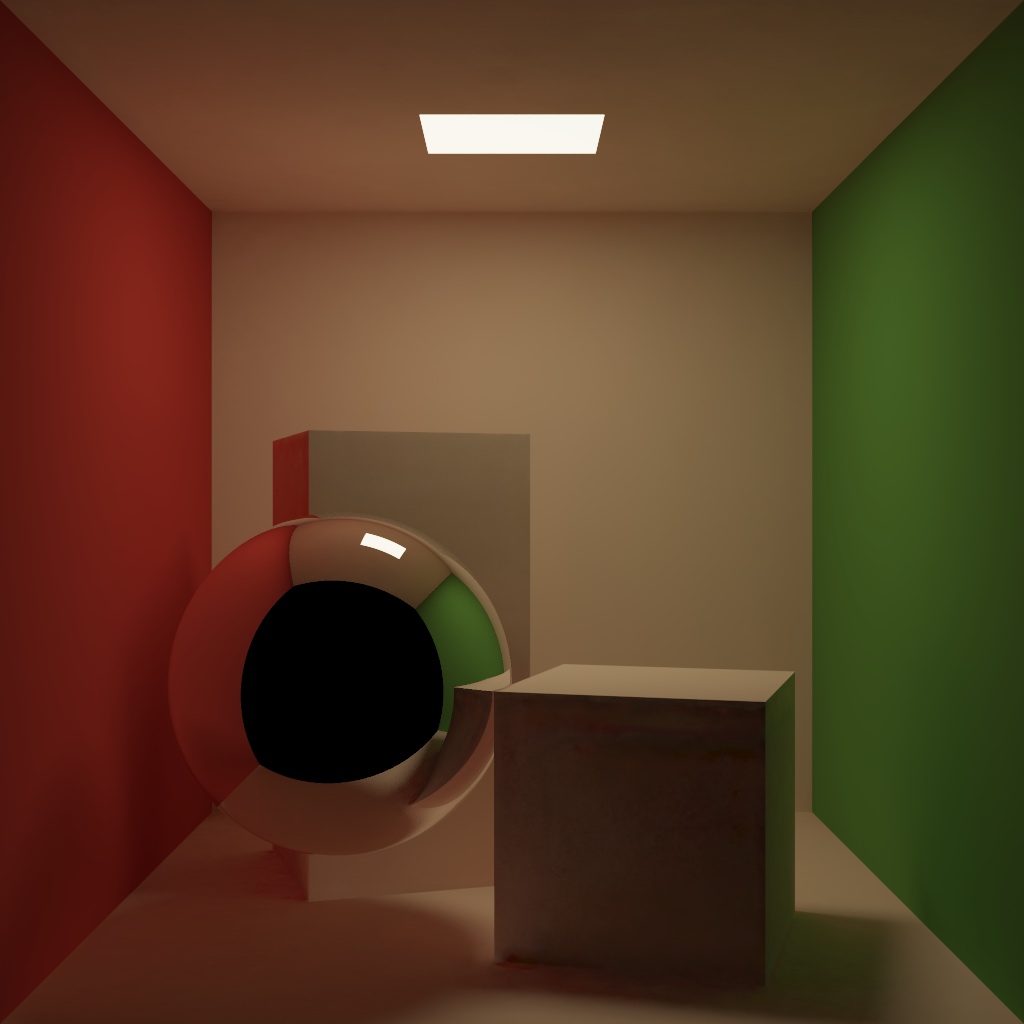}
    \caption{ NIV is trained on the static scene excluding the specular sphere. The modified render pass traces reflection rays from the sphere to query the NIV at diffuse intersection points.}
    \label{fig:specular_cbox}
\end{figure}

{
While NIV is designed for diffuse interactions, it can support specular or glossy materials by tracing secondary rays that query the NIV at the resulting diffuse intersection points. However, similar to other dynamic objects in a pre-baked volume, the specular object does not contribute to the global illumination of the scene; consequently, caustics and other light transport effects originating from the moving object are not captured. Figure~\ref{fig:specular_cbox} demonstrates a specular glass ball rendered within the Cornell Box.
}

 \section{{High Frequency Details}}

\begin{figure}[H]
    \centering
    \begin{subfigure}[b]{0.49\linewidth}
        \centering
        \includegraphics[width=\linewidth]{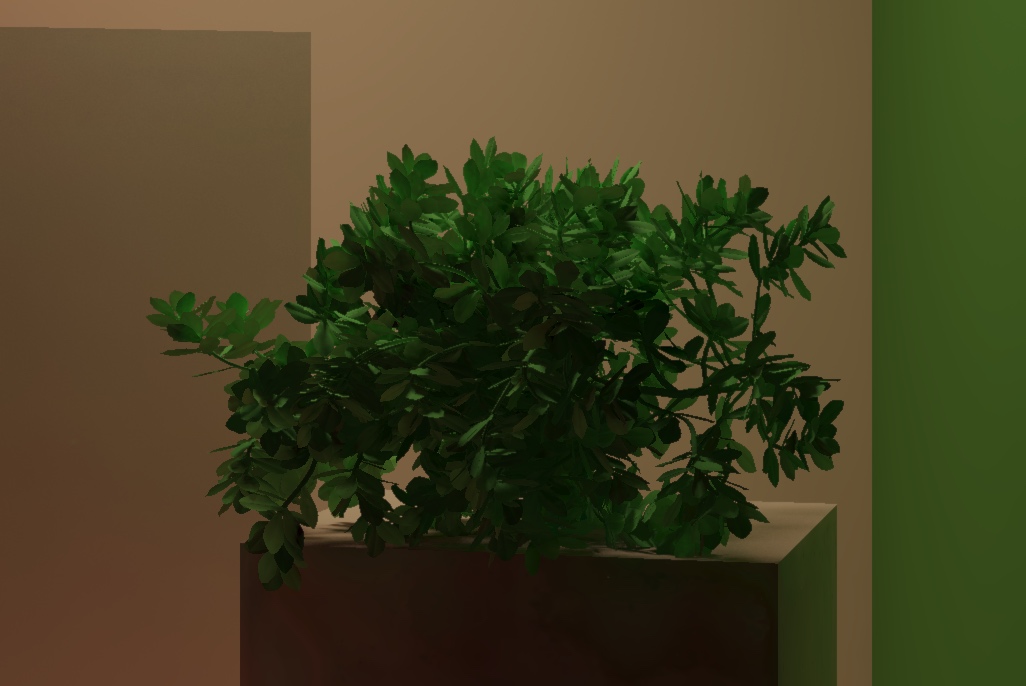}
        \caption{Ours (NIV, $\sim$5 MB)}
        \label{fig:method_supp}
    \end{subfigure}
    \hfill
    \begin{subfigure}[b]{0.49\linewidth}
        \centering
        \includegraphics[width=\linewidth]{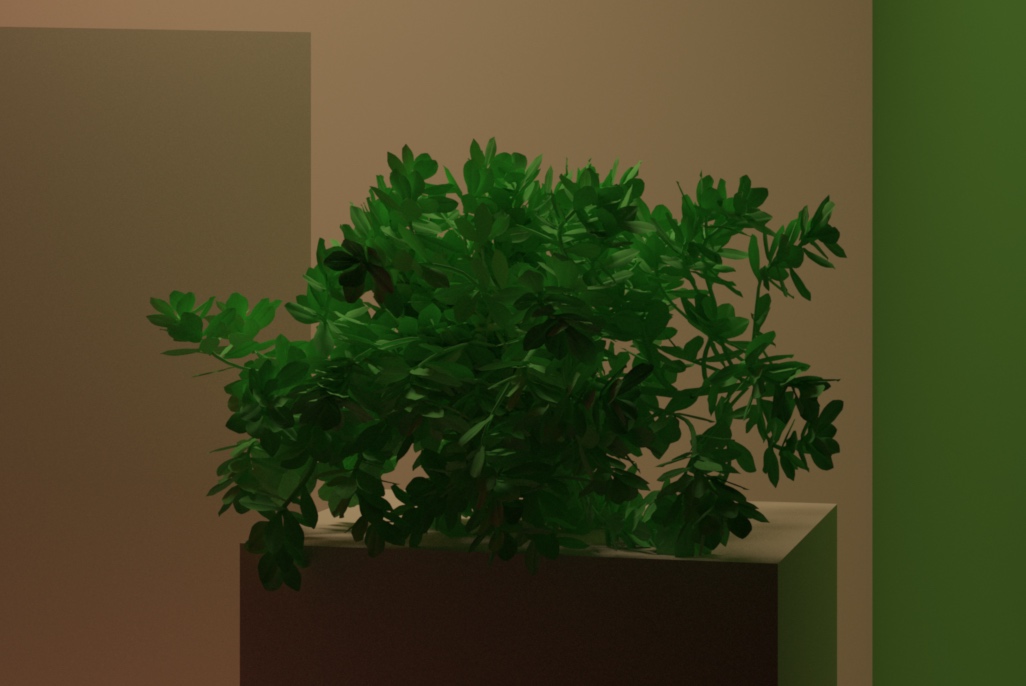}
        \caption{Reference}
        \label{fig:reference_supp}
    \end{subfigure}

    \caption{{ A modified Cornell Box containing a plant present during training. In such scenes with strong spatial variations, NIV does not match the reference solution as it fails to capture all complex higher order interreflections with its limited capacity.
    Increasing the encoding and model capacity resolves this issue at the cost of a higher inference time.
    }}
    \label{fig:main_label_supp}
\end{figure}

\section{Training Procedure Visualization}

\begin{figure}[h!]
    \centering
    \includegraphics[width=0.95\linewidth]{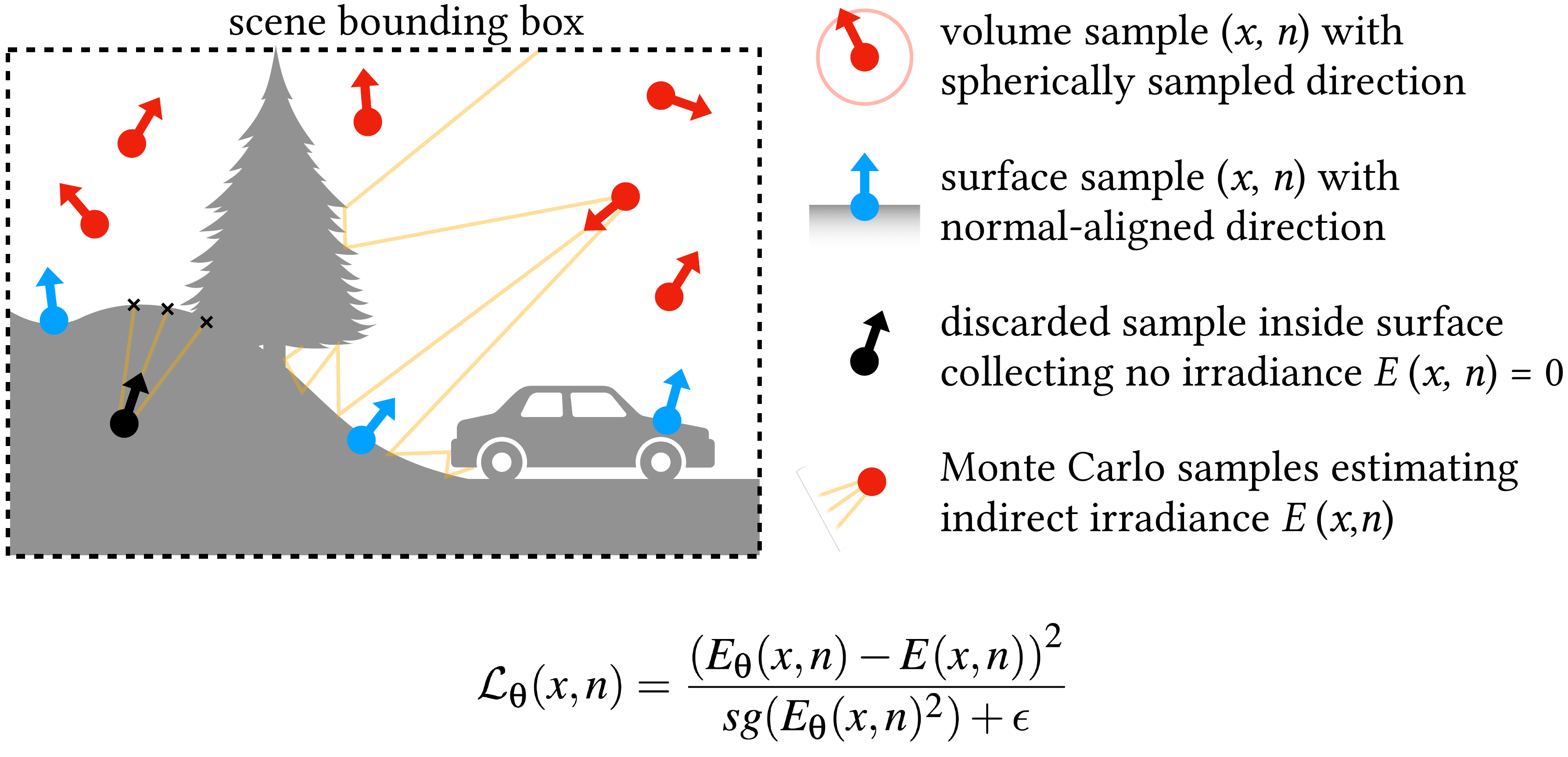}
    \caption{Visualization of the training procedure: position-direction pairs $(x, n)$ are uniformly sampled within the bounding-box of the scene, with direction sampled on the sphere. A fraction of points is sampled along surfaces, with the normal deterministically chosen as the surface normal. For each $(x,n)$ pair, the reference indirect irradiance $E$ is estimated via path tracing. Samples collecting null irradiance are discarded. The rest is use to compute the loss function and train the irradiance model $E_{\theta}$.}
    \label{fig:training_viz}
\end{figure}

\end{document}